\documentclass[twocolumn,preprintnumbers,superscriptaddress,amsmath,amssymb,10pt,a4paper]{revtex4-2}

\usepackage{geometry}
\usepackage{physics}
\usepackage{graphicx}
\usepackage{xcolor}

\usepackage{mathtools}
\usepackage{graphicx}
\usepackage{dcolumn}
\usepackage{bm}
\usepackage{makecell}
\usepackage{xcolor}
\usepackage{enumerate}
\usepackage[caption=false]{subfig}
\usepackage{float}
\usepackage{xr-hyper} 
\usepackage[unicode=true,pdfusetitle,
bookmarks=true,bookmarksnumbered=false,bookmarksopen=false,
 breaklinks=false,pdfborder={0 0 1},backref=false,colorlinks=true]
 {hyperref}
\hypersetup{linkcolor=blue,urlcolor=blue,citecolor=blue}
\usepackage{natbib}
\usepackage{physics}
\usepackage{placeins}
\usepackage{titlesec}
\usepackage{multirow}

\newcommand{\avg}[1]{\left\langle{#1}\right\rangle}
\newcommand{\mbf}[1]{\mathbf{{#1}}}

\newcommand{\be}{\begin{equation}}
\newcommand{\ee}{\end{equation}}
\newcommand{\bd}{\begin{displaymath}}
\newcommand{\ed}{\end{displaymath}}
\newcommand{\BE}{\begin{eqnarray}}
\newcommand{\EE}{\end{eqnarray}}

\newcommand{\bu}{\ensuremath{\mathbf{u}}}

\newcommand{\bx}{\ensuremath{\mathbf{x}}}
\newcommand{\by}{\ensuremath{\mathbf{y}}}
\newcommand{\bz}{\ensuremath{\mathbf{z}}}

\newcommand{\bA}{\ensuremath{\mathbf{A}}}

\newcommand{\st}{\text{st}}

\newcommand{\bs}{\ensuremath{\mathbf{s}}}

\newcommand{\nocontentsline}[3]{}
\let\oldaddcontentsline=\addcontentsline
\let\addcontentsline=\nocontentsline

\begin{document} 

\title{Estimating transmission noise on networks from stationary local order}

\author{Christopher R. Kitching}
\email{christopher.kitching@manchester.ac.uk}
\affiliation{Department of Physics and Astronomy, School of Natural Sciences, The University of Manchester, Manchester M13 9PL, UK}

\author{Henri Kauhanen}
\email{henri.kauhanen@uni-konstanz.de}
\affiliation{Department of
Linguistics, University of Konstanz, Universitätsstraße 10, 78464 Konstanz,
Germany}

\author{Jordan Abbott}
\email{jordanabbott2013@hotmail.co.uk}
\affiliation{Department of Physics and Astronomy, School of Natural Sciences, The University of Manchester, Manchester M13 9PL, UK}

\author{Deepthi Gopal}
\email{deepthi.gopal@lingfil.uu.se}
\affiliation{Institutionen f\"or lingvistik och filologi, Uppsala Universitet, 751 26 Uppsala, Sweden}

\author{Ricardo Bermúdez-Otero}
\email{ricardo.bermudez-otero@manchester.ac.uk}
\affiliation{Department of Linguistics and English Language; School of Arts, Languages, and Cultures; The University of Manchester, Manchester M13 9PL, UK}

\author{Tobias Galla}
\email{tobias.galla@ifisc.uib-csic.es}
\affiliation{Instituto de F\'isica Interdisciplinar y Sistemas Complejos, IFISC (CSIC-UIB), Campus Universitat Illes Balears, E-07122 Palma de Mallorca, Spain}

\date{\today}

\begin{abstract}
In this paper we study networks of nodes characterised by binary traits that change both endogenously and through nearest-neighbour interaction. Our analytical results show that those traits can be ranked according to the noisiness of their transmission using only measures of order in the stationary state. Crucially, this ranking is independent of network topology. As an example, we explain why, in line with a long-standing hypothesis, the relative stability of the structural traits of languages can be estimated from their geospatial distribution. We conjecture that similar inferences may be possible in a more general class of Markovian systems. Consequently, in many empirical domains where longitudinal information is not easily available the propensities of traits to change could be estimated from spatial data alone.
\\~\\
\end{abstract}

\maketitle
Many complex systems can be represented as networks of nodes, each characterised by a set of traits which can change endogenously and through interaction. Examples are abundant in evolutionary biology, gene regulatory systems, and the dynamics of languages \cite{huson, alon, beckner,baxter}. The spreading of opinions or diseases are further instances of noisy transmission processes on networks \cite{castellano,kiss}.  

We may naturally wish to estimate the relative propensities of different traits to change. The most direct way is to resort to longitudinal data: in population genetics, for example, cladistics and genetic sequencing have enabled the reconstruction of phylogenetic trees reaching to the beginnings of evolutionary time~\cite{hey_machado}. More indirect approaches are, however, possible \cite{Dettmer2018}. Mutation rates can also be estimated using mechanistic models together with summary statistics observed in natural populations \cite{hey_machado, Nachmann, Foster, Rosche}. In fact, these indirect methods offer the only avenue in fields where access to longitudinal information is restricted: the historical evolution of languages, for instance, can be inferred only to a very shallow temporal depth. 

Against this background, Greenberg surmised fifty years ago that the geospatial distribution of the structural traits of languages carries useful information about those traits' temporal dynamics \cite{Greenberg1974}. Motivated by a recent implementation of this intuition \cite{kauhanen2021geospatial}, we suggest that Greenberg's hypothesis applies to Markovian processes more widely. One can conceive of many such processes as involving the interplay of transmission noise and a tendency to order. We propose that the relative strength of these two, the noise-to-order ratio, can be estimated from a small set of measurable quantities in the stationary state. We call this the {\em transmission-noise conjecture}. 

In this paper we show that the conjecture does indeed hold in networked systems with binary traits that change both
spontaneously within nodes and through nearest-neighbour interaction. Moreover, detailed knowledge of the network is not needed to rank traits according to their noise-to-order ratios.

\medskip
{\em Model.} Our model describes $N$ nodes $i$ of an undirected network. At time $t$, each node is in one of two states, indicated by variables $s_i(t)=\pm 1$. In the context of horizontal gene transfer, the binary state could stand for the presence or absence of a mutation in an organism. In a model of opinion dynamics the two states could represent two distinct views on a particular question.
\begin{figure}[t]
    \centering
    \includegraphics[scale=0.16]{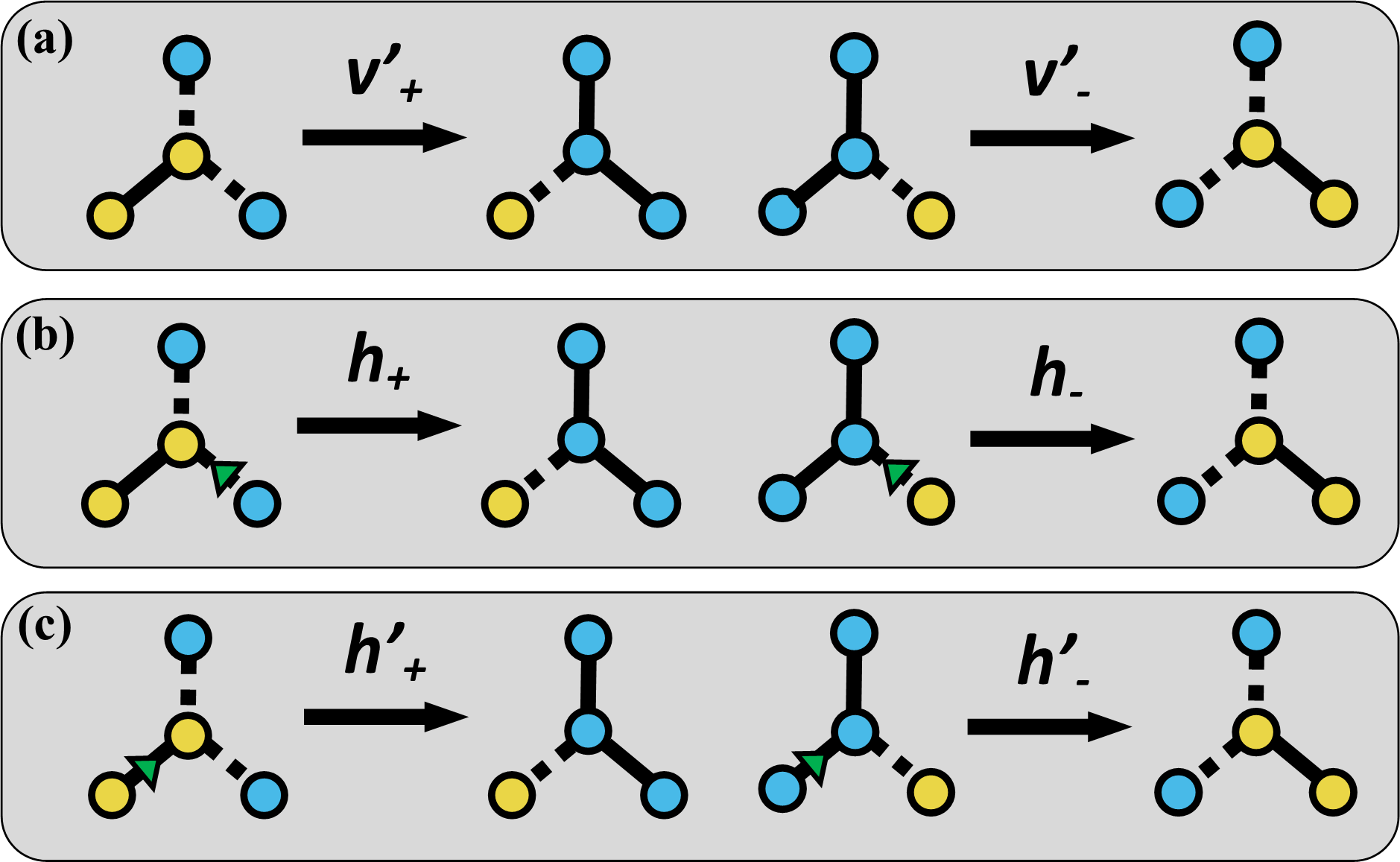}
    \caption{Illustration of the model dynamics. We show small networks of four nodes which possess (blue) or lack (yellow) a specific trait. Links that connect nodes in the same/different state are solid/dashed. The focal node is the node being considered for update. Panel (a): node changes state endogenously. Panel (b): node correctly copies a neighbour. Panel (c): node erroneously copies a neighbour.}
    \label{fig main: dynamics}
\end{figure}

The system evolves through several processes. (1) The state of a node can change endogenously from $s$ to $-s$ with rate $v_s'$, as shown in Fig.~\ref{fig main: dynamics}(a). This reflects transmission errors in time, for example from parent to offspring. (2) A node can change state by faithfully copying the state of a neighbour. The rate coefficients for these events are $h_\pm$ [Fig.~\ref{fig main: dynamics}(b)]. (3) Finally, a node can change state due to transmission error in an unfaithful copying event [Fig.~\ref{fig main: dynamics}(c)], this occurs with rate coefficients $h_\pm'$. A full definition of the dynamics can be found in the Supplemental Material (SM) \cite{supp}.

 Our starting point is deliberately general. The events in Fig.~\ref{fig main: dynamics} encompass all networked models with binary states, in which nodes can flip either spontaneously, or through pairwise interaction with one nearest neighbour. This generalises the model in \cite{kauhanen2021geospatial} as well as instances of the so-called `noisy voter model' \cite{granovsky1995noisy, carro2016noisy, peralta2018stochastic}. The subsequent analysis assumes $h_++h_-'=h_-+h_+'$. This is a mild requirement, indicating that nodes do not preferentially interact with nodes in a particular state \cite{supp}. We stress that, in contrast to most existing literature on the voter model, spontaneous flipping can occur asymmetrically in our model ($v_+'\neq v_-'$), and that the copying of a state from a neighbour can be subject to error ($h_+'\neq 0$, $h_-'\neq 0$). 

\medskip
{\em Summary statistics for local order.} We characterise the stationary state using two summary statistics. One is the trait frequency, $\rho$. This is the proportion of nodes which possess that trait (i.e. nodes with $s_i=+1$). The second is the proportion, $\sigma$, of links in the network that connect two nodes in opposite states. In line with voter model terminology we will call these active links \cite{vazquez2008analytical, REDNER2019}.

The average stationary trait frequency is given by
\be \label{eq main: steady-state rho}
    \avg{\rho}_{\rm st} = \frac{h_{+}'+v_{+}'}{h_{+}'+h_{-}'+v_{+}'+v_{-}'}. 
\ee
We can show that this expression is valid for any undirected network \cite{supp}.

We also define the ratio
\be \label{eq main: H}
    \mathbb{H}\equiv \frac{\sigma}{2\rho(1-\rho)}. 
\ee
If, for a given trait frequency $\rho$, traits were distributed randomly across the nodes, without correlations, one would have $\sigma=2\rho(1-\rho)$, and thus $\mathbb{H}=1$. When $\mathbb{H}<1$ there are fewer active links than random. Thus, $\mathbb{H}$ quantifies the amount of scatter in the system relative to a random configuration with the same value of $\rho$. We note that $\mathbb{H}$ can take values larger than one for configurations in which the states of neighbouring spins are anti-correlated \cite{supp}.

\begin{figure}[t]
    \centering
    \captionsetup{justification=justified}
    \includegraphics[width=0.48\textwidth]{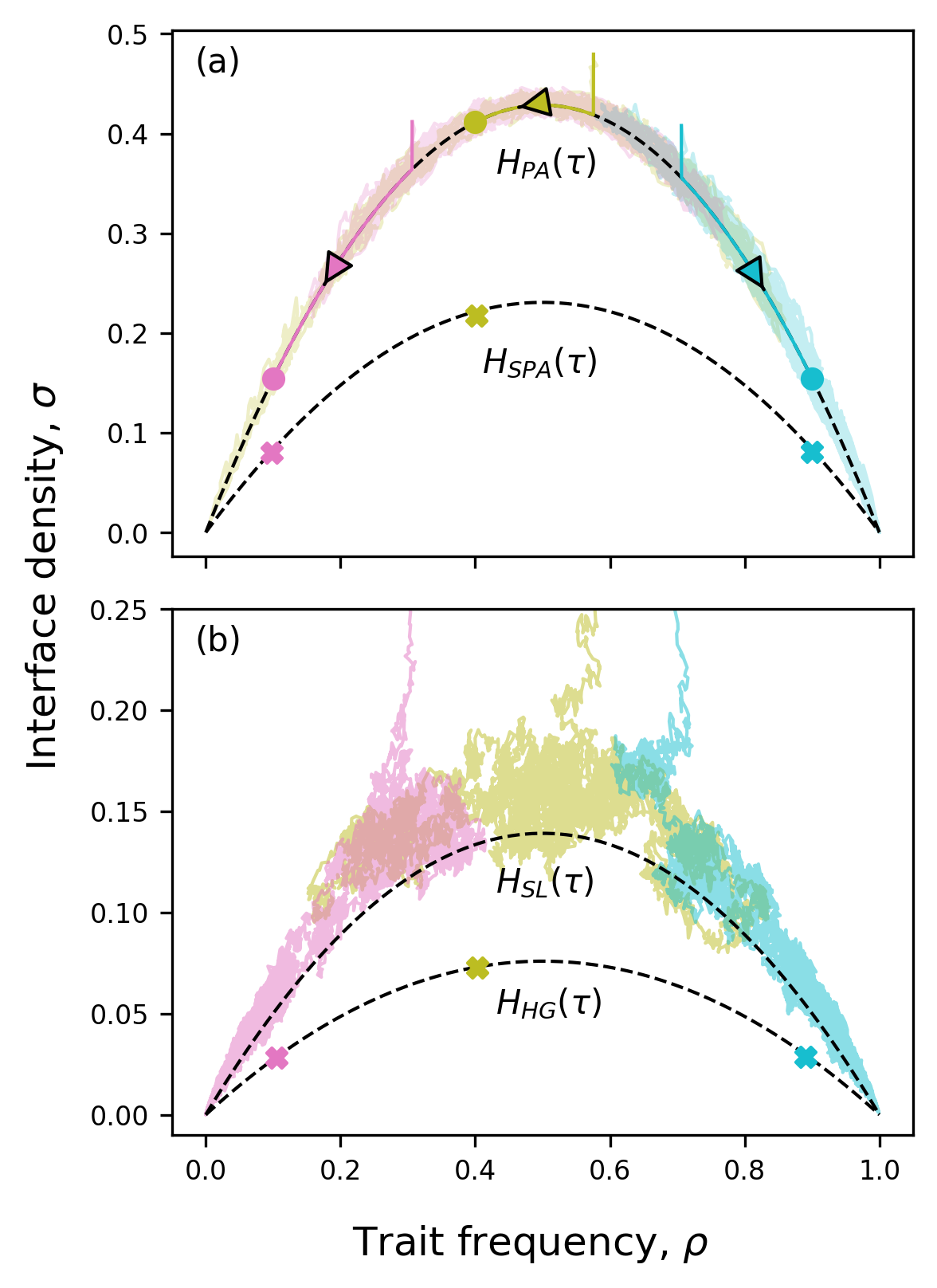}
    \caption{Model dynamics in the plane spanned by trait frequency $\rho$ and density of active interfaces $\sigma$. (a) is a Barab\'asi--Albert network ($N=2,500$), (b) is the periodic square lattice ($N=40,000$). The wiggly lines show different realisations in simulations. Each line is for a different set of model parameters, but with a common value of $\tau=1\times 10^{-3}$ in (a), and $\tau=1\times 10^{-4}$ in (b) [Eq.~(\ref{eq:tau})]. The solid lines in (a) show the trajectories within the pair approximation, derived in the limit $N\to\infty$ [Eqs.~(\ref{eq supp: complete network analytic magnetisation}) and (\ref{eq supp: pair approximation interface differential})]. The fixed points of these equations are shown by filled circles. The crosses in each panel indicate the points $(\avg{\rho}_{\st}, \avg{\sigma}_{\rm st})$ for the different parameter sets. Dashed lines are parabolas of constant $\mathbb{H}$, obtained from different analytical results, as indicated.}
        \label{fig main: parabolas}
\end{figure}

The typical time course from  simulations of the model is illustrated in the $(\rho,\sigma)$-plane in Fig.~\ref{fig main: parabolas}(a). As indicated by the wiggly lines, the system approaches a parabolic curve of constant $\mathbb{H}$ quickly, and then fluctuates in the region near the parabola. This has previously been pointed out for voter models in \cite{vazquez2008analytical, carro2016noisy}. For finite systems, the point defined by the time averages $\avg{\rho}_{\rm st}$ and $\avg{\sigma}_{\rm st}$ in the stationary state may not lie on this parabola. This will be discussed further below. 

\medskip
{\em Transmission noise determines local order.} Our aim is to demonstrate the validity of the transmission noise conjecture for our model. 

As a first step  we use the correlation function of spin states at different nodes to show that the stationary density of active interfaces $\avg{\sigma}_{\rm st}$ can be expressed in terms of the stationary trait frequency $\avg{\rho}_{\rm st}$ and the parameter combination \cite{supp}
\begin{equation}\label{eq:tau}
    \tau \equiv \frac{(h_{+}'+h_{-}')+(v_{+}'+v_{-}')}{\frac{1}{2}\Big[(h_{+}+h_{-})-(h_{+}'+h_{-}')\Big]}.
\end{equation}
This holds for any network. Further, for a wide range of networks we have  $\avg{\sigma}_{\rm st}=2H(\tau) \avg{\rho}_{\rm st} \left(1-\avg{\rho}_{\rm st}\right)$, where  $H(\tau)$ is dependent upon the network. Importantly, the model parameters enter only through the combination $\tau$. In simulations we confirm that $\avg{\mathbb{H}}_{\rm st}\approx H(\tau)$, provided $\tau$ is above some cutoff set by the network size \cite{supp}. 

The numerator in Eq.~(\ref{eq:tau}) describes the average rate of events in the system starting from a fully ordered state ($\rho=0$ or $\rho=1$). The only state changes that are possible in such a configuration are those due to transmission noise. Therefore the numerator can be interpreted as a strength of transmission noise. The denominator in the expression for $\tau$ is the rate with which correlations build up between neighbouring spins, starting from a random initial condition with $\rho=1/2$ \cite{supp}. Thus, we can think of $\tau$ as a noise-to-order ratio, and hence a measure of instability \footnote{The quantity $\tau$ can take negative values. This describes systems with anti-correlations between neighbouring spins. For details see Sec.~\ref{appendix: tau} in the SM}. 

This interpretation, together with the observation that local order is set by $\tau$ (i.e., $H$ is a function of $\tau$), confirms the transmission noise conjecture in our model.

\medskip
{\em Inference of $\tau$ from observed data: analytical results.} We now turn to the question of inferring the noise-to-order ratio $\tau$. This is possible if $\avg{\mathbb H}_{\rm st}$ can be estimated from observations, and, if the functional form of $H(\tau)$ can be calculated either exactly or as an approximation. We now discuss how to do the latter, and then turn to an empirical example.

The pair approximation (PA) is a standard tool for the analysis of interacting dynamics on networks. The method is known to work well on uncorrelated networks \cite{vazquez2008analytical,carro2016noisy,peralta2018stochastic,pugliese}. The main result of a PA analysis is a set of coupled differential equations for $\avg{\rho(t)}$ and $\avg{\sigma(t)}$, whose fixed point describes the stationary state. Adapting the calculation of \cite{carro2016noisy} we find \cite{supp}
\be \label{eq main: H_PA}
     H_{\rm PA}(\tau) = \frac{\mu-2 - \mu\tau+\sqrt{(\tau+1)^{2}\mu^{2}-4(\mu-1)}}{2(\mu-1)},
\ee
where $\mu$ is the mean degree of the network.

The conventional PA leading to Eq.~(\ref{eq main: H_PA}) assumes an infinite network. It is possible to compute the leading-order correction in the inverse network size adapting the stochastic pair approximation (SPA) \cite{peralta2018stochastic}. For our model, we find \cite{supp}
\be
    H_{\text{SPA}}(\tau)=H_{\text{PA}}(\tau)\left[1-\frac{\mu_{2}}{N\mu^{2}}\frac{\tau+H_{\text{PA}}(\tau)}{\tau}\right]+{\cal O}(N^{-2}),
    \label{eq main: H_SPA}
\ee
where $\mu_{2}$ is the second moment of the degree distribution.  

We have further used the annealed approximation (AA) \cite{carro2016noisy} for our model. Similar to the SPA, this technique targets large uncorrelated networks. The resulting expression $H_{\rm AA}(\tau)$ involves higher-order moments of the degree distribution, and is reported in the SM \cite{supp}.

In addition to these approximations, we can calculate  $H(\tau)$ exactly for networks which have the following properties for all fixed integers $\ell$: The number of distinct walks of length $\ell$ starting at any node is the same. Additionally, the fraction $\Omega^{(\ell)}$ of those walks ending at the starting point is also the same for all nodes. For such `homogeneous' networks (HG) we show that \cite{supp}
\begin{equation}\label{eq main: H_HG}
    H_{\rm HG}(\tau) = \frac{1+\tau}{\sum_{\ell=0}^{\infty}\Omega^{(\ell)}/(1+\tau)^{\ell}}-\tau. 
\end{equation}
In practice this expression approximates simulation results well for networks  which have a sufficiently tight degree distribution. Examples are shown in the SM.

Closed-form expressions for the $\Omega^{(\ell)}$ can be found for some homogeneous networks, including finite complete networks, and infinite Bethe and $d$-dimensional hyper-cubic lattices \cite{supp}. In particular, the expression $H_{\rm SL}(\tau)$ for infinite square lattices (SL) reported in \cite{kauhanen2021geospatial} for a restricted set of parameters is valid for the more general setup in Fig.~\ref{fig main: dynamics}.
  
We find that $\Omega^{(\ell)}\rightarrow 1/N$ for large $\ell$ for a range of networks, including Barab\'asi-Albert networks. When an exact calculation of $\Omega^{(\ell)}$ is not possible, we can thus enumerate walks up to some cutoff $\ell=L$, and set $\Omega^{(\ell)}= 1/N$ for $\ell > L$ in Eq.~(\ref{eq main: H_HG}). As shown in the SM this produces satisfactory results.

\medskip
{\em Interpretation and test in simulations.} Figure~\ref{fig main: parabolas}(a) illustrates the relation between the different approximations for Barab\'asi--Albert networks. We show simulations for different parameter combinations, but keeping $\tau$ in Eq.~(\ref{eq:tau}) fixed. The trajectories in $(\rho,\sigma)$-space obtained from the PA for infinite systems converge to fixed points on the parabola $\sigma=2H_{\rm PA}(\tau)\rho(1-\rho)$. Trajectories from simulations of finite systems fluctuate about these fixed points and remain near the parabola. The time averages in the stationary state, $\avg{\rho}_{\rm st}$ and $\avg{\sigma}_{\rm st}$, are indicated by crosses in Fig.~\ref{fig main: parabolas}(a), and lie below the parabola set by $H_{\rm PA}(\tau)$. This deviation is a consequence of finite-size fluctuations. As seen in the figure, the SPA captures this effect to good accuracy.

A similar effect is found for square lattices [Fig.~~\ref{fig main: parabolas}(b)]. Trajectories for finite systems fluctuate near the parabola $\sigma=2H_{\rm SL}(\tau)\rho(1-\rho)$. The time average $(\avg{\rho}_{\rm st},\avg{\sigma}_{\rm st})$, however, is not on this parabola, but is captured by Eq.~(\ref{eq main: H_HG}).
 
\begin{figure}[t]
    \centering
    \captionsetup{justification=justified}
    \includegraphics[width=0.48\textwidth]{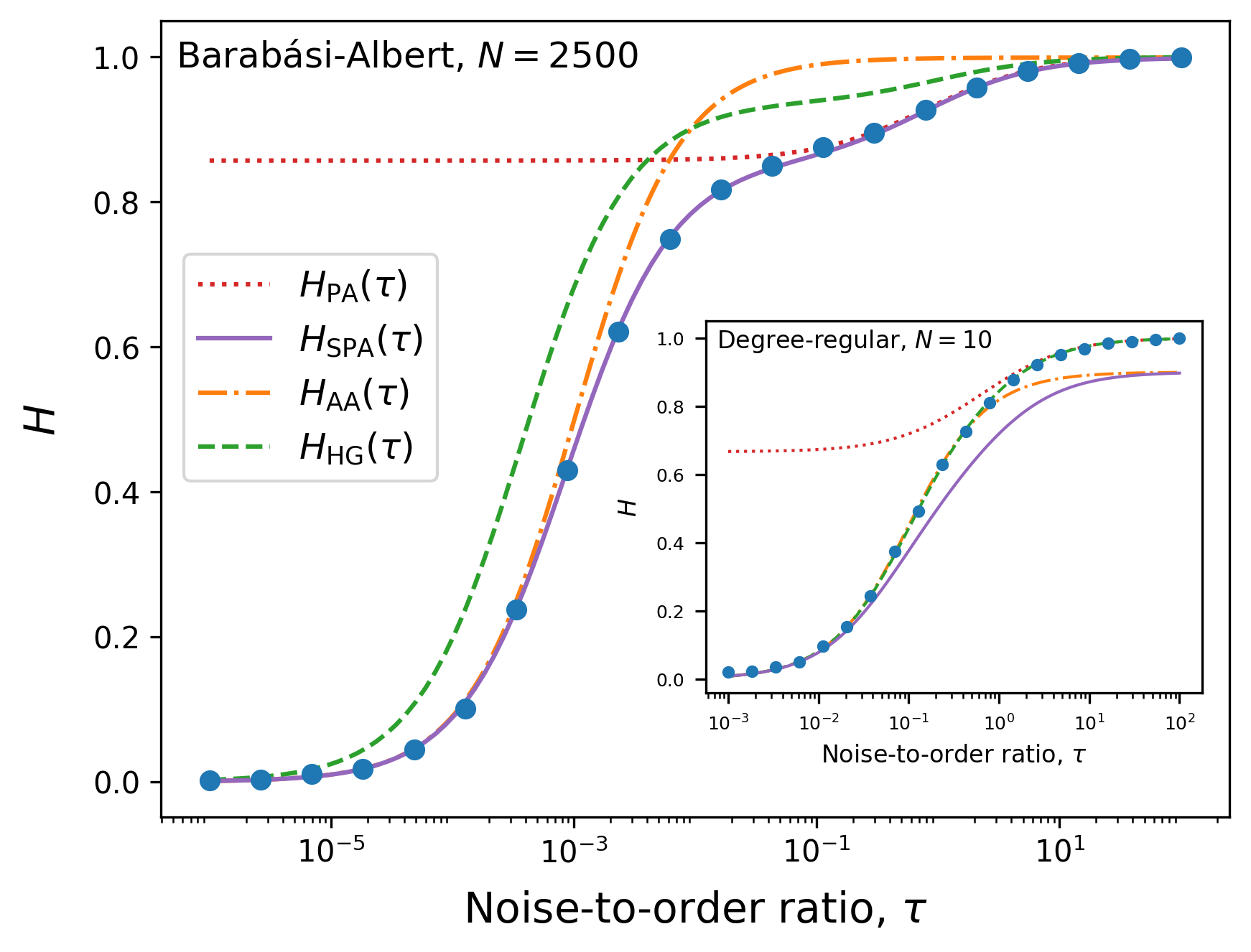}
    \caption{Order parameter $H(\tau)$ within the different theories discussed in the text. Main axis: Barab\'asi--Albert network ($N=2,500$, mean degree $\mu\approx 8$). Insert: Degree-regular network ($N=10$, $\mu=4$). Lines show the analytical results as indicated. Blue markers are from simulations, averaged over $100-10000$ independent samples in the steady state. Model parameters for a given choice of $\tau$ can be randomly generated via the algorithm described in the SM.}
    \label{fig main: theory comparison}
    \end{figure}
    
We test the different approximations further in Fig.~\ref{fig main: theory comparison}, focusing on Barab\'asi--Albert networks of size $N=2500$, and a small degree-regular network ($N=10$). The PA fails for $\tau\to 0$, because complete ordering can only occur on finite networks. We find $\lim_{\tau\to 0} H_{\text{PA}}= \frac{\mu-2}{\mu-1}$, describing long-lived partially ordered states of the voter model \cite{vazquez2008analytical}. For the Barab\'asi--Albert network the assumption of homogeneity is not valid and Eq.~(\ref{eq main: H_HG}) is inaccurate. The SPA, on the contrary, describes simulations well in the main panel, but becomes inaccurate for large $\tau$ on small networks (inset) due to finite-size effects beyond leading order. The result for homogeneous networks in Eq.~(\ref{eq main: H_HG}) on the other hand produces satisfactory results for all $\tau$. 

\medskip
{\em Stability of language traits.} The World Atlas of Language Structures (WALS) \cite{wals} contains geolocated data on 2,662 of the approximately 7,000 existing human languages. We extract $35$ binary traits, such as the presence or absence of a definiteness marker (e.g. the English definite article {\em the}). 

For each trait, we measure $\rho$ and, after constructing a putative interaction network, $\sigma$. To assess the uncertainty due to the incompleteness of WALS we use bootstrap sampling. This allows us to obtain a distribution for $\mathbb{H}=\sigma/[2\rho(1-\rho)]$ for each trait. We can use the different functional forms $H_X(\tau)$ to convert this into a distribution of $\tau_X$. The subscript $X$ denotes the different theories SPA, AA, SL or HG. 

\begin{figure}[t]
    \centering
    \captionsetup{justification=justified}
    \includegraphics[width=0.48\textwidth]{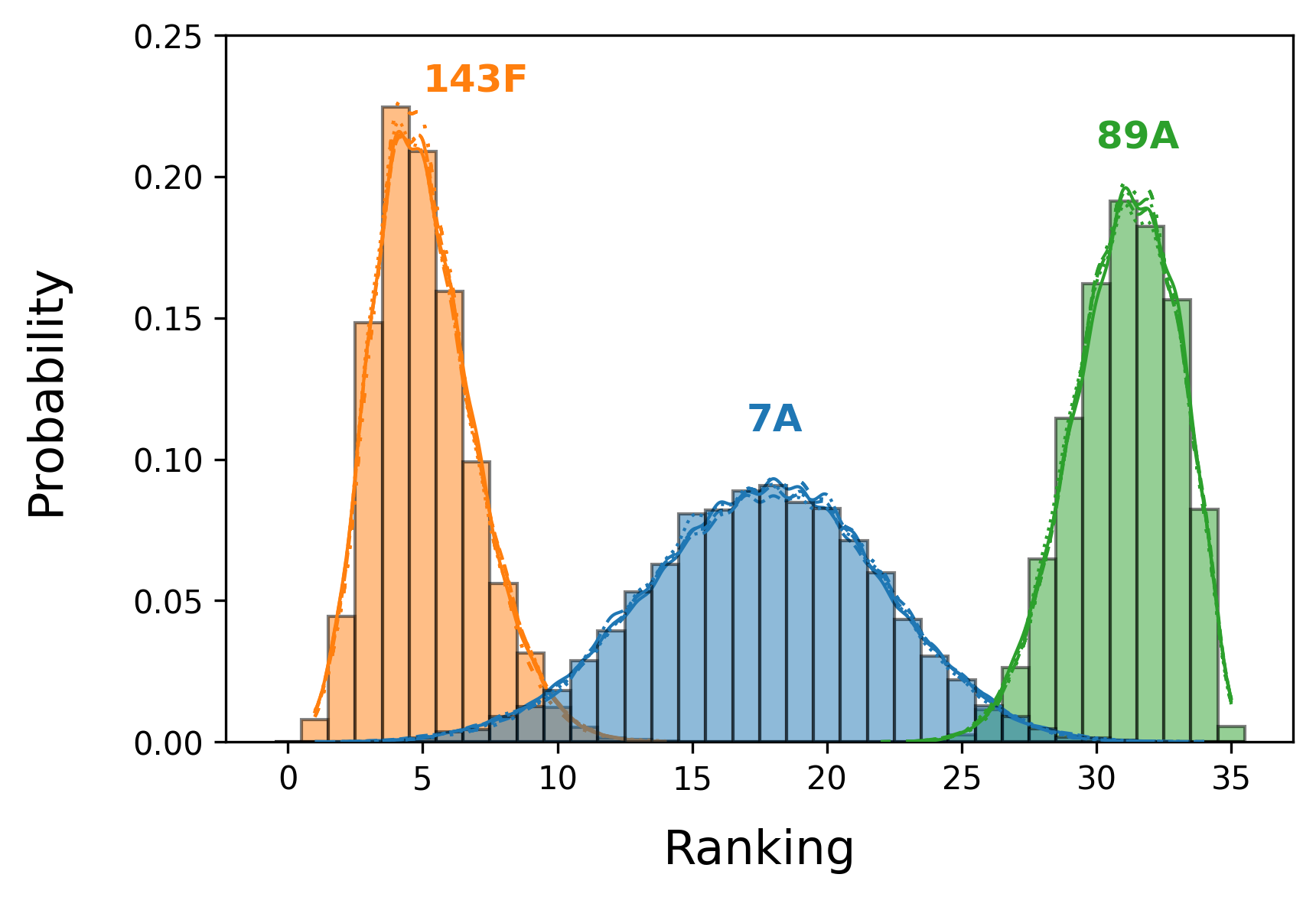}
    \caption{Rankograms for three linguistic traits 143F, 7A, and 89A from \cite{wals}, each obtained from $10,000$ bootstrap samples. For each trait we show rankograms obtained from $\mathbb{H}$ directly, and from $\tau$ obtained using the functions $H_{\rm AA}(\tau), H_{\rm SL}(\tau), H_{\rm HG}(\tau)$ and $H_{\rm SPA}(\tau)$. These are virtually indistinguishable from each other on the network. For clarity we show Gaussian kernel density estimates, for $H_{\rm HG}$ we show the full histogram. Further details are given in the SM along with rankograms for the remaining traits.}
    \label{fig main: rankogram}
    \end{figure}

In a second step we can, for a given $X$, rank-order the $35$ traits according to their inferred noise-to-order ratios $\tau_{X}$. In \cite{kauhanen2021geospatial} the ranking was based on median values of $\tau_{\rm SL}$. Here we go further and obtain the probability that the trait is ranked $r^{\rm th}$. This enables the construction of rankograms, as shown in Fig.~\ref{fig main: rankogram}, and the use of rank metrics \cite{supp}. The width of a trait's rankogram reflects the confidence we can have in its relative instability based on the data.

Strikingly, the results are independent of the functional form of $H(\tau)$ used to infer $\tau$. This robustness explains the surprising success of the stability estimates in \cite{kauhanen2021geospatial} despite the use of a square lattice geometry. Ranking on $\mathbb{H}
$ itself gives similar results \cite{supp}, thus confirming Greenberg's hypothesis that low scattering indicates relative stability, whereas highly scattered traits would be comparatively unstable.

{\em Discussion.} This paper has demonstrated the transmission-noise conjecture for two-state Markovian dynamics with nearest-neighbour interaction on networks. Traits can be ranked by stability from purely spatial information; neither longitudinal data nor detailed knowledge of the interaction network are required, opening up the prospect of applications in the many empirical fields where such information is not easily available. 

When more is known about the interaction network, an appropriate form of $H(\tau)$ can be selected, and our analytical results make it possible to infer values of $\tau$ for each trait. One can then not only rank traits, but also quantify their relative stability.

Further work should show how generally it is the case that the noise-to-order ratio in the dynamics of a complex system can be inferred from configurations in the stationary state.

\section*{Acknowledgements}
We acknowledge support from the Agencia Estatal de Investigaci\'on and Fondo Europeo de Desarrollo Regional (FEDER, UE) under project APASOS (PID2021-122256NB-C21, PID2021-122256NB-C22), the Mar\'ia de Maeztu programme for Units of Excellence, CEX2021-001164-M funded by  MCIN/AEI/10.13039/501100011033, ERC grant no. 851423, and the Government of the Balearic Islands CAIB, fund ITS2017-006 under project CAFECONMIEL (PDR2020/51). We further acknowledge an EPSRC studentship (grant NUMBER), and support by the Royal Society UK (APEX Award APX\textbackslash R1\textbackslash 211253). We thank Antonio Fern\'andez Peralta for useful discussions regarding the stochastic pair approximation.

\newpage

\setcounter{section}{0}		
\setcounter{page}{1}		
\setcounter{equation}{0}	
\setcounter{figure}{0}		
\setcounter{table}{0}		
\renewcommand{\thesection}{S\arabic{section}} 		
\renewcommand{\thepage}{S\arabic{page}} 			
\renewcommand{\theequation}{S\arabic{equation}}  	
\renewcommand{\thefigure}{S\arabic{figure}}  		
\renewcommand{\thetable}{S\arabic{table}}  

\onecolumngrid

{\begin{center}{\Large ------ Supplemental Material ------}\end{center}}
\setlength{\parskip}{-0.6pt}
\setlength{\parindent}{0pt}
\tableofcontents

\setlength{\parskip}{8pt}
\setlength{\parindent}{0pt}
\let\addcontentsline=\oldaddcontentsline
\let\nocontentsline=\oldaddcontentsline

\clearpage
\newpage
\section*{Overview}
In this supplement we provide the technical details of the results presented in the main paper. This document is structured as follows: 

Firstly, in Sec.~\ref{appendix: identities} we give a full definition of the model and some identities that will be used throughout the supplement in order to simplify the calculations.

The next several sections focus on deriving equations for the steady-state magnetisation and interface density on various topologies. In Sec.~\ref{appendix: complete graph} we focus on the model on complete networks. In Sec.~\ref{appendix: pair approximation} we then study the model on infinite uncorrelated networks using the pair approximation. This is where the noise-to-order ratio, $\tau$, is first introduced. In Sec.~\ref{appendix: annealed approximation} we then look at the model on finite uncorrelated networks using the annealed approximation. Sec.~\ref{appendix: square lattice} focuses on the model on an infinite square lattice. 

In Sec.~\ref{appendix: random walks} we present an approach based on counting walks. We are able to derive the steady-state magnetisation exactly for any network, and the steady-state interface density for homogeneous networks of any size. We present closed form solutions for a number of specific networks such as finite size complete networks, $d$-dimensional lattices, and infinite regular trees.

In Sec.~\ref{appendix: SPA} we use the stochastic pair approximation for our model, improving on the conventional pair approximation in Sec.~\ref{appendix: pair approximation}. This is valid for finite uncorrelated networks.

In Sec.~\ref{appendix: H and tau} we discuss the interpretations of the quantities $\mathbb{H}$, $H(\tau)$ and $\tau$, and the connection between them. This includes an analysis of cases with negative $\tau$.

In Sec.~\ref{appendix: linguistic analysis} we go into detail on how the theory can be applied to  linguistic data. This includes details about the bootstrapping method and ranking statistics that are presented in the main paper.

Finally, Sec.~\ref{appendix: proofs} contains miscellaneous proofs and further details of numerical algorithms. 

\newpage
\section{Full model definition and notation} \label{appendix: identities}

\subsection{Model definition} \label{appendix: model defn}
\subsubsection{Microscopic configurations of the model}
Throughout, we use the mathematical notion of networks, i.e, nodes connected by links which determine which nodes can interact. We only consider undirected networks.

We consider various `traits', and the nodes in the network can either process or lack a given trait. In this way, we assign node $i$ a property $s_i\in\{-1,1\}$. We will use physics terminology and call this a \textit{spin}. A node with $s_i=+1$ is referred to as `spin-up' or the node being in the `up-state'. This means that the node has the particular trait that we are considering. A node with $s_i=-1$, referred to as `spin-down' or the node being in the `down-state', means that the that the node lacks the particular trait we are considering. It is important to note that we only consider a single trait at any one time, there is no interaction between traits.

\subsubsection{Trait frequency and density of active links}
We adopt the voter model terminology in that links connecting nodes in states $s=+1$ and $s=-1$ nodes are referred to as \textit{active} links. Links that connect same spin nodes, i.e. $s=\pm1$ to $s=\pm1$ nodes are referred to as \textit{inactive} links. This convention from the voter model comes from the fact that the only existing processes in the voter model are error-free copying processes between neighbours. These can only occur between opposite spin nodes, hence those links are called active \cite{vazquez2008analytical}.

The two quantities of interest are the \textit{trait frequency}, $\rho$, and \textit{interface density}, $\sigma$. The trait frequency is defined as the proportion of nodes in state $s=+1$. Sometimes it is mathematically simpler to work in terms of the \textit{magnetisation}, $m$, defined as $m=\frac{1}{N}\sum_{i} s_{i}$, where we are summing over all nodes in the network. The magnetisation and trait frequency are related via $m=2\rho-1$. The interface density $\sigma$ is defined as the proportion of active links among all links in the network.

\subsubsection{Model dynamics}
There are two `modes' in which the dynamics can act:
\begin{enumerate}
    \item endogenous changes of a node,
    \item changes due to interactions between nodes.
\end{enumerate}
The first is modelled by the spontaneous flipping of a nodes' state. This is what is referred to as `noise' in conventional noisy voter models \cite{granovsky1995noisy, carro2016noisy}. However, in our model the rates for spontaneous flips from $+1$ to $-1$ and those in the reverse direction may differ from each other. In \cite{kauhanen2021geospatial} endogenous processes are refereed to as `vertical' events. In our model any vertical event leads to the state change of one spin, that is the spontaneous acquisition or loss of a trait in that node. This can be understood as `unfaithful' vertical transmission from one generation to the next, i.e., the next generation of the entity represented by the node incorrectly copies the state of the previous generation at that node. We do not include faithful vertical events, as these do not lead to any state changes.

In the second mode a spin may change state due to an interaction with a neighbouring node. This will be detailed below. In \cite{kauhanen2021geospatial} such processes are referred to as `horizontal' events. A copying process from a neighbour can be \textit{faithful} or \textit{unfaithful}. Faithful horizontal transmission means that the node correctly copies the state of the neighbour. Unfaithful horizontal transmission means that an error occurs in a copying event, so that one node adopts the state opposite to that of the neighbour it attempts to copy. 

Six model parameters are needed to define the above dynamics. They are defined as follows (rates associated with unfaithful events are indicated by a dash):
\begin{equation} \label{eq supp: model params}
\begin{aligned}
    v'_{+} &= \text{rate with which a node gains the trait through} \\
    &\quad \text{unfaithful vertical transmission}, \\
    v'_{-} &= \text{rate with which a node loses the trait through} \\
    &\quad \text{unfaithful vertical transmission}, \\
    h_{+}  &= \text{rate coefficient for events in which a node} \\
    &\quad \text{gains the trait through faithful horizontal transmission}, \\
    h_{-}  &= \text{rate coefficient for events in which a node} \\
    &\quad \text{loses the trait through faithful horizontal transmission}, \\
    h'_{+} &= \text{rate coefficient for events in which a node} \\
    &\quad \text{gains the trait through unfaithful horizontal transmission}, \\
    h'_{-} &= \text{rate coefficient for events in which a node} \\
    &\quad \text{loses the trait through unfaithful horizontal transmission}.
\end{aligned}
\end{equation}

The exact mathematical interpretation of these rates is as follows:

We focus on a node which is connected to $k$ other nodes, $n(t)$ of which are in the opposite state to that of the focal node at time $t$.
\par

If the focal node is the state $+1$, then the rate at which it flips to $-1$ is
\begin{align}
    T_{n,k,t}^- &= \underbrace{h_{-}\frac{n(t)}{k}}_{\substack{\text{Correctly copy absence} \\ \text{from a neighbour}}} + \underbrace{h_{-}'\frac{k-n(t)}{k}}_{\substack{\text{Incorrectly copy} \\ \text{presence as absence}}} + \underbrace{v_{-}'}_{\substack{\text{Spontaneously lose} \\ \text{existing trait}}}, \label{eq supp: dyn1}
\end{align}
where $n(t)$ is the number of neighbours of the focal node in the down-state.

If the focal node is in state $-1$ then the rate at which nodes flip to state $+1$ is
\begin{align}
    T_{n,k,t}^+ &= \underbrace{h_{+}\frac{n(t)}{k}}_{\substack{\text{Correctly copy presence} \\ \text{from a neighbour}}} + \underbrace{h_{+}'\frac{k-n(t)}{k}}_{\substack{\text{Incorrectly copy} \\ \text{absence as presence}}} + \underbrace{v_{+}'}_{\substack{\text{Spontaneously acquire} \\ \text{the trait}}}, \label{eq supp: dyn2}
\end{align}
noting that $n(t)$ is now the number of neighbours of the focal node in the up-state.

Fig.~\ref{fig main: dynamics} in the main paper shows a pictorial representation of these dynamics.

\subsection{Notation}
We will see that the six model parameters in Eq.~(\ref{eq supp: model params}) ultimately only appear in specific combinations. Here we define shorthands for these combinations in order to reduce the notation later on,
\begin{align} \label{eq:shorthands}
\begin{split}
    \alpha &= (h_{+}'-h_{-}')+(v_{+}'-v_{-}'), \\
    \beta &= (h_{+}'+h_{-}')+(v_{+}'+v_{-}'), \\
    \gamma_{-} &= h_{-}-h_{-}', \\
    \gamma_{+} &= h_{+}-h_{+}', \\
    \gamma &= \frac{1}{2}(\gamma_{+}+\gamma_{-}), \\
    \delta &= h_{+}'+v_{+}', \\
    \kappa &= h_{-}' + v_{-}', \\
    \eta &= (v_{+}'+v_{-}') + \frac{1}{2}\Big[(h_{+}+h_{-})+(h_{+}'+h_{-}')\Big].
\end{split}
\end{align}
Throughout most of the mathematical analysis we will make the restriction
\begin{align}
    \gamma_{+} = \gamma_{-}. \label{eq supp: gamma restriction}
\end{align}
This is required to proceed analytically. The restriction means that we assume
\be
h_++h_-'=h_- + h_+'.
\ee
The coefficients on the left are those for faithful and unfaithful copying from a node in the $+1$ state. The coefficients on the right are those for events in which faithful or unfaithful copying occurs from a node in state $-1$. Broadly speaking, the constraint means that in the horizontal processes, the total rate of copying from $+1$ nodes is equal to the total rate of copying from $-1$ nodes. 

\subsection{Relation of the present setup to the model in \texorpdfstring{\cite{kauhanen2021geospatial}}{}} \label{appendix: relation to SA}
In \cite{kauhanen2021geospatial} time is discrete and the model parameters are $q, p_I, p_E, p_I'$ and $p_E'$. More precisely a spin is chosen at random from all spins (with equal probability). Then a vertical event is attempted with probability $1-q$, or a horizontal event with probability $q$. 

If a vertical event occurs and the spin is in state $-1$, then it flips to $+1$ with probability $p_I$. If the spin is in state $+1$, then it flips to $-1$ with probability $p_E$ in the vertical event.

In a horizontal event the focal spin attempts to copy the state of a random nearest neighbour. If that neighbour is in state $-1$, then a transmission error occurs with probability $p_I'$, the focal spin then takes value $+1$. If the neighbour is in state $+1$, then a copying error occurs with probability $p_E'$ and the focal spin takes value $+1$. Otherwise the horizontal copying is faithful.

This leads to transition probabilities (once the spin to be updated has been picked) with analogous structure to Eqs.~(\ref{eq supp: dyn1}) and (\ref{eq supp: dyn2}):
\begin{subequations}
    \begin{align}
         T_{n,k,t}^- &=q(1-p_I')\frac{n(t)}{k}+ qp_E'\frac{k-  n(t)}{k} + (1-q)p_E, \label{eq supp: ddyn1} \\
         T_{n,k,t}^+ &=q(1-p_E')\frac{n(t)}{k}+ qp_I'\frac{k-n(t)}{k} + (1-q)p_I,  \label{eq supp: ddyn2}
    \end{align}
\end{subequations}
where $n(t)$ is the number of neighbours in the opposite state to the focal node at time $t$.

Comparing Eqs.~(\ref{eq supp: ddyn1}) and (\ref{eq supp: ddyn2}) with (\ref{eq supp: dyn1}) and (\ref{eq supp: dyn2}), we find the following mapping between the two parameter sets:
\begin{align} \label{eq supp: science advances model params0}
\begin{split}
    (1-q)p_{I} \leftrightarrow v'_{+}, \\
    (1-q)p_{E} \leftrightarrow v'_{-}, \\
    qp_{I}' \leftrightarrow h'_{+}, \\
    qp_{E}' \leftrightarrow h'_{-}, \\
    q(1-p_{I}') \leftrightarrow h_{-}, \\
    q(1-p_{E}') \leftrightarrow h_{+}. \\
\end{split}
\end{align}
Is it clear that a re-scaling of all rates $v_\pm', h_\pm, h_\pm'$ by a common positive factor only changes the units of time, but not the actual model dynamics. Therefore, we can re-write Eq.~(\ref{eq supp: science advances model params0}) as
\begin{align} \label{eq supp: science advances model params2}
\begin{split}
    (1-q)p_{I} =\chi v'_{+}, \\
    (1-q)p_{E}  =\chi v'_{-}, \\
    qp_{I}'  =\chi h'_{+}, \\
    qp_{E}'  =\chi h'_{-}, \\
    q(1-p_{I}')  =\chi h_{-}, \\
    q(1-p_{E}')  =\chi h_{+}, \\
\end{split}
\end{align}
with $\chi>0$.

We note that the constraint $h_++h_-'=h_- + h_+'$ in Eq.~(\ref{eq supp: gamma restriction}) is automatically fulfilled by the setup in \cite{kauhanen2021geospatial}, as both sides add up to $q/\chi$ using the replacements in Eq.~(\ref{eq supp: science advances model params2}).

Assume now we are given non-negative rates $v_\pm', h_\pm, h_\pm'$ such that $h_++h_-'=h_- + h_+'$. We now show that we can always find $p_E, p_I, p_E', p_I', q\in[0,1]$ and $\chi>0$ such that Eq.~(\ref{eq supp: science advances model params2}) holds. 

To do this, define
\be\label{eq:sigma}
    \Sigma\equiv h_+ + h_-'=h_- + h_+'.
\ee
Then set
\begin{subequations}
\begin{align}
    p_I'&= \frac{h_+'}{\Sigma}, \label{eq supp: pI'}\\
    p_E'&= \frac{h_-'}{\Sigma}. \label{eq supp: pE'}
\end{align}
\end{subequations}
Because of Eq.~(\ref{eq:sigma}) these quantities both take values between zero and one. 

We also choose $q$ as follows:
\be
    q=\chi\Sigma. \label{eq supp: q chi sigma relation}
\ee
The factor $\chi$ is not determined at this point, but has to be chosen such that $\chi<1/\Sigma$ (to ensure $q<1$). 

One can directly check that the combination of Eqs.~(\ref{eq supp: pI'}), (\ref{eq supp: pE'}) and (\ref{eq supp: q chi sigma relation}) ensures that the last four relations in Eqs.~(\ref{eq supp: science advances model params2}) are fulfilled.

We now further fix $p_I$ and $p_E$ as follows:
\begin{subequations}
\begin{align}
    p_I &= \frac{\chi}{1-\chi\Sigma} v_+', \label{eq supp: pI} \\
    p_E &= \frac{\chi}{1-\chi\Sigma} v_-'. \label{eq supp: pE}
\end{align}
\end{subequations}
Using $q=\chi\Sigma$ [Eq.~(\ref{eq:sigma})], this ensures that the first two relations in Eqs.~(\ref{eq supp: science advances model params2}) are fulfilled. For given $\Sigma$ and $v_{\pm}'$ we can always choose $\chi>0$ small enough so that $p_I$ and $p_E$ in Eqs.~(\ref{eq supp: pI}) and (\ref{eq supp: pE}) are less than one.
 
\newpage
\section{Complete graph} \label{appendix: complete graph}
We consider the model on a complete network (all-to-all interaction) with $N$ nodes. There is then no notion of space, and any node is the nearest neighbour of any other node. Each node $i$ can be in one of two states $s_{i}\in\{-1,1\}$, and at each time the overall configuration of the model is given by $\mathbf{s}=(s_{1},...,s_{N})$. The system only has one degree of freedom, the number of nodes $n$ in state $+1$,
\begin{equation}
    n = \sum_{i}\frac{1+s_{i}}{2}, \label{eq supp: complete network n and m relation}
\end{equation}
where we sum over all nodes $i$ in the network.

\subsection{Transition rates in continuous-time and master equation}
 The dynamics proceeds through transitions $n\to n\pm 1$, with the following rates:
\begin{subequations}
\begin{align}
    T_{n}^{+} &= \underbrace{h_{+}\frac{n(N-n)}{N}}_{\text{Correctly copy presence}} +\underbrace{h_{+}'\frac{(N-n)^{2}}{N}}_{\substack{\text{Incorrectly copy} \\ \text{absence as presence}}} + \underbrace{v_{+}'(N-n)}_{\substack{\text{Spontaneously acquire} \\ \text{the trait}}} \nonumber \\
    &= \delta N + (\gamma_{+}-\delta)n-\gamma_{+}\frac{n^2}{N}, \label{eq supp: T+} \\
    T_{n}^{-} &= \underbrace{h_{-}\frac{n(N-n)}{N}}_{\text{Correctly copy absence}} + \underbrace{h_{-}'\frac{n^{2}}{N}}_{\substack{\text{Incorrectly copy} \\ \text{presence as absence}}} + \underbrace{v_{-}'n}_{\substack{\text{Spontaneously lose an} \\ \text{existing trait}}} \nonumber \\
    &= (h_{-}+v_{-}')n - \gamma_{-}\frac{n^2}{N}, \label{eq supp: T-}
\end{align}
\end{subequations}
where we have used the expressions in Eq.~(\ref{eq:shorthands}). Here $T_n^+$ is the rate of the event $n\to n+1$, and $T_n^-$ that for the event $n\to n-1$. 

Writing $P(n,t)$ for the probability of finding the system in state $n$ at time $t$, one has the master equation
\begin{equation}
    \frac{\dd}{\dd t}P(n,t) = T^{+}_{n-1}P(n-1,t)+T^{-}_{n+1}P(n+1,t)-\left[T^{+}_{n}+T_n^{-}\right]P(n,t). \label{eq supp: complete network master eqn}
\end{equation}

\subsection{Magnetisation} \label{appendix: complete network magnetisation}
We define the magnetisation as
\begin{equation}
    m = \frac{1}{N}\sum_{i}s_{i}. \label{eq supp: complete network magnetisation definition}
\end{equation}
Using the master equation from Eq.~(\ref{eq supp: complete network master eqn}) we can derive a differential equation for the average magnetisation $\avg{m(t)}$, where $\avg{\cdots}$ represents an average over $P(n,t)$, i.e. realisations of the dynamics. To do this, we first see from Eqs.~(\ref{eq supp: complete network n and m relation}) and (\ref{eq supp: complete network magnetisation definition}),
\begin{equation}
    m = 2\frac{n}{N}-1 \implies \frac{\dd \avg{m(t)}}{\dd t} = \frac{2}{N}\frac{\dd \avg{n(t)}}{\dd t}. \label{eq supp: complete graph m and n/N}
\end{equation}
Then, from the master equation for the probability distribution $P(n,t)$ [Eq.~(\ref{eq supp: complete network master eqn})], 
\begin{align}
    \frac{\dd \avg{n(t)}}{\dd t} &= \frac{\dd}{\dd t}\sum_{n}n P(n, t) = \sum_{n}n\frac{\dd P(n,t)}{\dd t} \nonumber \\
    &= \sum_{n}n\Big\{T^{+}_{n-1}P(n-1,t)+T^{-}_{n+1}P(n+1,t)-\left[T^{+}_{n}+T_n^{-}\right]P(n,t)\Big\} \nonumber \\
    &= \sum_{n}\Big\{(n+1)T^{+}_{n}P(n, t)+(n-1)T^{-}_{n}P(n,t)-\left[T^{+}_{n}+T^{-}_{n}\right]P(n,t)\Big\} \nonumber \\
    &= \sum_{n}\left[T^{+}_{n}-T^{-}_{n}\right]P(n,t) \nonumber \\
    &= \avg{T^{+}_{n(t)}}-\avg{T^{-}_{n(t)}}. \label{eq supp: complete network mag diff initial}
\end{align}
Thus,
\begin{equation}
    \frac{\dd \avg{m(t)}}{\dd t}=\frac{2}{N}\Bigg[\left<T_{n(t)}^{+}\right> - \left<T_{n(t)}^{-}\right>\Bigg]. \label{eq supp: complete network dm/dt general rates}
\end{equation}
Assuming the population is of infinite size, we can ignore  fluctuations, and make the following {\em deterministic approximation}:  
\begin{equation}\label{eq:det_approx}
    \avg{T^{\pm}_{n(t)}} = T^{\pm}_{\avg{n(t)}}.
\end{equation}
We then obtain the following rate equation for the magnetisation by substituting the explicit expressions for the rates from Eqs.~(\ref{eq supp: T+}) and (\ref{eq supp: T-}) into Eq.~(\ref{eq supp: complete network dm/dt general rates}),
\begin{equation}
    \frac{\dd \avg{m(t)}}{\dd t} = 2\delta + 2(\gamma_{+}-\delta - h_{-}-v_{-}')\frac{\avg{n(t)}}{N}-2(\gamma_{+}-\gamma_{-})\frac{\avg{n(t)}^{2}}{N^{2}}.
\end{equation}
We now make the restriction from Eq.~(\ref{eq supp: gamma restriction}), $\gamma_{+}=\gamma_{-}$, which closes the above equation and gives
\begin{align}
    \frac{\dd \avg{m(t)}}{\dd t} &= 2\delta  - 2\beta \avg{n(t)} \nonumber \\
    &= 2\delta  - 2\beta\frac{1+\avg{m(t)}}{2} \nonumber \\
    &= \alpha - \beta \avg{m(t)}. \label{eq supp: complete network magnetisation differential equation}
\end{align}
Thus (assuming $\beta\neq 0$) the average steady-state magnetisation is
\begin{equation}
    \avg{m}_{\rm st} = \frac{\alpha}{\beta} = \frac{(h_{+}'-h_{-}')+(v_{+}'-v_{-}')}{(h_{+}'+h_{-}')+(v_{+}'+v_{-}')}. \label{eq supp: complete network steady-state magnetisation}
\end{equation}
We note that the case $\beta=0$ corresponds to all unfaithful model parameters being 0. This is the standard voter model where the average steady-state magnetisation is simply the initial average magnetisation, $\avg{m}_{\rm st} = \avg{m}_{0}$. 

Equation (\ref{eq supp: complete network magnetisation differential equation}) can also be solved to give the full solution for $m(t)$,
\begin{equation}
    \avg{m(t)} = \avg{m}_{\rm st} - \big(\avg{m}_{\rm st}-\avg{m}_{0}\big)e^{-\beta t}. \label{eq supp: complete network analytic magnetisation}
\end{equation}

\subsection{Interface density}
We define interface density as the proportion of active links, i.e the proportion of links connecting two nodes in different states. For the complete graph we have
\begin{equation}
    \sigma = \frac{\text{\# active links}}{\text{\# links}} = \frac{n(N-n)}{\frac{N(N-1)}{2}} = \frac{1}{2}(1-m^{2}), \label{eq supp: complete graph sigma defn}
\end{equation}
where $m$ is given in Eq.~(\ref{eq supp: complete graph m and n/N}). The definition in Eq.~(\ref{eq supp: complete graph sigma defn}) holds for all values of $N$. For any time $t$ we thus have, again for all $N$,
\be
\avg{\sigma(t)} = \frac{1}{2} \left(1-\avg{m(t)^2}\right)
\ee
Under the deterministic approximation, Eq.~(\ref{eq:det_approx}), we have $\avg{m(t)^{2}}=\avg{m(t)}^{2}$, thus
\begin{align}
    \avg{\sigma(t)} = \frac{1}{2}\left(1-\avg{m(t)}^{2}\right), \label{eq supp: complete network interface full solution}
\end{align}
with $\avg{m(t)}$ as given in Eq.~(\ref{eq supp: complete network analytic magnetisation}). The average steady-state interface density for infinite complete graphs is then given by 
\begin{equation}
    \avg{\sigma}_{\rm st} = \frac{1}{2}\left(1-\avg{m}^{2}_{\rm st}\right). \label{eq supp: complete network steady-state interface}
\end{equation}

\FloatBarrier
\newpage
\section{Pair approximation} \label{appendix: pair approximation}
Next, we analyse the model in the so-called \textit{pair approximation}. This follows the lines of \cite{vazquez2008analytical}. In interacting-agent models of the type we are considering here the pair approximation usually captures the behaviour of the model to good accuracy on infinite uncorrelated networks \cite{gleeson2013binary}.

By uncorrelated networks, we mean networks where there is no preference for nodes to attach to nodes with similar degree \cite{Dorogovtsev}. This is quantified by the assortativity coefficient $r$, i.e. the Pearson correlation coefficient of degree between pairs of linked nodes, being zero \cite{newman2002assortative}. Erd\"os–R\'enyi networks have $r=0$ exactly, since nodes are connected randomly without regard for degree. Barab\'asi-Albert are not strictly uncorrelated, they show an assortative/disassortative behaviour as nodes of small degree tend to connect to nodes of large degree (disassortative), while nodes of large degree tend to connect to nodes of large degree (assortative) . However, for these networks it has been shown that $r\to 0$ as $\frac{\log^{2}({N})}{\sqrt{N}}$, as $N$ becomes large \cite{bertotti2019configuration}, so large Barab\'asi-Albert networks are approximately uncorrelated.

\subsection{Transition rates for individual nodes}
We start from the rates in Eqs.~(\ref{eq supp: dyn1}) and (\ref{eq supp: dyn2}), for a focal node of degree $k$ and with $n(t)$ neighbours in the opposite state to that of the focal node at time $t$. 

We can write
\begin{align}
    T_{n,k,t}^- &=  h_{-}\frac{n(t)}{k} +  h_{-}'\frac{k-n(t)}{k} + v_{-}'  \nonumber \\
    &= (h_{-}-h_{-}')\frac{n(t)}{k}+(h_{-}'+v_{-}') \nonumber \\
    &= \gamma_{-}\frac{n(t)}{k} + \kappa, \label{eq supp: p up simple}
\end{align}
and
\begin{align}
    T_{n,k,t}^+ &=  h_{+}\frac{n(t)}{k} +  h_{+}'\frac{k-n(t)}{k} +  v_{+}'\nonumber \\
    &= (h_{+}-h_{+}')\frac{n(t)}{k}+(h_{+}'+v_{+}') \nonumber \\ 
    &= \gamma_{+}\frac{n(t)}{k} + \delta. \label{eq supp: p down simple}
\end{align}
We have used the shorthands in Eq.~(\ref{eq:shorthands}).

\subsection{Magnetisation} \label{appendix: pair approximation magnetisation}
The change in the average magnetisation is described by the  differential equation,
\begin{align}
    \frac{\dd \avg{m(t)}}{\dd t} 
    &= \sum_{k}\sum_{s=\pm} N P_k\avg{\rho_{s}(t)}\sum_{n=0}^{k}T_{n,k,t}^{-s}B(n|k, s, t)\Delta m(s). \label{eq supp: uncorrelated network initial mag differential}
\end{align}
We use $P_{k}$ to denote the degree distribution of the network, with normalisation $\sum_{k}P_{k}=1$. The quantity $\avg{\rho_{s}(t)}$ is the average proportion of nodes in state $s$. Thus $N P_k \avg{\rho_s(t)}$ is the expected number of nodes with degree $k$ and in state $s$ at time $t$. Further, $B(n|k,s,t)$ is the probability that a node of degree $k$ and in state $s\in\{-1,1\}$ has $n$ neighbours in the opposite state at time $t$. Finally, $\Delta m(s)=(-2s)/N$ is the amount by which the magnetisation changes if a node in state $s$ flips to $-s$. 

Substituting in the expression for $\Delta m(s)$ and expanding the sum over $s$, Eq.~(\ref{eq supp: uncorrelated network initial mag differential}) becomes
\begin{equation}
    \frac{\dd \avg{m(t)}}{\dd t} = \sum_{k}P_{k}\sum_{n=0}^{k}\Bigg\{\Big[1-\avg{m(t)}\Big]T_{n,k,t}^+B(n|k,-,t) - \Big[1+\avg{m(t)}\Big]T_{n,k,t}^-B(n|k,+,t)\Bigg\}. \label{eq supp: uncorrelated network mag differential}
\end{equation}
The pair approximation consists of assuming that $B(n|k,s,t)$ is a binomial distribution. This means we assume that the states of nearest neighbours of a focal node are uncorrelated. The single-event probability is $P(-s|s,t)$, which is the conditional probability that a neighbour of a node is in state $-s$ given that the node is in state $s$ at time $t$. We have,
\begin{equation}
    B(n|k,s,t)=\binom{k}{n} P(-s|s,t)^{n}\Big[1-P(-s|s,t)\Big]^{k-n}. \label{eq supp: pair approx}
\end{equation}
Writing $\mu$ for the average degree of the network, the quantity $P(-s|s,t)$ can be obtained as follows,
\begin{equation}
    P(-s|s,t)  = \frac{\avg{\sigma(t)} \cdot \frac{\mu N}{2}}{\avg{\rho_{s}(t)}\cdot \mu N} = \frac{\avg{\sigma(t)}}{2\avg{\rho_{s}(t)}} = \frac{\avg{\sigma(t)}}{1+s\avg{m(t)}},\label{eq supp: aux}
\end{equation}
The numerator in Eq.~(\ref{eq supp: aux}) is the average total number of active links in the network (on average there are $\frac{\mu N}{2}$ links overall, and a fraction $\avg{\sigma(t)}$ of these are active). The denominator ($\avg{\rho_{s}(t)}\cdot \mu N$) is the average total number of links emanating from all nodes in state $s$.

The first and second moments of the binomial distribution in Eq.~(\ref{eq supp: pair approx}) are then \cite{knoblauch2008closed}
\begin{subequations}
\begin{align}
    E[n]_{k,s,t} &= \frac{k\left<\sigma(t)\right>}{1+s\avg{m(t)}}, \label{eq supp: n moment}\\
    E[n^{2}]_{k,s,t} 
    &= \frac{k\left<\sigma(t)\right>}{1+s\avg{m(t)}}\left(1-\frac{\left<\sigma(t)\right>}{1+s\avg{m(t)}}\right)+\frac{k^{2}\left<\sigma(t)\right>^{2}}{(1+s\avg{m(t)})^{2}} \nonumber \\
    &= \frac{k\left<\sigma(t)\right>}{1+s\avg{m(t)}}+\frac{k(k-1)\left<\sigma(t)\right>^{2}}{(1+s\avg{m(t)})^{2}}. \label{eq supp: n squared moment}
\end{align}
\end{subequations}
Using these moments, and the rates from Eqs.~(\ref{eq supp: p up simple}) and (\ref{eq supp: p down simple}), we can simplify Eq.~(\ref{eq supp: uncorrelated network mag differential}) to obtain
\begin{align}
    \frac{\dd \avg{m(t)}}{\dd t} 
    &= \sum_{k}P_{k}\Bigg\{\Big(1-\avg{m(t)}\Big)\left[\gamma_{+}\frac{E[n]_{k,-,t}}{k}+\delta\right]-\Big(1+\avg{m(t)}\Big)\left[\gamma_{-}\frac{E[n]_{k,+,t}}{k}+\kappa\right]\Bigg\} \nonumber \\
    &= \alpha - \beta\avg{m(t)} + (\gamma_{+}-\gamma_{-})\avg{\sigma(t)}. \label{eq supp: pair approximation magnetisation differential}
\end{align}
We now again make the restriction from Eq.~(\ref{eq supp: gamma restriction}), $\gamma_{+}=\gamma_{-}$, after which Eq.~(\ref{eq supp: pair approximation magnetisation differential}) is identical to Eq.~(\ref{eq supp: complete network magnetisation differential equation}) for the complete network. Thus the solution for $\avg{m(t)}$ within the pair approximation is as in Eq.~(\ref{eq supp: complete network analytic magnetisation}), and the steady-state value is given by Eq.~(\ref{eq supp: complete network steady-state magnetisation}).

\subsection{Interface density} \label{appendix: uncorrelated network - interface density}
We consider again a node of degree $k$ and with $n(t)$ active links to its neighbours at time $t$, i.e. $n(t)$ of its neighbours are of in the opposite state. If this focal node changes state, the change in the number of active links is
\begin{equation}
    \Delta n(t) = n(t+1) - n(t) = [k-n(t)]-n(t) = k-2n(t),
\end{equation}
Given that the total number of links in the network is $\frac{\mu N}{2}$ on average, the expected change in interface density associated with the flip of the focal node is
\begin{equation}
    \Delta \sigma(n,k, t) = \frac{k-2n(t)}{\frac{\mu N}{2}} = \frac{2[k-2n(t)]}{\mu N}.
\end{equation}
Similar to the magnetisation, the rate of change of the interface density can be written
\begin{align}
    \frac{\dd \avg{\sigma(t)}}{\dd t} 
    &= \sum_{k}\sum_{s=\pm} NP_k\avg{\rho_{s}(t)}\sum_{n=0}^{k}T_{n,k,t}^{-s}B(n|k,s,t)\Delta\sigma(n,k,t)\nonumber \\
    &= \frac{1}{\mu}\sum_{k}P_k\sum_{n=0}^{k}\Bigg\{\Big[1-\avg{m(t)}\Big]T_{n,k,t}^{+}B(n|k,-,t) \nonumber \\
    &\qquad\qquad\qquad\qquad+\Big[1+\avg{m(t)}\Big]T_{n,k,t}^{-}B(n|k,+,t)\Bigg\}[k-2n(t)].
\end{align}
As in Sec.~\ref{appendix: pair approximation magnetisation} we use Eqs.~(\ref{eq supp: p up simple}), (\ref{eq supp: p down simple}), (\ref{eq supp: n moment}) and (\ref{eq supp: n squared moment}) to find
\begin{align}
    \frac{\dd \avg{\sigma(t)}}{\dd t} 
    &= \beta-\alpha\avg{m(t)}+2\avg{\sigma(t)}\left[\gamma\left(1-\frac{2}{\mu}\right)-\beta\right] \nonumber \\
    &\qquad -2\left(1-\frac{1}{\mu}\right)\avg{\sigma(t)}^{2}\left[\frac{2\gamma+(\gamma_{+}-\gamma_{-})\avg{m(t)}}{1-\avg{m(t)}^{2}}\right]. \label{eq supp: pair approximation isogloss differential general}
\end{align}
We now use the restriction from Eq.~(\ref{eq supp: gamma restriction}), $\gamma_{+}=\gamma_{-}$, to write
\begin{equation}
    \frac{\dd \avg{\sigma(t)}}{\dd t} 
    = \beta-\alpha\avg{m(t)}+2\avg{\sigma(t)}\left[\gamma\left(1-\frac{2}{\mu}\right)-\beta\right] -4\gamma\left(1-\frac{1}{\mu}\right)\frac{\avg{\sigma(t)}^{2}}{1-\avg{m(t)}^{2}}. \label{eq supp: pair approximation interface differential}
\end{equation}
It is possible to solve this equation numerically [this is how the analytical trajectories of Fig.~\ref{fig main: parabolas}(a) in the main paper were obtained].

We are only interested in the steady-state solution which can be found by equating the right-hand side of Eq.~(\ref{eq supp: pair approximation interface differential}) to zero, and, based on Eq.~(\ref{eq main: H}), substituting in the following ansatz,
\begin{equation}
    \avg{\sigma}_{\rm st} = \frac{1}{2}H_\text{PA}( \tau)\left(1-\avg{m}^{2}_{\rm st}\right). \label{eq supp: pair approx interface}
\end{equation}
Here $H_\text{PA}(\tau)$ is, at this stage, an unknown function of the network and the model parameters (the subscript `PA' stands for `pair approximation'). This results in the following quadratic equation for $H_{\text{PA}}(\tau)$:
\begin{equation}
    (\mu-1)H_{\text{PA}}(\tau)^{2}+\left[\mu\tau-(\mu-2)\right]H_{\text{PA}}(\tau)-\mu\tau = 0,
\end{equation}
which can be solved to give
\begin{equation}
    H_{\text{PA}}(\tau) = \frac{\mu-2 - \mu\tau\pm\sqrt{(\tau+1)^{2}\mu^{2}-4(\mu-1)}}{2(\mu-1)}, \label{eq supp: H pair}
\end{equation}
where 
\begin{equation}
    \tau = \frac{\beta}{\gamma} = \frac{(h_{+}'+h_{-}')+(v_{+}'+v_{-}')}{\frac{1}{2}\left[(h_{+}-h_{+}')+(h_{-}-h_{-}')\right]}. \label{eq supp: tau}
\end{equation}
We refer to $\tau$ as the \textit{noise-to-order} ratio, see Sec.~\ref{appendix: H and tau} for a more detailed discussion of the general interpretation of $\tau$ and the order parameter $H(\tau)$.

\subsection{Further discussion of the function \texorpdfstring{$H_{\rm PA}(\tau)$}{} obtained from the pair approximation}

\subsubsection{Regions of validity of the two branches of the solution}
We note that, Eq.~(\ref{eq supp: H pair}) has both a positive and negative branch for a given value of $\mu$. We denote these $H_{\rm PA}^{\pm}(\tau)$. Given that $H(\tau)$ must be real we require the object under the square root in Eq.~(\ref{eq supp: H pair}) to be non-negative. This means that we must either have $\tau \leq -1-\frac{2}{\mu}\sqrt{\mu-1}$ or $\tau \geq -1+\frac{2}{\mu}\sqrt{\mu-1}$.  An interpretation of $\tau$, and a discussion of the range of values this parameter can take in our model can be found in Sec.~\ref{appendix: tau}. This includes the possibility of negative values of $\tau$.

In the limit $\tau\to-\infty$ we find $H_{\rm PA}^{+}(\tau)\to \infty$, whereas $H_{\rm PA}^{-}(\tau)\to 1$. In simulations we observe $H(\tau)\to 1$ as $\tau \to -\infty$. Therefore we use $H_{\rm PA}^{-}(\tau)$ in the region $\tau \leq -1-\frac{2}{\mu}\sqrt{\mu-1}$. 

In the limit $\tau\to\infty$ we find $H_{\rm PA}^{-}(\tau)\to -\infty$, whereas $H_{\rm PA}^{+}(\tau)\to 1$. In simulations we observe $H(\tau)\to 1$ as $\tau\to \infty$. Therefore, we use $H_{\rm PA}^{+}(\tau)$ in the region $\tau \geq -1+\frac{2}{\mu}\sqrt{\mu-1}$.

\begin{figure}[htbp]
    \includegraphics[scale = 0.95]{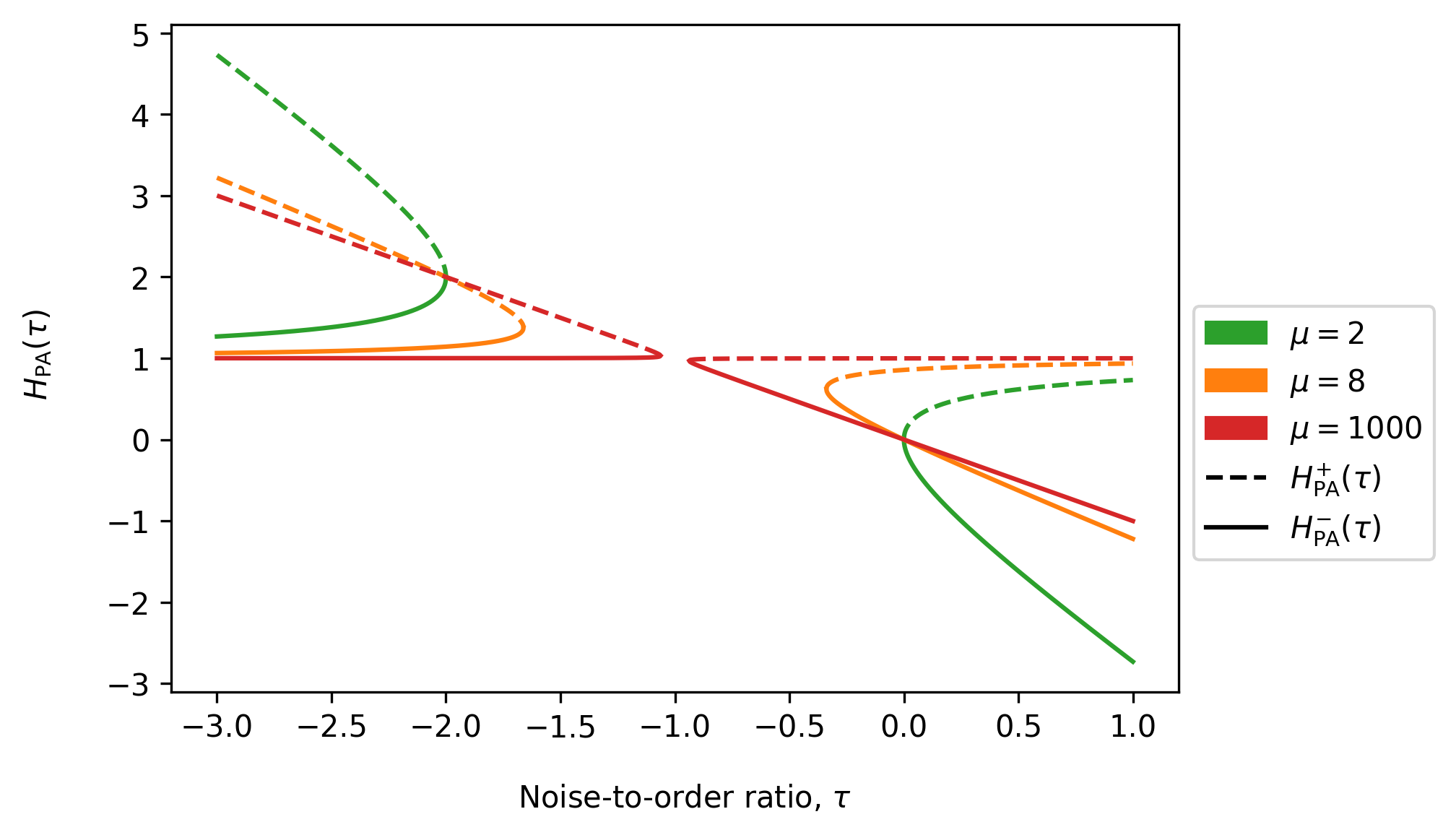}
    \caption{Plots of $H^\pm_{\rm PA}(\tau)$ from Eq.~(\ref{eq supp: H pair}) as a function of the noise-to-order ratio, $\tau$, for different values of $\mu$. $H_{\rm PA}(\tau)$ only takes real values in the regions $\tau \leq -1-\frac{2}{\mu}\sqrt{\mu-1}$ or $\tau \geq -1+\frac{2}{\mu}\sqrt{\mu-1}$. Solid lines are $H^-_{\rm PA}(\tau)$ \Big(applicable for $\tau \leq -1-\frac{2}{\mu}\sqrt{\mu-1}$\Big), dashed lines are $H^+_{\rm PA}(\tau)$ \Big(applicable for $\tau \geq -1+\frac{2}{\mu}\sqrt{\mu-1}$\Big).} 
    \label{fig supp: pair approx branches}
    \end{figure}

In Fig.~\ref{fig supp: pair approx branches} we show both branches of $H_{\rm PA}(\tau)$ for different values of $\mu$. 

\subsubsection{Comparison against simulations}
\begin{figure}[htbp]
    \includegraphics[scale = 0.95]{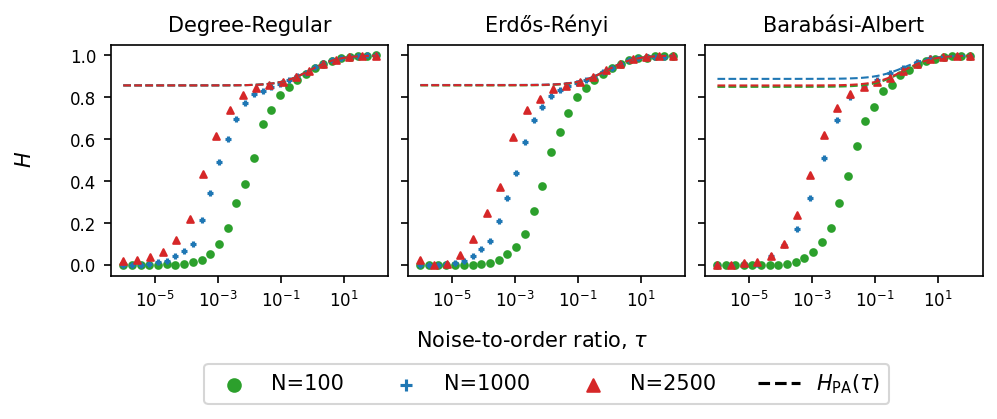}
    \caption{Plots of $H(\tau)=\avg{\sigma}_{\rm st}/[\frac{1}{2}(1-\avg{m}_{\rm st}^{2})]$, which characterises the scattering, against the noise-to-order ratio $\tau$ for networks of size $N$. Degree-regular with $\mu=8$ (left), Erd\"os--R\'enyi with $p=8/N$ (middle), Barab\'asi--Albert with $4$ new links established upon adding a new node (right). For all three types of network we have $\mu\approx8$ (deviations are due to the finite size of the graphs). The dashed lines are the results $H^+_{\rm PA}(\tau)$ from the pair approximation, Eq.~(\ref{eq supp: H pair}). Note that there are 3 lines on each plot, as varying $N$ can change $\mu$ which results in a different $H_{\rm PA}(\tau)$. For degree-regular networks, $\mu=8$ exactly always so the lines are identical, however for Barab\'asi--Albert networks the difference is noticeable. Each marker is the result from averaging $1000$ independent simulations in the steady-state at a particular $\tau$ value. Model parameters for a particular $\tau$ are generated via the algorithm detailed in Sec.~\ref{appendix: determining model parameters tau and m}. }
    \label{fig supp: H pair}
    \end{figure}

We are principally interested in $\tau\geq 0$, so we do not consider the negative branch of Eq.~(\ref{eq supp: H pair}) further. In Fig.~\ref{fig supp: H pair} we demonstrate how the positive branch of Eq.~(\ref{eq supp: H pair}) compares to simulation for different size degree-regular (left), Erd\"os--R\'enyi (middle), and Barab\'asi--Albert (right) networks. 

As $\tau\rightarrow\infty$ we see that both $H(\tau)\to 1$ both in simulations and that $H^+_{\rm PA}(\tau)\to 1$. For sufficiently low values of $\tau$, simulation results deviate from the theory. This is because simulations are performed on finite size networks and/or because of inaccuracies in the pair approximation. 

As $\tau\to 0$ we observe in simulations on finite networks that $H(\tau)\to 0$. However $H_{\rm PA}(\tau)\to \frac{\mu-2}{\mu-1}$ as $\tau\to 0$. We note that this is equivalent to the long-lived plateau of the interface density reported in \cite{vazquez2008analytical}. We find that the agreement between simulation results and $H_{\rm PA}(\tau)$ improves with increasing system size $N$ as expected.

\subsubsection{Limits on the inference of \texorpdfstring{$\tau$}{} from \texorpdfstring{$H(\tau)$}{}}
It is not always possible to map a value of $H(\tau)=\avg{\sigma}_{\rm st}/[\frac{1}{2}(1-\avg{m}_{\rm st}^{2})]$ from simulations (or data) to a value of $\tau$ within the pair approximation. For example, for $\mu=8$ no value of $\tau$ is associated with values of $H(\tau)$ between zero and approximately $0.85$ [the limiting value of $H_{\rm PA}(\tau)$ for $\tau\to 0$ in Fig.~\ref{fig supp: H pair}]. The fact that not all values of $H(\tau)$ are reached by the applicable branches of $H_{\rm PA}(\tau)$ can also been seen in Fig.~\ref{fig supp: pair approx branches}.

\FloatBarrier
\subsection{Limit of all-to-all interaction} \label{appendix: pair approx limit to complete graph}
As a sanity check we verify that the results in the pair approximation reduce to those for the complete network in the limit $\mu\to\infty$. Taking this limit in Eq.~(\ref{eq supp: pair approximation interface differential}) we find
\begin{align}
    \frac{\dd \avg{\sigma(t)}}{\dd t} = \beta-\alpha m(t)+2\sigma(\gamma-\beta)-4\gamma\frac{\avg{\sigma(t)}^{2}}{1-\avg{m(t)}^{2}}. \label{eq supp: pair approx interface differential complete network limit}
\end{align}
We now verify that the result in Eq.~(\ref{eq supp: complete network interface full solution}) is a solution of Eq~(\ref{eq supp: pair approx interface differential complete network limit}). Substituting Eq.~(\ref{eq supp: complete network interface full solution}) into the right-hand side of Eq.~(\ref{eq supp: pair approx interface differential complete network limit}) gives
\begin{align}
    \frac{\dd \avg{\sigma(t)}}{\dd t} = -\Big[\alpha-\beta \avg{m(t)}\Big]\avg{m(t)}. \label{eq supp: pair approx interface differential in complete network limit with substitution}
\end{align}
On the other hand, differentiating the solution in Eq.~(\ref{eq supp: complete network interface full solution}) with respect to time gives
\begin{equation}
    \frac{\dd \avg{\sigma(t)}}{\dd t} = -\avg{m(t)}\frac{\dd \avg{m(t)}}{\dd t}. \label{eq supp: complete network interface differentiated}
\end{equation}
Comparing Eqs.~(\ref{eq supp: pair approx interface differential in complete network limit with substitution}) and (\ref{eq supp: complete network interface differentiated}) it must then be true that 
\begin{equation}
    \frac{\dd \avg{m(t)}}{\dd t} = \alpha-\beta\avg{m(t)}.
\end{equation}
This is the same differential equation we obtained on the complete network, i.e. Eq.~(\ref{eq supp: complete network magnetisation differential equation}), and we have already established in Eq.~(\ref{eq supp: pair approximation magnetisation differential}) that $\avg{m(t)}$ in the pair approximation is the same as that on the complete network. Therefore we have shown that the result for the interface density from the pair approximation reduces to that for the complete network in the limit $\mu\to\infty$.

We can also take the limit $\mu\to\infty$ of $H_{\rm PA}(\tau)$ from Eq.~(\ref{eq supp: H pair}), and we find $H_{\rm PA}(\tau)\to 1$ for all $\tau$, i.e, we obtain Eq.~(\ref{eq supp: complete network steady-state interface}).

\FloatBarrier
\newpage
\section{Annealed approximation} \label{appendix: annealed approximation}
We now analyse the model in the so-called \textit{annealed approximation}, which works well for large but finite-size uncorrelated networks. We broadly follow the steps of \cite{carro2016noisy}, but we note a difference in notation. While this earlier reference uses $s_i\in\{0,1\}$ for the binary states of the nodes, we use $s_{i}\in\{-1,1\}$, consistent with the rest of our paper. \par

\subsection{Nature of the approximation}
The annealed approximation centers around replacing an adjacency matrix $\bA$ (representing a network with a given degree sequence) with a weighted adjacency matrix $\tilde{\bA}$. The elements $\tilde{A}_{ij}$ are the probabilities that nodes $i$ and $j$ are connected over an ensemble of networks generated from the configuration model using the degree sequence of the original network. 

We first explain the configuration model, and how to derive the elements of the adjacency matrix $\tilde{\bA}$, as well as the limits in which the theory is valid.

\subsubsection{Configuration model}

Given a degree sequence $\{k_{1}, k_{2},...,k_{N}\}$ we can generate an ensemble of networks using the configuration model \cite{barabasi2013network}. The procedure is as follows. In a first step, $k_{i}$ `half-edges' are attached to each node $i$. The total number of stubs must be even, i.e. $\sum_{i}k_{i}=2m$ where $m\in \mathbb{Z}^{+}$. Two half-edges are then chosen at random and connected. The process repeats until no half-edges remain. The resulting network will have the specified degree sequence. This can be repeated many times to generate an ensemble of different networks all having the same degree sequence.

We now wish to determine the probability that two nodes $i$ and $j$ are connected over the ensemble of networks generated by the configuration model. First, we see that in total there are $\mu N$ half edges in the network, where $\mu$ is the first moment of the degree distribution $P_{k}$ of the network. Thus, the probability that a given stub of node $i$ connects to one of the stubs of node $j$ is $k_j/(\mu N-1)$. Assuming independent events, the expected number of connections from $i$ to $j$ is therefore 
\be\label{eq supp: annealed prob}
p_{ij}=\frac{k_i k_j}{\mu N-1}\approx \frac{k_i k_j}{\mu N},
\ee
where the last step is valid for $N\gg 1$. For large $N$, this number is small, and equal to the probability that $i$ and $j$ become connected when constructing the network.

\subsubsection{Double-links and self-loops}
The above procedure does not exclude the possibility of self-loops or double-links. For example the final two stubs might be connected to the same node, or belong to a pair of nodes which already have an edge. However for large networks we can show that the relative number of these edges vanishes \cite{newman2018networks}.

The probability that there exists a double edge between nodes $i$ and $j$ can be calculated from Eq.~(\ref{eq supp: annealed prob}). When constructing the network, the probability that $i$ and $j$ become connected is $k_{i}k_{j}/\mu N$. After one edge is formed, the probability that another edge forms becomes $(k_{i}-1)(k_{j}-1)/\mu N$. Thus the probability that there are two edges between nodes $i$ and $j$ once the network is constructed is
\be
    \frac{k_{i}k_{j}(k_{i}-1)(k_{j}-1)}{(\mu N)^{2}}.
\ee
From this we can calculate the expected number of double edges by summing over $i$ and $j$, remembering to divide by $2$ to avoid double counting the edges:
\begin{align}
    \frac{1}{2(\mu N)^{2}}\sum_{i,j}k_{i}k_{j}(k_{i}-1)(k_{j}-1) &= \frac{1}{2(\mu N)^{2}}\sum_{i}k_{i}(k_{i}-1)\sum_{j}k_{j}(k_{j}-1) \nonumber \\
    &= \frac{1}{2}\left(\frac{\mu_{2}}{\mu}-1\right)^{2}, \label{eq supp: annealed number double}
\end{align}
where $\mu_{2}$ is the second moment of $P_{k}$. The proportion of double-links among all links in the system can be calculated by dividing the number in Eq.~(\ref{eq supp: annealed number double}) by the total number of edges $\frac{\mu N}{2}$. Provided that $\mu$ and $\mu_{2}$ remain finite as $N$ increases, this ratio vanishes as $\frac{1}{N}$.  

Regarding self-loops, the number of ways of pairing up stubs from node $i$ is $\binom{k_{i}}{2}=\frac{1}{2}k_{i}(k_{i}-1)$.  With $\mu N$ half-edges total, the probability of a self-loop is
\be
    p_{ii} \approx \frac{k_{i}(k_{i}-1)}{2\mu N}.
\ee
We can use this to determine the expected number of self-loops:
\be
    \sum_{i}p_{ii} = \frac{1}{2}\left(\frac{\mu_{2}}{\mu}-1\right). \label{eq supp: annealed number loops}
\ee
Again assuming that $\mu$ and $\mu_{2}$ remain constant as $N$ increases, the average number of self-loops among all links scales as $\frac{1}{N}$, and is therefore negligible when $N\gg 1$.

\begin{figure}[htbp]
    \includegraphics[scale=0.6]{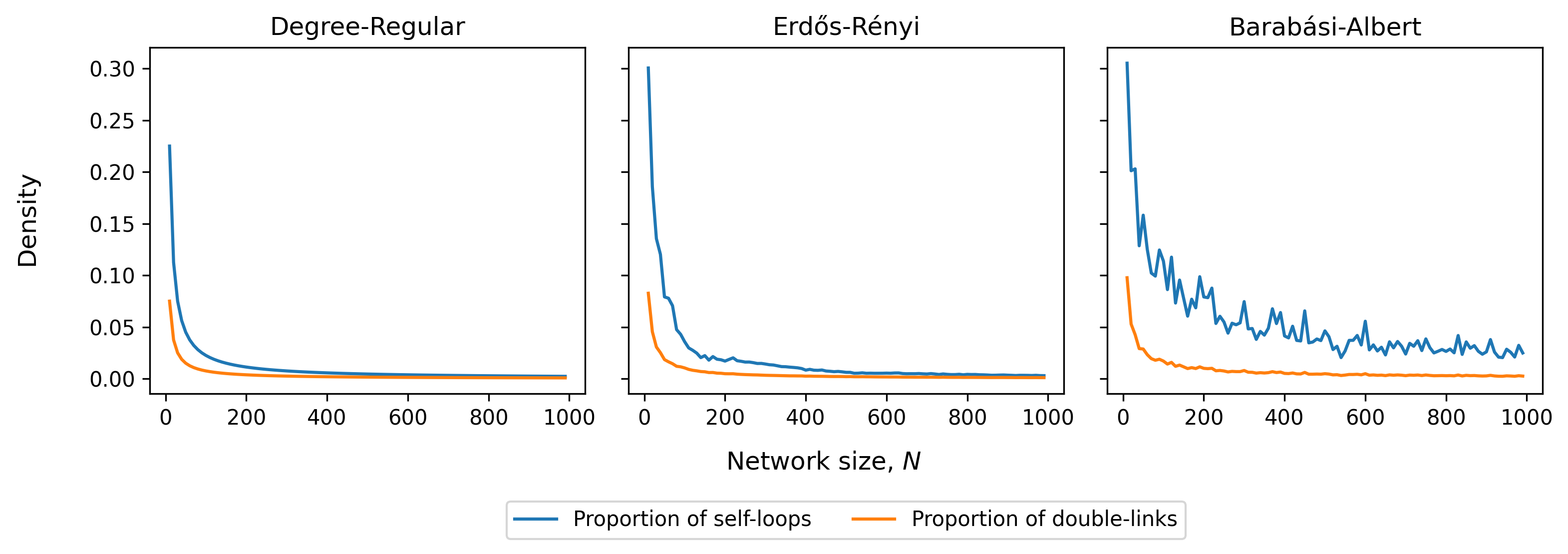}
    \caption{Plots of the proportion of self-loops and double-links calculated as the ratio of Eqs.~(\ref{eq supp: annealed number double}) and (\ref{eq supp: annealed number loops}) to the total number of links, $\mu N/2$, for different networks as the number of nodes, $N$, increases. The networks shown have degree sequences which follow that of degree-regular (left), Erd\"os--R\'enyi (middle) and Barab\'asi--Albert (right) networks. They were created using the configuration model. All have approximately $\mu\approx4$ (deviations are due to the finite size of the graphs).}
    \label{fig supp: annealed moment convergence}
    \end{figure}

The assumption that $\mu$ and $\mu_{2}$ remain constant is not generally valid. For example, networks with a power-law distribution, $P_{k}\sim k^{-\gamma}$, where $\gamma < 3$ have a finite first moment, $\mu$, but a divergent second moment, $\mu_{2}$, as $N\to\infty$ \cite{barabasi2013network}. Many networks such as degree-regular, Erd\"os--R\'enyi or Barab\'asi--Albert networks do satisfy the condition. In Fig.~\ref{fig supp: annealed moment convergence} we show that estimates of the proportion of self-loops and double-links, calculated using Eqs.~(\ref{eq supp: annealed number double}) and (\ref{eq supp: annealed number loops}), vanish as $N$ becomes large on different networks which were created using the configuration model.

\subsubsection{Rate equations}

We can now define the elements of the matrix $\tilde{\bA}$ to be the probabilities from Eq.~(\ref{eq supp: annealed prob}), i.e.
\begin{equation}
    \tilde{A}_{ij} \approx \frac{k_{i}k_{j}}{N\mu} \label{eq supp: annealed weighted adj}
\end{equation}

With this we can replace any summation over nearest neighbours of a node as follows (we write $j\in i$ if $j$ is a nearest neighbour of $i$):
\begin{equation}
    \sum_{j\in i}s_{j} \rightarrow \frac{k_{i}}{\mu N}\sum_{j}k_{j}s_{j}. \label{eq supp: annealed approx}
\end{equation}
We will refer to this as the \textit{annealed approximation}. Note that such an approximation conserves the initial degree sequence, 
\begin{equation}
    k_{i} = \sum_{j=1}^{N}\tilde{A}_{ij} = k_{i}\frac{1}{\mu}\left(\frac{1}{N}\sum_{j=1}^{N}k_{j}\right) = k_{i},
\end{equation}
and thus also the total number of links. 

If node $i$ is in in state $s_i=-1$, then it will flip to state $s_i=+1$ with rate
\begin{align}
    T_{i}^{+} &= v_{+}' + \frac{h_{+}}{k_{i}}\sum_{j\in i}\frac{1+s_{j}}{2} + \frac{h_{+}'}{k_{i}}\sum_{j\in i}\frac{1-s_{j}}{2} \nonumber \\
    &= v_{+}'+\frac{1}{2}(h_{+}+h_{+}')+\frac{(h_{+}-h_{+}')}{2k_{i}}\sum_{j \in i}s_{j} \nonumber \\
    &= (h_{+}'+v_{+}')+\frac{1}{2}(h_{+}-h_{+}')+\frac{(h_{+}-h_{+}')}{2k_{i}}\sum_{j \in i}s_{j} \nonumber \\
    &= \delta + \frac{\gamma_{+}}{2} + \frac{\gamma_{+}}{2k_{i}}\sum_{j\in i}s_{j} \nonumber \\
    &= \delta + \frac{\gamma_{+}}{2} + \frac{\gamma_{+}}{2\mu N}\sum_{m}k_{m}s_{m},\label{eq supp: annealed rate up}
\end{align}
and if the node is in state $s_i=+1$ it will flip to $s_i=-1$ with rate
\begin{align}
    T_{i}^{-} &= v_{-}' + \frac{h_{-}}{k_{i}}\sum_{j\in i}\frac{1+s_{j}}{2} + \frac{h_{-}'}{k_{i}}\sum_{j \in i}\frac{1-s_{j}}{2} \nonumber \\
    &= v_{-}'+\frac{1}{2}(h_{-}+h_{-}')+\frac{(h_{-}-h_{-}')}{2k_{i}}\sum_{j\in i}s_{j} \nonumber \\
    &= (h_{-}'+v_{-}')+\frac{1}{2}(h_{-}-h_{-}')+\frac{(h_{-}-h_{-}')}{2k_{i}}\sum_{j\in i}s_{j} \nonumber \\
    &= \kappa + \frac{\gamma_{-}}{2} - \frac{\gamma_{-}}{2k_{i}}\sum_{j\in i}s_{j} \nonumber \\
    &= \kappa + \frac{\gamma_{-}}{2} - \frac{\gamma_{-}}{2\mu N}\sum_{m}k_{m}s_{m}, \label{eq supp: annealed rate down}
\end{align}
where we have used the annealed approximation in Eq.~(\ref{eq supp: annealed approx}). We also introduced expressions from Eq.~(\ref{eq:shorthands}).

With these definitions we then have the overall flip rate for spin $i$ which depends on the spin configuration $\bs = (s_{1},\dots,s_{N})$,
\be\label{eq supp: annealed W_i}
    W_i(\bs)\equiv \frac{1+s_i}{2}T_i^- + \frac{1-s_i}{2}T_i^+.
\ee

\subsection{Master equation}
To formulate the master equation for the model, it is useful to define the flip operators $F_i$ acting on functions of $\bs$ by flipping the $i$\textsuperscript{th} spin,
\be
F_i f(s_1,\dots,s_i,\dots s_N)\equiv f(s_1,\dots, -s_i,\dots, s_N).
\ee
The master equation for $P(\bs)$, i.e. the probability that $N$ nodes have a particular spin configuration, is then given by
\begin{equation}
    \frac{\dd P(\bs)}{\dd t} = \sum_{i=1}^{N}\Bigg\{ F_i\Big[W_i(\bs)P(\bs)\Big]-W_i(\bs)P(\bs)\Bigg\}. \label{eq supp: annealed master}
\end{equation}

\subsection{Magnetisation} \label{appendix: annealed magnetisation}
Following \cite{carro2016noisy} we first introduce the following summation notation: $\sum_{\bs}$ refers to a sum over all possible spin combinations, i.e
\begin{equation}
    \sum_{\bs} = \sum_{s_{1}}\sum_{s_{2}}...\sum_{s_N}, 
\end{equation}
and $\sum_{\bs_{j}}$ refers to a sum over all possible spin combinations excluding node $j$,
\begin{equation}
    \sum_{\bs_{j}} \equiv \sum_{s_{1}}...\sum_{s_{j-1}}\sum_{s_{j+1}}...\sum_{s_{N}}. \label{eq supp: annealed partial sum}
\end{equation}
These two summation definitions are related by
\begin{equation}
    \sum_{\bs} = \sum_{\bs_{j}}\sum_{s_{j}}. \label{eq supp: annealed combine sum}
\end{equation}
This allows us to split summations over all system configurations into a summation over a specific nodes spin and a summation over the spin configuration of the remaining system. We denote the average over the spin configuration $\bs$ as
\begin{equation}
    \avg{f(\bs)} = \sum_{\bs}f(\bs)P(\bs). \label{eq supp: annealed average}
\end{equation}

Using the master equation, Eq.~(\ref{eq supp: annealed master}), we have
\begin{align}
    \frac{\dd \avg{s_{i}}}{\dd t} &= \sum_{\bs}s_{i}\frac{\dd P(\bs)}{\dd t} \nonumber \\
    &= \sum_{\bs}s_i\sum_{j=1}^{N}\Big\{F_j[W_jP]- W_j P\Big\},
\end{align}
where we omit the argument $\bs$ of the $W_j$ and $P$ on the right for simplicity. We now split the sum over $j$ into summations of $j=i$ and $j\neq i$,
\begin{align}
    \frac{\dd \avg{s_{i}}}{\dd t} &= \sum_{\bs}s_{i}\Bigg\{\Big(F_{i}[W_{i}P]-W_{i}P\Big) + \sum_{j\neq i}^{N}\Big(F_{j}[W_{j}P]-W_{j}P\Big)\Bigg\} \nonumber \\
    &= \sum_{\bs_{i}}\sum_{s_{i}}s_{i}\Big(F_{i}[W_{i}P]-W_{i}P\Big) + \sum_{j\neq i}^{N}\sum_{\bs_{j}}s_{i}\sum_{s_{j}}\Big(F_{j}[W_{j}P]-W_{j}P\Big) \nonumber \\
    &= \sum_{\bs}\big(-2s_{i}W_{i}P\big),
\end{align}
where in the second line, the second term vanishes after performing the $s_{j}$ summation, and the first term simplifies after the $s_{i}$ sum. Evaluating the sum with Eq.~(\ref{eq supp: annealed average}) we have
\be  
    \frac{\dd\avg{s_{i}}}{\dd t} = -2 \avg{W_i s_i}.
\ee
 This is also clear intuitively. The quantity $W_i$ on the right is the probability with which spin $i$ flips from $s_i$ to $-s_i$. The change of $s_i$ in such an event is $-2s_i$.

Using the definition of $W_i$ in Eq.~(\ref{eq supp: annealed W_i}), we then have
\begin{align}
    \frac{\dd \avg{s_{i}}}{\dd t} 
    &= \avg{T_{i}^{+}-T_{i}^{-}}-\avg{\left(T_{i}^{+}+T_{i}^{-}\right)s_{i}}.
\end{align}

We now use the explicit expressions for the rates defined in Eqs.~(\ref{eq supp: annealed rate up}) and (\ref{eq supp: annealed rate down}) to write
\be
    \frac{\dd \avg{s_{i}}}{\dd t} = \alpha +\frac{1}{2}(\gamma_{+}-\gamma_{-})-(\beta + \gamma)\avg{s_{i}}+\frac{\gamma}{\mu N}\sum_{m}k_{m}\avg{s_{m}} - \frac{(\gamma_{+}+\gamma_{-})}{2 \mu N}\sum_{m}k_{m}\avg{s_{i}s_{m}}.
\ee
Then we make the restriction from Eq.~(\ref{eq supp: gamma restriction}), $\gamma_{+}=\gamma_{-}$, and find
\begin{equation}
    \frac{\dd \avg{s_{i}}}{\dd t} = \alpha -(\beta + \gamma)\avg{s_{i}}+\frac{\gamma}{\mu N}\sum_{m}k_{m}\avg{s_{m}}.\label{eq supp: dsi/dt}
\end{equation}
In the steady-state, this gives the following set of $N$ simultaneous equations:
\be
    \alpha -(\beta+\gamma)\avg{s_{i}} + \frac{\gamma}{\mu N}\sum_{m}k_{m}\avg{s_{m}} = 0. \label{eq supp: annealed steady-state dsi/st}
\ee 
One can directly verify that
\begin{equation}
    \avg{s_{i}}_{\rm st} = \frac{-\alpha}{-(\gamma+\beta)+\sum_{i=1}^{N}A_{i}} = \frac{\alpha}{\beta}
\end{equation}
for all $i$ is a solution. This then also means 
\be
    \avg{m}_{\rm st}=\frac{1}{N}\sum_{i=1}^{N}\avg{s_{i}}_{\rm st}  = \frac{\alpha}{\beta}. \label{eq supp: annealed steady mag}
\ee 
We now show that this is also the only solution of Eq.~(\ref{eq supp: annealed steady-state dsi/st}). If we define 
\begin{subequations}
\begin{align}
    \mathcal{A}_{i} &\equiv \frac{\gamma k_{i}}{\mu N}, \label{eq supp: annealed A} \\
    \mathcal{B}_{i} &\equiv A_{i}- (\gamma + \beta), \label{eq supp: annealed B}
\end{align}
\end{subequations}
then we can cast Eq.~(\ref{eq supp: annealed steady-state dsi/st}) as a matrix equation,
\begin{equation}
    \begin{bmatrix}
    \mathcal{B}_{1} & \mathcal{A}_{2} & \cdots & \mathcal{A}_{N} \\
  \mathcal{A}_{1} & \mathcal{B}_{2} & \cdots & \mathcal{A}_{N} \\
     \vdots & \vdots & \ddots & \vdots \\
     \mathcal{A}_{1} & \mathcal{A}_{2} & \cdots & \mathcal{B}_{N} \\
    \end{bmatrix}
    \begin{bmatrix}    
    \avg{s_{1}}_{\rm st} \\
    \avg{s_{2}}_{\rm st} \\
    \vdots \\
    \avg{s_{N}}_{\rm st}
    \end{bmatrix}
    = \begin{bmatrix}
    -\alpha \\
       -\alpha \\
    \vdots \\
      -\alpha
    \end{bmatrix}. \label{eq supp: annealed matrix}
\end{equation}
To prove the uniqueness of the above solution, we need to show that the matrix on the left has full rank, i.e. that its columns are linearly independent. We show this by contradiction. 

Assume there are multiple solutions, i.e., there are coefficients $c_1,\dots,c_N$ (which are not all equal to zero) such that
\begin{equation}
    \begin{bmatrix}
    \mathcal{B}_{1} & \mathcal{A}_{2} & \cdots & \mathcal{A}_{N} \\
  \mathcal{A}_{1} & \mathcal{B}_{2} & \cdots & \mathcal{A}_{N} \\
     \vdots & \vdots & \ddots & \vdots \\
     \mathcal{A}_{1} & \mathcal{A}_{2} & \cdots & \mathcal{B}_{N} \\
    \end{bmatrix}
    \begin{bmatrix}    
    c_1 \\
    c_2 \\
    \vdots \\
   c_N
    \end{bmatrix}
    = \begin{bmatrix}
   0\\
      0 \\
    \vdots \\
     0
    \end{bmatrix}. \label{eq supp: annealed matrix 2}
\end{equation} 
 Looking at the $i$\textsuperscript{th} and $j$\textsuperscript{th} row together, one concludes
\be
c_i(\mathcal{B}_i-\mathcal{A}_i)=c_j(\mathcal{B}_j-\mathcal{A}_j)
\ee
for all pairs $i,j$. Given that $\mathcal{B}_i-\mathcal{A}_i=-(\gamma+\beta)$ for all $i$, this then implies that all $c_i$ must take the same value (we discard the special case $\gamma+\beta=0$). This common value can be non-zero only if the sum of elements in each row of the matrix in Eq.~(\ref{eq supp: annealed matrix 2}) sum to zero, i.e. if
\begin{gather}
    \mathcal{B}_{i} + \sum_{j\neq i}^{N}\mathcal{A}_{j} = 0 \notag \\
    \implies -(\gamma+\beta) + \sum_{i=1}^{N}\mathcal{A}_{i}  = 0 \notag \\
    \implies -(\gamma+\beta) + \frac{\gamma}{\mu N}\sum_{i=1}^{N}k_{i} = 0 \notag \\
    \implies \beta = 0,
\end{gather}
where we have used the definitions in Eqs.~(\ref{eq supp: annealed A}) and (\ref{eq supp: annealed B}). Therefore we are left with the condition $\beta=0$. Keeping in mind the definition of $\beta$ in Eq.~(\ref{eq:shorthands}) this corresponds to a `pathological' scenario, in which there is no spontaneous spin flipping (no vertical dynamics), and in which horizontal copying is without error. This is the standard voter model, which, for finite systems, will always end up in a consensus state. We note that $\alpha=0$ in this case as well.

In summary, for all cases of interest ($\beta \neq 0$), the vector $(c_{1},\dots,c_{N})$ is non-zero. Thus Eq.~(\ref{eq supp: annealed steady mag}) is the only non-trivial solution to Eq.~(\ref{eq supp: annealed steady-state dsi/st}).

\subsection{Interface density} \label{appendix: annealed interface density}
The average interface density can be expressed in the following form
\begin{equation}
    \avg{\sigma} = \frac{\frac{1}{2}\sum_{ij}A_{ij}\left[\frac{1+\avg{s_{i}}}{2}\frac{1-\avg{s_{j}}}{2}+\frac{1-\avg{s_{i}}}{2}\frac{1+\avg{s_{j}}}{2}\right]}{\frac{1}{2}\sum_{ij}A_{ij}},
\end{equation}
where the denominator is the total number of links and the numerator is the number of links connecting opposite spins. Using the annealed approximation, Eq.~(\ref{eq supp: annealed weighted adj}), this can be written as
\begin{equation}
    \avg{\sigma} = \frac{1}{2(\mu N)^{2}}\sum_{ij}k_{i}k_{j}\left[1-\avg{s_{i}s_{j}}\right].
\end{equation}
In the steady-state we therefore have
\begin{equation}
    \avg{\sigma}_{\text{st}} = \frac{1}{2(\mu N)^{2}}\sum_{ij}k_{i}k_{j}\left[1-\avg{s_{i}s_{j}}_{\text{st}}\right]. \label{eq supp: annealed steady-state interface initial}\\
\end{equation}
We now define the correlation matrix $\Upsilon_{ij}$,
\begin{equation}
    \Upsilon_{ij} = \avg{s_{i}s_{j}} - \avg{s_{i}}\avg{s_{j}}, \label{eq supp: annealed correlation matrix}
\end{equation}
which in the steady-state, since $\avg{s_{i}}_{\text{st}}=\avg{m}_{\text{st}}$ for all $i$ [see Eq.~(\ref{eq supp: annealed steady mag})], can be written as
\begin{equation}
    (\Upsilon_{ij})_{\rm st} = \avg{s_{i}s_{j}}_{\rm st}-\avg{m}_{\rm st}^{2}.
\end{equation}
Eq.~(\ref{eq supp: annealed steady-state interface initial}) then becomes
\begin{equation}
    \avg{\sigma}_{\rm st} = \frac{1}{2}\left(1-\avg{m}_{\rm st}^{2}\right)-\frac{1}{2(\mu N)^{2}}\sum_{ij}k_{i}k_{j}(\Upsilon_{ij})_{\text{st}}. \label{eq supp: annealed steady-state interface intermediate}
\end{equation}
The first term in this expression is the density of active links if the states of neighbouring spins were entirely uncorrelated, which is the same as the formula for the complete network [see Eq.~(\ref{eq supp: complete network steady-state interface})]. The second term represents the reduction in interface density due to correlations introduced by the properties of the network. These appear as terms proportional to $k_{i}k_{j}$ from the weighted adjacency matrix, Eq.~(\ref{eq supp: annealed weighted adj}).

To find $(\Upsilon_{ij})_{\text{st}}$, we first formulate a  differential equation for $\avg{s_{i}s_{j}}$. This is done in a similar manner to Sec.~\ref{appendix: annealed magnetisation}. We use the master equation, Eq.~(\ref{eq supp: annealed master}), to write
\begin{align}
    \frac{\dd \avg{s_{i}s_{j}}}{\dd t} &= \sum_{\bs}s_{i}s_{j}\frac{\dd P(\bs)}{\dd t}  \nonumber \\
    &= \sum_{\bs}s_{i}s_{j}\sum_{k=1}^{N}\Big\{F_k[W_kP]- W_k P\Big\}.
\end{align}
Again, we have dropped the explicit dependence of $W_{k}$ and $P$ on $\bs$. We first consider the term $k\neq i, k\neq j$. One has
\begin{equation}
     \sum_{s_{k}}\Big(F_{k}[W_{k}P]-W_{k}P\Big)  = 0,
\end{equation}
for any fixed $s_1,\dots,s_{k-1},s_{k+1},\dots, s_N$. Thus this term does not contribute. If $k=i\neq j$ we have
\be
    \sum_{\bs_{i}}s_{j}\Bigg\{\sum_{s_{i}}s_{i}\Big(F_{i}[W_{i}P]-W_{i}P\Big)\Bigg\} = -2\sum_{\bs}s_{i}s_{j}W_{i}P = -2\avg{s_{i}s_{j}W_{i}}.
\ee
Analogously the case $k=j\neq i$ gives $-2\avg{s_{i}s_{j}W_{j}}$. Finally for $k=i=j$, we note $s_{i}^{2}=1$, and therefore
\be
    \frac{\dd \avg{s_{i}s_{i}}}{\dd t} = 0. 
\ee
Combining all of this, and substituting the expression for $W_{i}$ from Eq.~(\ref{eq supp: annealed W_i}), gives
\begin{equation} \label{eq supp: annealed dsidj/dt cases}
\begin{aligned}
    \frac{\dd \avg{s_{i}s_{j}}}{\dd t} =
    \begin{cases}
        \avg{(T_{i}^{+}-T_{i}^{-})s_{j}} + \avg{(T_{j}^{+}-T_{j}^{-})s_{i}} - \avg{q_{ij}s_{i}s_{j}} & \text{if $i\neq j$}  \\
        0 & \text{if $i= j$}
    \end{cases},
\end{aligned}
\end{equation}
where $q_{ij} = T_{i}^{+}+T_{i}^{-}+T_{j}^{+}+T_{j}^{-}$. Using the Kronecker delta, Eq.~(\ref{eq supp: annealed dsidj/dt cases}) can be written
\begin{align} \label{eq supp: dsisj/dt initial}
    \frac{\dd \avg{s_{i}s_{j}}}{\dd t} &= \avg{\left(T_{i}^{+}-T_{i}^{-}\right)s_{j}}+\avg{\left(T_{j}^{+}-T_{j}^{-}\right)s_{i}} \nonumber \\
    &\qquad - \avg{q_{ij}s_{i}s_{j}}+2\delta_{ij}\Big[\avg{T_{i}^{+}+T_{i}^{-}}-\avg{\left(T_{i}^{+}-T_{i}^{-}\right)s_{i}}\Big]. 
\end{align}

We now substitute the explicit expressions for the rates, Eqs.~(\ref{eq supp: annealed rate up}) and (\ref{eq supp: annealed rate down}), into Eq.~(\ref{eq supp: dsisj/dt initial}) which gives
\begin{align}
    \frac{\dd \avg{s_{i}s_{j}}}{\dd t} &= \left[\alpha+\frac{1}{2}(\gamma_{+}-\gamma_{-})\right]\big(\avg{s_{i}}+\avg{s_{j}}\big) + \frac{\gamma}{\mu N}\sum_{m}k_{m}\Big[\avg{s_{m}s_{j}}+\avg{s_{m}s_{i}}\Big] \nonumber \\
    &\qquad -2(\beta+\gamma)\avg{s_{i}s_{j}}+\frac{(\gamma_{+}-\gamma_{-})}{\mu N}\sum_{m}k_{m}\avg{s_{m}s_{i}s_{j}} \nonumber \\
    &\qquad +2\delta_{ij}\Bigg\{\beta+\gamma+\frac{(\gamma_{+}-\gamma_{-})}{\mu N}\sum_{m}k_{m}\avg{s_{m}}-\left[\alpha+\frac{1}{2}(\gamma_{+}-\gamma_{-})\right]\avg{s_{i}} \nonumber \\
    &\qquad\qquad\qquad -\frac{\gamma}{\mu N}\sum_{m}k_{m}\avg{s_{m}s_{i}}\Bigg\}.
\end{align}
Then we again make the restriction from Eq.~(\ref{eq supp: gamma restriction}), $\gamma_{+}=\gamma_{-}$, and obtain
\begin{align}
    \frac{\dd \avg{s_{i}s_{j}}}{\dd t} &= \alpha\big(\avg{s_{i}}+\avg{s_{j}}\big) + \frac{\gamma}{\mu N}\sum_{m}k_{m}\Big[\avg{s_{m}s_{i}}+\avg{s_{m}s_{j}}\Big] -2(\beta+\gamma)\avg{s_{i}s_{j}} \nonumber \\
    &\qquad +2\delta_{ij}\Bigg[\beta+\gamma-\alpha\avg{s_{i}}-\frac{\gamma}{\mu N}\sum_{m}k_{m}\avg{s_{i}s_{m}}\Bigg].\label{eq supp: dsisj/dt full}
\end{align}
Using Eqs.~(\ref{eq supp: dsi/dt}), (\ref{eq supp: annealed correlation matrix}) and (\ref{eq supp: dsisj/dt full}) we can then write a differential equation for elements of the correlation matrix,
\begin{align}
    \frac{\dd \Upsilon_{ij}}{\dd t} &= \frac{\gamma}{\mu N}\sum_{m}k_{m}\left(\Upsilon_{mi} + \Upsilon_{mj}\right)-2(\beta+\gamma)\Upsilon_{ij} \nonumber \\
    &\qquad + 2\delta_{ij}\Big[\beta+\gamma-\alpha\avg{s_{i}}-\frac{\gamma}{\mu N}\sum_{m}k_{m}\Upsilon_{im}-\avg{s_{i}}\frac{\gamma}{\mu N}\sum_{m}k_{m}\avg{s_{m}}\Big]. \label{eq supp: annealed upsilon diff}
\end{align}
Using the facts that $\tau = \frac{\beta}{\gamma}$ and $\avg{s_{i}}_{\text{st}}=\avg{m}_{\text{st}}=\frac{\alpha}{\beta}$, from Eqs.~(\ref{eq supp: tau}) and (\ref{eq supp: annealed steady mag}) respectively, $(\Upsilon_{ij})_{\text{st}}$ is found from Eq.~(\ref{eq supp: annealed upsilon diff})  to be
\begin{equation}
    (\Upsilon_{ij})_{\rm st} = \frac{\frac{1}{\mu N}\sum_{m}k_{m}\left[(\Upsilon_{mi})_{\rm st}+(\Upsilon_{mj})_{\rm st}\right]+2\delta_{ij}\left[(1+\tau)\left(1-\avg{m}^{2}_{\rm st}\right)-\frac{1}{\mu N}\sum_{m}k_{m}(\Upsilon_{mi})_{\text{st}}\right]}{2(1+\tau)}. \label{eq supp: annealed upsilon}
\end{equation}
This equation for $(\Upsilon_{ij})_{\text{st}}$ involves terms such as $\sum_{m}k_{m}(\Upsilon_{mi})_{\text{st}}$ and from Eq.~(\ref{eq supp: annealed steady-state interface intermediate}) we know we ultimately need to calculate terms such as $\sum_{ij}k_{i}k_{j}(\Upsilon_{ij})_{\text{st}}$. Motivated by this and following the notation in \cite{carro2016noisy}, we introduce a new variable $S_{x}$,
\be
    S_{x} = \sum_{ij}k_{i}^{x}k_{j}(\Upsilon_{ij})_{\rm st}, \label{eq supp: annealed Sx initial}
\ee
where $x$ is an integer such that $k_{i}^{x}$ is simply the degree of node $i$ raised to the $x$\textsuperscript{th} power. Substituting Eq.~(\ref{eq supp: annealed upsilon}) into Eq.~(\ref{eq supp: annealed Sx initial}) allows us to write $S_{x}$ as
\be
    S_{x} = \frac{S_{x}+\frac{1}{\mu}\overline{k^{x}}S_{1}+2N\overline{k^{x+1}}(1+\tau)\left(1-\avg{m}^{2}_{\text{st}}\right)-\frac{2}{\mu N}S_{x+1}}{2(1+\tau)},
\ee
where the overbar stands for an average over the degree distribution of the network,
\begin{equation}
    \overline{f(k)} = \frac{1}{N}\sum_{i=1}^{N}f(k_{i}). \label{eq supp: overline definition}
\end{equation}
We then obtain the following recurrence relation,
\begin{align}
    S_{x+1} &= \frac{-\mu N}{2}(1+2\tau)S_{x} + \frac{N}{2}\Bigg[\overline{k^{x}}S_{1}+2\mu N \overline{k^{x+1}}(1+\tau)\left(1-\avg{m}^{2}_{\text{st}}\right)\Bigg] \nonumber \\
    &\equiv {\cal C}S_{x} + {\cal D}_{x}, \label{eq supp: annealed recurrance to solve}
\end{align}
where, 
\begin{align}
    {\cal C} &\equiv \frac{-\mu N}{2}(1+2\tau),\nonumber \\
    {\cal D}_{x} &\equiv \frac{N}{2}\Bigg[\overline{k^{x}}S_{1}+2\mu N \overline{k^{x+1}}(1+\tau)\left(1-\avg{m}^{2}_{\text{st}}\right)\Bigg].
\end{align}

Eq.~(\ref{eq supp: annealed recurrance to solve}) has the solution
\begin{align}
    S_{x+1} &= {\cal C}^{x}S_{1} + \sum_{m=1}^{x}{\cal C}^{x-m}{\cal D}_{m}\nonumber \\
    &= \left[\frac{-\mu N(1+2\tau)}{2}\right]^{x}S_{1} \nonumber \\
    &\qquad + \sum_{m=1}^{x}\left[\frac{-\mu N(1+2\tau)}{2}\right]^{x-m}\frac{N}{2}\Bigg[\overline{k^{m}}S_{1}+2\mu N\overline{k^{m+1}}(1+\tau)\left(1-\avg{m}^{2}_{\rm st}\right)\Bigg]. \label{eq supp: annealed recurrence solution}
\end{align}
Ultimately we want to determine $S_{1}$, as this is the term that appears in Eq.~(\ref{eq supp: annealed steady-state interface intermediate}). With this in mind, we can isolate the $S_{1}$ term in Eq.~(\ref{eq supp: annealed recurrence solution}) by dividing both sides by ${\cal C}^{x}$, and taking the limit $x\to\infty$. The LHS is
\begin{align}
    \lim_{x\to\infty}\frac{S_{x+1}}{\left[\frac{-\mu N(1+2\tau)}{2}\right]^{x}} &= \lim_{x\to\infty}\frac{\sum_{ij}k_{i}^{x+1}k_{j}\Upsilon_{ij}}{\left[\frac{-\mu N(1+2\tau)}{2}\right]^{x}} \nonumber \\
    &= \left[\frac{-\mu N(1+2\tau)}{2}\right]\lim_{x\to\infty}\sum_{ij}\left[\frac{-2k_{i}}{\mu N(1+2\tau)}\right]^{x+1}k_{j}\Upsilon_{ij}. 
\end{align}
The limit on the right hand side will be zero provided 
\be
    \abs{\frac{-2k_{i}}{\mu N(1+2\tau)}} < 1. \label{eq supp: annealed bound}
\ee
The only possible ranges of $\tau$ are $\tau\geq 0$ and $\tau\leq -2$ [see Sec.~\ref{sec:range_tau}]. In each of these we have $\left|\frac{1}{1+2\tau}\right|\leq 1$, and thus the inequality in Eq.~(\ref{eq supp: annealed bound}) is fulfilled if $k_i<\mu N/2$. This is the case for $\mu\geq 2$. All networks we consider here have a mean degree higher than two. 

We note that since $k_i\leq N$, the inequality in Eq.~(\ref{eq supp: annealed bound}) is actually always fulfilled for any $\tau\neq -\frac{1}{2}$ provided that $\mu$ is sufficiently large. But due to the restrictions on $\tau$ [Sec.~\ref{sec:range_tau}] we do not focus on such cases.

Having proven the limit in Eq.~(\ref{eq supp: annealed bound}) vanishes, we can find $S_{1}$ from Eq.~(\ref{eq supp: annealed recurrence solution}) to be
\begin{equation}
    S_{1} = \frac{-\mu N^{2}(1+\tau)\left(1-\avg{m}^{2}_{\rm st}\right)\sum_{m=1}^{\infty}\left[\frac{-\mu N(1+2\tau)}{2}\right]^{-m}\overline{k^{m+1}}}{1+\frac{N}{2}\sum_{m=1}^{\infty}\left[\frac{-\mu N(1+2\tau)}{2}\right]^{-m}\overline{k^{m}}}.
\end{equation}
Now notice that the summations in the above equation can be evaluated as follows,
\begin{equation}
    \sum_{m=1}^{\infty}\mathcal{C}^{-m}\overline{k^{m+z}} = \overline{k^{z}\sum_{m=1}^{\infty}\mathcal{C}^{-m}k^{m}} = \overline{\frac{\mathcal{C}^{-1}k^{z+1}}{1-\mathcal{C}^{-1}k}},
\end{equation}
where we have used the infinite geometric series formula under the condition that $\abs{\mathcal{C}^{-1}k} < 1$, which is exactly the same condition as in Eq.~(\ref{eq supp: annealed bound}). After some manipulation we have the final form for $S_{1}$,
\begin{equation}
    S_{1} = \frac{\mu N^{2}(1+\tau)(1+2\tau)\left(1-\avg{m}^{2}_{\rm st}\right)\overline{\left(\frac{k^{2}}{(1+2\tau)\mu N+2k}\right)}}{\tau + \frac{1}{\mu}\overline{\left(\frac{k^{2}}{(1+2\tau)\mu N+2k}\right)}}. \label{eq supp: annealed S1}
\end{equation}
This is then substituted back into Eq.~(\ref{eq supp: annealed steady-state interface intermediate}) and we find a parabolic relationship between $\avg{\sigma}_{\text{st}}$ and $\avg{m}_{\text{st}}$,
\begin{align}
    \avg{\sigma}_{\text{st}} &= \frac{1}{2}H_{\text{AA}}(\tau)\left(1-\avg{m}_{\text{st}}^{2}\right), \label{eq supp: annealed approx interface}
\end{align}
with
\begin{align}
    H_{\text{AA}}(\tau) = 1-\frac{(1+\tau)(1+2\tau)\overline{\left(\frac{k^{2}}{(1+2\tau)\mu N + 2k}\right)}}{\tau\mu + \overline{\left(\frac{k^{2}}{(1+2\tau)\mu N + 2k}\right)}}. \label{eq supp: H annealed}
\end{align}
The subscript `AA' stands for `annealed approximation'. See Sec.~\ref{appendix: H and tau} for detailed discussion on $\tau$ and $H(\tau)$.

We see that in the limit $\tau\to 0$, $H_{\text{AA}}(\tau)\to 0$. This matches what we observe in simulation.
In the limit $\tau\to\infty$ Eq.~(\ref{eq supp: H annealed}) reduces to 
\be
    \lim_{\tau\to\infty}H_{\text{AA}}(\tau) = 1-\frac{\overline{k^{2}}}{\mu^{2} N}. \label{eq supp: annealed tau to inf}
\ee
Simulations in this limit give $H(\tau)\to 1$. Thus Eq.~(\ref{eq supp: annealed tau to inf}) captures this up to corrections of order $1/N$.

\subsection{Comparison against simulations}
\begin{figure}[htbp]
    \includegraphics[scale = 0.95]{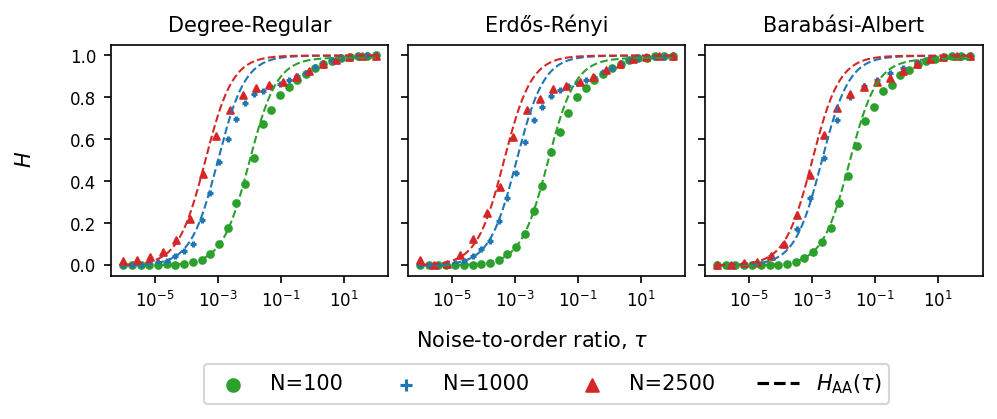}
    \caption{Plots of $H(\tau)=\avg{\sigma}_{\rm st}/[\frac{1}{2}(1-\avg{m}_{\rm st}^{2})]$, which characterises the scattering, against the noise-to-order ratio $\tau$ for networks of size $N$. Degree-regular with $\mu=8$ (left), Erd\"os--R\'enyi with $p=8/N$ (middle), Barab\'asi--Albert with $4$ new links established upon adding a new node (right). For all three types of network we have $\mu\approx8$ (deviations are due to the finite size of the graphs). The dashed lines are the results from the annealed approximation, Eq.~(\ref{eq supp: H annealed}). Each marker is the result from averaging $1000$ independent simulations in the steady-state at a specific value of $\tau$. Model parameters for specific values of $\tau$ are generated via the algorithm detailed in Sec.~\ref{appendix: determining model parameters tau and m}.}
    \label{fig supp: H annealed}
    \end{figure}

In Fig.~\ref{fig supp: H annealed} we demonstrate how the approximation in Eq.~(\ref{eq supp: H annealed}) compares to simulations for different size degree-regular networks (left), Erd\"os--R\'enyi networks (middle), and Barab\'asi--Albert networks (right). As discussed above the theory matches simulation in the low $\tau$ limit, and approximately so in the large $\tau$ limit up to a $\frac{1}{N}$ correction. There are discrepancies with simulation data for intermediate $\tau$. This is because in such a regime the structural properties of the network are most relevant. The nature of the annealed approximation is to replace the network with a weighted complete network, thus a lot of information about the original network structure is lost. 

\FloatBarrier
\newpage
\section{Analytical treatment of the model on an infinite square lattice} \label{appendix: square lattice} 
Next, we consider an infinite square lattice with $N=L\times L$ sites. The calculation is based on that of \cite{kauhanen2021geospatial}, but we correct a typo that previously prohibited solving the equations for a general choice of the model parameters. The notation in \cite{kauhanen2021geospatial} is different and a detailed discussion is given in Sec.~\ref{appendix: relation to SA}.

\subsection{Setup and spin flip rates}
The state of the node at lattice site $\textbf{x}=(x_{1}, x_{2})$ is written $s(\textbf{x})\in\{-1,1\}$. The average state (over realisations of the dynamics) at site $\textbf{x}$ is written $\left<s(\textbf{x})\right>$. The global mean magnetisation over the whole lattice is then $\avg{m}=\sum_{\textbf{x}}\avg{s(\mbf{x})}/N$. The pair correlation between two spins $s(\textbf{x})$ and $s(\textbf{y})$ will be written $\left<s(\textbf{x})s(\mathbf{y})\right>$. In summations we will use the notation $\mathbf{y} \in \mathbf{x}^{(1)}$ to denote the set of von Neumann neighbours of $\textbf{x}$, i.e. the set $\{(x_{1}-1,x_{2}),(x_{1}+1,x_{2}), (x_{1},x_{2}-1),(x_{1},x_{2}+1)\}$.

The spin flip probability (in discrete-time) is the probability with which the spin at site $\textbf{x}$ changes its state if it is selected for update. In continuous-time we can think of spin flip rates. The flip probability (or rate) for the spin at lattice site $\bx$ takes the form,
\begin{equation}
    w(\textbf{x}) = A(\textbf{x})+B(\textbf{x}), \label{eq supp: 2D lattice spin flip basic}
\end{equation}
where $A(\textbf{x})$ is the contribution from the vertical process
\begin{align}
    A(\bx) &= v_{+}'\frac{1-s(\bx)}{2}+v_{-}'\frac{1+s(\bx)}{2} \nonumber \\
    &= \frac{1}{2}\Big[v_{+}'+v_{-}'-(v_{+}'-v_{-}')s(\bx)\Big],
\end{align}
and $B(\textbf{x})$ is the contribution from the horizontal process
\begin{align}
    B(\bx) &= \frac{1}{4}\sum_{\by \in \bx^{(1)}}\Bigg[h_{-}'\frac{1+s(\bx)}{2}\frac{1+s(\by)}{2} + h_{+}\frac{1-s(\bx)}{2}\frac{1+s(\by)}{2} \nonumber \\
    &\qquad + h_{+}'\frac{1-s(\bx)}{2}\frac{1-s(\by)}{2} + h_{-}\frac{1+s(\bx)}{2}\frac{1-s(\by)}{2}\Bigg] \nonumber \\
    &= \frac{1}{4}\Big[(h_{+}+h_{-})+(h_{+}'+h_{-}')\Big] - \frac{1}{4}\Big[(h_{+}'-h_{-}')+(h_{+}-h_{-})\Big]s(\bx) \nonumber \\
    &\qquad + \frac{1}{16}\sum_{\by \in \bx^{(1)}}\Big[(\gamma_{+}-\gamma_{-})s(\by)-(\gamma_{+}+\gamma_{-})s(\bx)s(\by)\Big]. \label{eq supp: square lattice B(x)}
\end{align}
The total spin flip probability is then
\begin{align}
    w(\textbf{x}) &= A(\textbf{x})+B(\textbf{x}) \nonumber \\
    &= \frac{\eta}{2}-\frac{1}{2}\bigg\{(v_{+}'-v_{-}')+\frac{1}{2}\Big[(h_{+}'-h_{-}')+(h_{+}-h_{-})\Big]\bigg\}s(\bx) \nonumber \\
    &\qquad + \frac{1}{16}\sum_{\by \in \bx^{(1)}}\Big[(\gamma_{+}-\gamma_{-})s(\by)-(\gamma_{+}+\gamma_{-})s(\bx)s(\by)\Big], \label{eq supp: 2D lattice spin flip inter}
\end{align}
where we have introduced expressions from Eq.~(\ref{eq:shorthands}). We now make the restriction from Eq.~(\ref{eq supp: gamma restriction}), $\gamma_{+}=\gamma_{-}$, which allows us to re-write Eq.~(\ref{eq supp: 2D lattice spin flip inter}) as
\begin{equation}
    w(\bx) = \frac{\eta}{2} - \frac{\alpha}{2}s(\bx) - \frac{\gamma}{8}\sum_{\by \in \bx^{(1)}}s(\bx)s(\by). \label{eq supp: 2D lattice spin flip full}
\end{equation}

\subsection{Magnetisation}
We proceed with a discrete-time setup in mind. In a given time step, the spin at site $\textbf{x}$ changes state with probability $\frac{w(\bx)}{N}$ where the $\frac{1}{N}$ factor represents the probability of the spin being selected for update. If a spin flip occurs, $s(\bx)$ changes by an amount $-2s(\textbf{x})$. The change in the average spin, $\avg{s(\textbf{x})}$, is then
\begin{equation}
    \avg{s(\textbf{x}, t+\Delta t)} - \avg{s(\textbf{x}, t)} = \frac{1}{N}\left<-2w(\textbf{x})s(\textbf{x})\right>,
\end{equation}
where $\Delta t$ denotes a single time step. We now make the standard choice $\Delta t = \frac{1}{N}$ and take the limit $N\to\infty$ (i.e the continuous-time limit). Substituting in Eq.~(\ref{eq supp: 2D lattice spin flip full}) and evaluating the expectations leads to
\begin{align}
     \frac{\dd \avg{s(\textbf{x})}}{\dd t} & = \alpha - \eta \avg{s(\textbf{x})} + \frac{\gamma}{4}\sum_{\textbf{y} \in \textbf{x}^{(1)}}\avg{s(\textbf{y})}. \label{eq supp: 2d lattice dS/dt}
\end{align}
The same result is obtained if we start out in continuous-time, and think of $w(\bx)$ in Eq.~(\ref{eq supp: 2D lattice spin flip full}) as the rate for the spin at $\bx$ to flip. 

Now we perform a summation over all lattice sites $\textbf{x}$ to find a differential equation for the magnetisation $\avg{m(t)}$,
\begin{equation}
    \frac{\dd \avg{m(t)}}{\dd t} =  \alpha-\beta \avg{m(t)}. \label{eq supp: lattice differential equation magnetisation}
\end{equation}
This is exactly the same as Eq.~(\ref{eq supp: complete network magnetisation differential equation}) for the complete network, so the solution is identical.

\subsection{Interface density} \label{appendix: square lattice interface density}
To compute the interface density we start with the pair correlation $\left<s(\textbf{x})s(\textbf{y})\right>$. The quantity $s(\textbf{x})s(\textbf{y})$ ($\textbf{x}\neq\textbf{y}$) changes by an amount $-2s(\textbf{x})s(\textbf{y})$ when either the spin at $\textbf{x}$ or $\textbf{y}$ flips. Working in the continuous-time limit, we find
\begin{align}
    \frac{\dd \avg{s(\textbf{x})s(\textbf{y})}}{\dd t} &= \left<-2\left[w(\textbf{x})+w(\textbf{y})\right]s(\mbf{x})s(\mbf{y})\right> \nonumber \\
    & = -2\eta \avg{s(\textbf{x})s(\textbf{y})} - 2\alpha\Big[\avg{s(\mbf{x})}+\avg{s(\mbf{y})}\Big]-\frac{\gamma}{4}\left(\sum_{\mbf{z}\in\mbf{x}^{(1)}}\avg{s(\textbf{y})s(\textbf{z})}+\sum_{\mbf{z}\in\mbf{y}^{(1)}}\avg{s(\textbf{x})s(\textbf{z})}\right). \label{eq supp: 2d lattice dS(x,y)/dt}
\end{align}
We now assume translational invariance so that $\avg{s(\textbf{x})s(\textbf{y})}$ is a function of $\bx-\by$ only. We write $\avg{s(\textbf{x})s(\textbf{y})}=C(\textbf{x}-\textbf{y})$. For $\textbf{r}\equiv \bx-\by\neq 0$, we then have
\begin{align}
    \frac{\dd C(\textbf{r})}{\dd t} &= 2\beta\Big[\avg{m}_{\rm st}\big(\avg{s(\mbf{x})}+\avg{s(\mbf{y})}\big)-C(\mbf{r})\Big]+2\gamma\Delta C(\mbf{r}), \label{eq supp: lattice dC/dt}
\end{align}
where $\Delta$ is the lattice Laplacian, 
\begin{equation}
    \Delta f(\textbf{x}) = -f(\textbf{x}) + \frac{1}{4}\sum_{\textbf{y}\in\mbf{x}^{(1)}}f(\textbf{y}).
\end{equation}
All lattice sites are statistically equivalent to one another in the ensemble of  realisations (assuming initial conditions are independent and identical for each lattice site). In the stationary state we then have $\avg{s(\textbf{x})}_{\rm st} = \avg{s(\textbf{y})}_{\rm st} = \avg{m}_{\rm st}$ . We also have $\frac{\dd C(\textbf{r})}{\dd t} = 0$, so Eq.~(\ref{eq supp: lattice dC/dt}) becomes
\begin{align}
    0 &= \beta\Big[\avg{m}^{2}_{\rm st}-C(\mbf{r})\Big]+\gamma\Delta C(\mbf{r}),
\end{align}
which has the same form as equation (S75) in Ref.~\cite{kauhanen2021geospatial}, but where $2\lambda$ in the expression in the reference is now replaced by $p_{I}'+p_{E}'$ if we were to map back to the model parameters of that earlier reference via Eq.~(\ref{eq supp: science advances model params0}). We note a typographical error in the calculation in \cite{kauhanen2021geospatial}: In equation (S59) of that reference the final term in the last line is $s(\bx')$, whereas it should be $s(\bx)$. Correcting this means that a problematic term which could not be dealt with in this earlier work no longer appears in Eqs.~(\ref{eq supp: 2d lattice dS/dt}) and (\ref{eq supp: 2d lattice dS(x,y)/dt}) of the present paper. In Ref.~\cite{kauhanen2021geospatial}, the problematic term is removed by making the assumption $p_{I}'=p_{E}'$, but the present calculation shows that this  assumption is not actually required to proceed. 
\par

We omit the remainder of the derivation as it follows that in Ref.~\cite{kauhanen2021geospatial}. We find that the expression for the average stationary-state interface is given by
\begin{equation}
   \avg{\sigma}_{\text{st}}=\frac{1}{2}H_{\rm SL}(\tau)\left(1-\avg{m}^{2}_{\text{st}}\right), \label{eq supp: sigma steady-state 2d square lattice}
\end{equation}
the subscript `SL' stands for `square lattice'. This was obtained for a restricted set of model parameters in  \cite{kauhanen2021geospatial} but we now know this to be true when all model parameters are in free variation. The function $H_{\rm SL}(\tau)$ is given by
\begin{equation}
    H_{\rm SL}(\tau)=\frac{\pi(1+\tau)}{2K\left(\frac{1}{1+\tau}\right)}-\tau. \label{eq supp: H 2d square lattice}
\end{equation}
$K(\cdot)$ is the complete elliptical integral of the first kind defined as
\begin{equation}
    K(x) = \int_{0}^{\frac{\pi}{2}}\frac{\dd \theta}{\sqrt{1-x^{2}\text{sin}^{2}(\theta)}}.
\end{equation}
The noise-to-order ratio $\tau$ is defined as in Eq.~(\ref{eq supp: tau}),
\begin{equation}
    \tau = \frac{\beta}{\gamma} = \frac{(h_{+}'+h_{-}')+(v_{+}'+v_{-}')}{\frac{1}{2}\left[(h_{+}-h_{+}')+(h_{-}-h_{-}')\right]}.
\end{equation}
See Sec.~\ref{appendix: H and tau} for detailed discussion on $\tau$ and $H(\tau)$.

We note that $K(x)$ is symmetric about $x=0$ and defined only for $\abs{x} \leq 1$. When $x\to 0$, $K(x)\to \frac{\pi}{2}$, and when $x\to \pm 1$, $K(x)\to \infty$. The argument in Eq.~(\ref{eq supp: H 2d square lattice}) is $\frac{1}{1+\tau}$, thus $K$ is only defined for $\tau \leq -2$ and $\tau \geq 0$. A further discussion on the different limits of $\tau$ can be found in Sec.~\ref{appendix: behaviour in tau regimes}.

\begin{figure}[htbp]
    \centering
    \includegraphics[scale = 0.9]{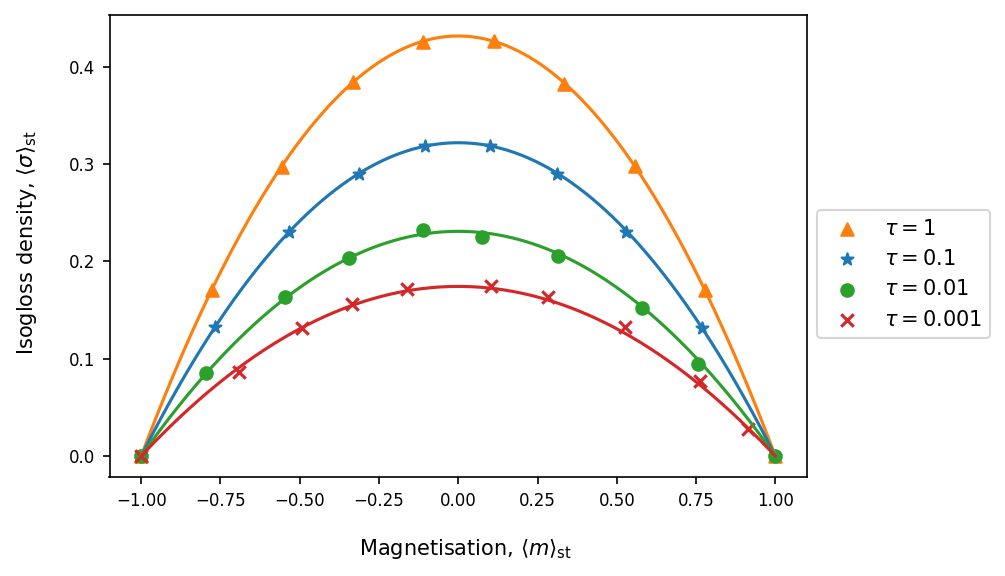}
    \caption{Plots of $\avg{\sigma}_{\rm st}$ against $\avg{m}_{\rm st}$. Solid lines are the analytic solutions, $H_{\rm SL}(\tau)$, from Eq.~(\ref{eq supp: sigma steady-state 2d square lattice}) for different values of the the noise-to-order ratio, $\tau$. These solutions apply for infinite 2D square lattices. Each marker is the result from averaging 100 independent simulations which were performed on 2D square lattices with $N=10,000$ and periodic boundary conditions. Model parameters for a given $\avg{m}_{\rm st}$ and $\tau$ can be generated via the method in Sec.~\ref{appendix: determining model parameters tau and m}.}
    \label{fig main: lattice parabolas}
    \end{figure}

Fig.~\ref{fig main: lattice parabolas} confirms the validity of these results in simulations, and demonstrates that Eq.~(\ref{eq supp: sigma steady-state 2d square lattice}) holds without the restrictions on model parameters that were needed in \cite{kauhanen2021geospatial}. 

\newpage
\FloatBarrier
\section{Approach based on network walks} \label{appendix: random walks}

\subsection{Useful definitions and identities for the further analysis}

\subsubsection{Definitions}
For any non-negative integer $\ell$ we define a walk of length $\ell$ as an ordered set of $\ell+1$ nodes $\bx_1,\bx_2,\dots,\bx_{\ell+1}$ on the network, such that $\bx_{i+1}$ is a nearest neighbour of $\bx_i$ for all $i=1,\dots,\ell$. We do not include networks with self-loops or multi-links, as these are not relevant in the context of the model. We note that the $\ell+1$ nodes do not need to be pairwise different. That is to say, the walker can visit the same node on the network multiple times. 

\begin{figure}[htbp]
    \centering
	\includegraphics[scale=0.4]{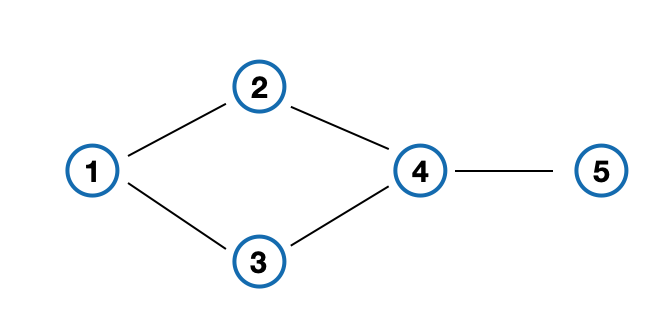}
	\caption{Illustration of walks on a simple network (see text).}
	\label{fig supp:walks}
    \end{figure}

An illustration of a simple network with five nodes is shown in Fig.~\ref{fig supp:walks}. There are four walks of length $\ell=2$ starting at node $1$. These are $(1,2,4), (1,3,4), (1,2,1)$ and $(1,3,1)$. Similarly, there are two walks of length $\ell=2$ starting at node $1$ and ending at node $4$, namely $(1,2,4)$ and $(1,3,4)$. As a further example, there are four walks of length $\ell=3$ starting at $1$ and ending at $2$, these are $(1,2,1,2)$, $(1,2,4,2)$, $(1,3,4,2)$ and $(1,3,1,2)$.  

We define
\BE
    W^{(\ell)}(\bx,\by)&=&\mbox{ number of distinct walks of length $\ell$ starting at $\bx$ and ending at $\by$},\nonumber \\ ~\nonumber \\
    \nu^{(\ell)}(\bx)&=&\mbox{number of distinct walks of length $\ell$ starting at $\bx$}.
\EE
For any node $\bx$ we also introduce 
\be
\bx^{(\ell)}=\big\{\by~|~\by~\mbox{is the endpoint of a walk of length $\ell$ starting at $\bx$}\big\}.
\ee
We note that $\bx^{(\ell)}$ is to be understood as a set, i.e. no node can appear in $\bx^{(\ell)}$ multiple times. Using again the example in Fig.~\ref{fig supp:walks}, and setting $\bx$ to be node $1$, we have
\begin{gather}
    \begin{split}
        \bx^{(1)}&=\{2,3\}, \\
        \bx^{(2)}&=\{1,4\},\\
        \bx^{(3)}&=\{2,3,5\}, \\
        \bx^{(4)}&=\{1,4\}.
    \end{split}
\end{gather}
We then have
\begin{equation}
    \nu^{(\ell)}(\bx)=\sum_{\by\in \bx^{(\ell)}}W^{(\ell)}(\bx,\by).\label{eq supp: nuW}
\end{equation}
We also note that $\nu^{(1)}(\bx)$ is simply the number of nearest neighbours of node $\bx$, i.e. its degree.

The quantity $W^{(\ell)}(\bx,\by)$ is given by the $\bx\by$ element of the $\ell$\textsuperscript{th} power of the adjacency matrix $\bA$ of the network,
\be
W^{(\ell)}(\bx,\by)=(A^\ell)_{\bx\by}.\label{eq supp: aux_id1}
\ee
This can be seen by realising that
\be
(A^\ell)_{\bx\by}=\sum_{\bz_1,\dots,\bz_{\ell-1}}A_{\bx\bz_1}A_{\bz_1\bz_2}...A_{\bz_{\ell-1}\by}, 
\ee
and keeping in mind that elements of the adjacency matrix only take values zero or one. Consequently we also have
\be
\nu^{(\ell)}(\bx)=\sum_\by (A^\ell)_{\bx\by}.\label{eq supp: aux_id2}
\ee

\subsubsection{Identities}
There are several identities we will use throughout in order to manipulate the summations. Recall, as per Sec.~\ref{appendix: model defn}, we only consider undirected networks. 

The first identity is
\BE
     \sum_{\bx}f(\bx)\sum_{\by\in\bx^{(\ell)}}g(\by) &=& \sum_{\bx}\sum_{\by \in \bx^{(\ell)}}f(\bx)g(\by) \nonumber \\
     &=& \sum_{\substack{\bx,\by \\ \mbox{\small $\ell$-connected}}}f(\bx)g(\by) \nonumber \\
     &=& \sum_{\by}\sum_{\bx \in \by^{(\ell)}}f(\bx)g(\by) \nonumber \\
     &=& \sum_{\bx}g(\bx)\sum_{\by \in \bx^{(\ell)}}f(\by), \label{eq supp: random walk identity 1}
\EE
where we say that two nodes $\bx$ and $\by$ are `$\ell$-connected' if there exists a walk of length $\ell$ connecting $\bx$ and $\by$, i.e. if $\by\in\bx^{(\ell)}$ or equivalently, $\bx\in\by^{(\ell)}$. In the last line we swap the labels $\bx$ and $\by$. This identity says that if we have a sum of a function over all nodes, and then an inner sum of a different function over the $\ell$ nearest neighbours of those nodes, we can swap the functions around. 

A similar identity, derived in the same way, valid for any function $f(\bx,\by)$,
\be
\sum_{\bx} \sum_{\by\in\bx^{(\ell)}}f(\bx,\by)=\sum_{\bx} \sum_{\by\in\bx^{(\ell)}}f(\by,\bx). \label{eq supp: random walk identity 1 special case 2}
\ee
A second identity is derived by first noting that for any fixed $\bx$ the following two statements for two nodes $\by$ and $\bz$ are equivalent:
\begin{enumerate}
    \item[(1)]$\by\in\bx^{(\ell)}$ and $\bz\in\by^{(1)}$; 
    \item[(2)] $\bz\in\bx^{(\ell+1)}$ and $\by\in\bx^{(\ell)}\cap\,\bz^{(1)}$.  
\end{enumerate}
Statement (2) implies that $\by\in\bx^{(\ell)}$ and that $\bz$ and $\by$ are nearest neighbours, so statement (1) follows. Further, if $\by\in\bx^{(\ell)}$, i.e. there is a walk of length $\ell$ from $\bx$ to $\by$, and if $\bz$ is a nearest neighbour of $\by$, then there exists a walk of length $\ell+1$ from $\bx$ to $\bz$.  This means that (2) is fulfilled. Using this equivalence we have
\begin{align}
    \sum_{\bx}\sum_{\by\in\bx^{(\ell)}}\sum_{\bz\in\by^{(1)}} f(\bx,\by) g(\bx,\bz) &= \sum_{\bx}\sum_{\bz\in\bx^{(\ell+1)}}\left(\sum_{\by\in\bx^{(\ell)}\cap\,\bz^{(1)}}f(\bx,\by)\right) g(\bx,\bz) \nonumber \\
    &= \sum_{\bx}\sum_{\by\in\bx^{(\ell+1)}}\left(\sum_{\bz\in\bx^{(\ell)}\cap\,\by^{(1)}}f(\bx,\bz)\right) g(\bx,\by), \label{eq supp: aux_id3}
\end{align}
for all functions $f(\cdot,\cdot)$ and $g(\cdot,\cdot)$. 

In the same way the following two statements are equivalent to one another,
\begin{enumerate}
    \item[(1)]$\by\in\bx^{(\ell)}$ and $\bz\in\bx^{(1)}$; 
    \item[(2)] $\by\in\bz^{(\ell+1)}$ and $\bx\in\by^{(\ell)}\cap\,\bz^{(1)}$,
\end{enumerate}
leading to the identity,
\begin{align}
    \sum_{\bx}\sum_{\by \in \bx^{(\ell)}}\sum_{\bz\in\bx^{(1)}}f(\bx, \by)g(\by, \bz) &= \sum_{\bz}\sum_{\by\in\bz^{(\ell + 1)}}\left(\sum_{\bx \in \by^{\ell} \cap \bz^{(1)}}f(\bx, \by)\right)g(\by, \bz) \nonumber \\
    &= \sum_{\bx}\sum_{\by\in\bx^{(\ell + 1)}}\left(\sum_{\bz\in \by^{(\ell)}\cap\bx^{(1)}}f(\bz, \by)\right)g(\by, \bx), \label{eq supp: aux_id4}
\end{align}
for all functions $f(\cdot,\cdot)$ and $g(\cdot,\cdot)$.

\subsection{Magnetisation} \label{appendix: random walks magnetisation} 
The spin-flip probability, $w(\bx)$, in the case of a general network is constructed in exactly the same way as that of the infinite square lattice in Sec.~\ref{appendix: square lattice}. The only difference is that the pre-factor $1/4$ in Eq.~(\ref{eq supp: square lattice B(x)}) becomes $1/\nu^{(1)}(\bx)$. This leads to an equation for $w(\bx)$ analogous to Eq.~(\ref{eq supp: 2D lattice spin flip full}),
\begin{align}
    w(\bx) = \frac{\eta}{2} - \frac{\alpha}{2}s(\bx)-\frac{\gamma}{2}\frac{1}{\nu^{(1)}(\bx)}\sum_{\by\in\bx^{(1)}}s(\bx)s(\by), \label{eq supp: random walk spin flip full}
\end{align}
where we have used the shorthands from Eq.~(\ref{eq:shorthands}). We note that we have already made the restriction from Eq.~(\ref{eq supp: gamma restriction}), $\gamma_{+}=\gamma_{-}$. Using this we can write an equation which is analogous to Eq.~(\ref{eq supp: 2d lattice dS/dt}),
\begin{align}
    \frac{\dd S(\bx)}{\dd t} &= \alpha -\eta S(\bx) + \gamma \frac{1}{\nu^{(1)}(\bx)}\sum_{\by \in \bx^{(1)}}S(\by).\label{eq supp: random walk dS/dt mag}
\end{align}
We then take a sum over all nodes $\bx$ and find the following differential equation for the magnetisation,
\begin{align}
    \frac{\dd \avg{m}}{\dd t} &=  \alpha -\eta \avg{m} + \gamma \frac{1}{N}\sum_{\bx}\frac{1}{\nu^{(1)}(\bx)}\sum_{\by \in \bx^{(1)}}S(\by) \nonumber \\
    &= \alpha -\eta \avg{m} + \gamma \frac{1}{N}\sum_{\bx}S(\bx)\sum_{\by\in\bx^{(1)}}\frac{1}{\nu^{(1)}(\by)},\label{eq supp: random walk magnetisation differential equation initial}
\end{align}
where we have made use of Eq.~(\ref{eq supp: random walk identity 1}).

For each node $\bx$ we introduce
\begin{equation}\label{eq:lambda_1}
    \lambda_{\bx, 1} = \sum_{\by\in\bx^{(1)}}\frac{1}{\nu^{(1)}(\by)},
\end{equation}
and we further define
\begin{equation}
    \avg{m_{1}} = \frac{1}{N}\sum_{\bx}\lambda_{\bx, 1}S(\bx).\label{eq supp: random walk lambda 1}
\end{equation}

We can write Eq.~(\ref{eq supp: random walk magnetisation differential equation initial}) as
\begin{align}
    \frac{\dd \avg{m}}{\dd t} 
    &= \alpha -\eta \avg{m} + \gamma \frac{1}{N}\sum_{\bx}\lambda_{\bx, 1}S(\bx) \nonumber \\
    &= \alpha -\eta \avg{m} + \gamma \avg{m_{1}}. \label{eq supp: random walk magnetisation differential equation_aux}
\end{align}
Next, using Eq.~(\ref{eq supp: random walk dS/dt mag}), we have
\begin{align}
    \frac{\dd \avg{m_{1}}}{\dd t} &= \alpha \Lambda_{1} -\eta \avg{m_{1}}+\gamma \frac{1}{N}\sum_{\bx}\lambda_{\bx, 1}\frac{1}{\nu^{(1)}(\bx)}\sum_{\by\in\bx^{(1)}}S(\by), \label{eq supp: general network dm1/dt}
\end{align}
where we have defined
\be
 \Lambda_{1} = \frac{1}{N}\sum_{\bx}\lambda_{\bx, 1}.
\ee
We can make the following simplification,
\begin{align} 
    \Lambda_{1} &= \frac{1}{N}\sum_{\bx}\lambda_{\bx, 1} \nonumber \\
    & = \frac{1}{N}\sum_{\bx}\sum_{\by\in\bx^{(1)}}\frac{1}{\nu^{(1)}(\by)} \nonumber \\
    &= \frac{1}{N}\sum_{\bx}\frac{1}{\nu^{(1)}(\bx)}\sum_{\by \in \bx^{(1)}}1 \nonumber \\
    &= \frac{1}{N}\sum_{\bx}\frac{1}{\nu^{(1)}(\bx)}\cdot\nu^{(1)}(\bx) \nonumber \\
    &= 1, \label{eq:Lambda1eq1}
\end{align}
where we have used Eq.~(\ref{eq supp: random walk identity 1}) in going from the second to third line. Thus, we have
\begin{align}
    \frac{\dd \avg{m_{1}}}{\dd t} &= \alpha -\eta \avg{m_{1}}+\gamma \frac{1}{N}\sum_{\bx}\lambda_{\bx, 1}\frac{1}{\nu^{(1)}(\bx)}\sum_{\by\in\bx^{(1)}}S(\by) \nonumber \\
    &= \alpha -\eta \avg{m_{1}}+\gamma \avg{m_{2}},
\end{align}
with the definition
\be
  \avg{m_{2}}\equiv \frac{1}{N}\sum_{\bx}\lambda_{\bx, 1}\frac{1}{\nu^{(1)}(\bx)}\sum_{\by\in\bx^{(1)}}S(\by).
\ee
This expression can be simplified as follows,
\begin{align}
     \avg{m_{2}}&=\frac{1}{N}\sum_{\bx}\lambda_{\bx, 1}\frac{1}{\nu^{(1)}(\bx)}\sum_{\by\in\bx^{(1)}}S(\by) \nonumber \\
     &= \frac{1}{N}\sum_{\bx}S(\bx)\sum_{\by\in\bx^{(1)}}\frac{1}{\nu^{(1)}(\by)}\lambda_{\by, 1} \nonumber \\
    & = \frac{1}{N}\sum_{\bx}S(\bx)\sum_{\by\in\bx^{(1)}}\frac{1}{\nu^{(1)}(\by)}\sum_{\bz\in\by^{(1)}}\frac{1}{\nu^{(1)}(\bz)} \nonumber \\
    &\equiv \frac{1}{N}\sum_{\bx}\lambda_{\bx, 2}S(\bx),
\end{align}
where we again have used Eq.~(\ref{eq supp: random walk identity 1}) in going from the first to second line, and where we have defined
\be
\lambda_{\bx, 2}\equiv \sum_{\by\in\bx^{(1)}}\frac{1}{\nu^{(1)}(\by)}\sum_{\bz\in\by^{(1)}}\frac{1}{\nu^{(1)}(\bz)}.
\ee
We now make the following recursive definition:
\begin{equation}
    \begin{aligned}
        \lambda_{\bx, n} = 
        \begin{cases}
             \sum_{\bx_{1}\in \bx^{(1)}}\frac{1}{\nu^{(1)}(\bx_{1})}\lambda_{\bx_{1}, n-1} \hspace{5mm} &n = 1,2,3,... \\
             1 \hspace{5mm} &n = 0
        \end{cases}.
    \end{aligned}
\end{equation}
Further we write
\be
    \avg{m_{n}}\equiv \frac{1}{N}\sum_{\bx} \lambda_{\bx,n}S(\bx). \label{eq supp: general mag n}
\ee
We also define
 \be
 \Lambda_n\equiv \frac{1}{N}\sum_{\bx}\lambda_{\bx, n},
 \ee
and note that
\begin{align}
    \Lambda_n&=\frac{1}{N}\sum_{\bx}\lambda_{\bx, n} \nonumber \\
    &= \frac{1}{N}\sum_{\bx}\sum_{\bx_{1}\in\bx^{(1)}}\frac{1}{\nu^{(1)}(\bx_{1})}\lambda_{\bx_{1}, n-1} \nonumber \\
    &= \frac{1}{N}\sum_{\bx}\frac{\lambda_{\bx, n-1}}{\nu^{(1)}(\bx)}\sum_{\bx_{1}\in\bx^{(1)}} 1 \nonumber \\
    &= \frac{1}{N}\sum_{\bx}\frac{\lambda_{\bx, n-1}}{\nu^{(1)}(\bx)}\cdot\nu^{(1)}(\bx) \nonumber \\
    &= \frac{1}{N}\sum_{\bx}\lambda_{\bx, n-1} = \Lambda_{n-1}.
\end{align}
Thus, all $\Lambda_n$ take the same value, and using Eq.~(\ref{eq:Lambda1eq1}), we find $\Lambda_n=1$ for all $n$.

From Eq.~(\ref{eq supp: random walk dS/dt mag}) we can derive differential equations for general $\avg{m_{n}}$, $n=1, 2, 3,\dots$, leading to the following hierarchy, 
\begin{equation}
    \frac{\dd \avg{m_{n}}}{\dd t} = \alpha -\eta \avg{m_{n}} + \gamma \avg{m_{n+1}}. \label{eq supp: general network dmn/dt}
\end{equation}
Next, we take the steady-state limit, and multiply the equation through by $\frac{\gamma^{n}}{\eta^{n}}$, and sum both sides over $n$ (from $0$ to $\infty$),
\begin{align}
    0 &= \alpha\sum_{n=0}^{\infty}\frac{\gamma^{n}}{\eta^{n}}-\sum_{n=0}^{\infty}\left(\frac{\gamma^{n}}{\eta^{n-1}}\avg{m_{n}}_{\rm st}-\frac{\gamma^{n+1}}{\eta^{n}}\avg{m_{n+1}}_{\rm st}\right) \nonumber \\
    &= \alpha\frac{1}{1-\frac{\gamma}{\eta}}-\eta \avg{m}_{\rm st}+\lim_{n\to\infty}\left(\frac{\gamma^{n+1}}{\eta^{n}}\avg{m_{n+1}}_{\rm st}\right).\label{eq:summed_recursion}
\end{align}
The first term results from an infinite geometric series, which only converges provided $\abs{\gamma/\eta}=\abs{1/(1+\tau)}<1$ [where we have used the definitions in Eqs.~(\ref{eq:shorthands}) and (\ref{eq supp: tau})]. This is only the case provided that $\tau < -2$ or $\tau >0$. The second and third terms in the second line of Eq.~(\ref{eq:summed_recursion}) result from the telescopic sum in the first line. We note that $\avg{m_{0}}=\avg{m}$ by construction.

The limit in the second line of Eq.~(\ref{eq:summed_recursion}) vanishes. Recall that all $\lambda_{\bx,n}\geq 0$ and $\Lambda_n\equiv \frac{1}{N}\sum_{\bx}\lambda_{\bx, n}=1$, so we have $0\leq \lambda_{\bx,n}\leq 1$. Then from the definition in Eq.~(\ref{eq supp: general mag n}), we always have $-1 \leq \avg{m_{n}}_{\rm st} \leq 1$. Since $\abs{\gamma/\eta}<1$ the limit then vanishes.

Ultimately then we can solve for $\avg{m}_{\rm st}$. We find
\begin{equation}
    \avg{m}_{\rm st}=\frac{\alpha}{\eta-\gamma} = \frac{\alpha}{\beta}, \label{eq supp: random walk steady-state mag}
\end{equation}
where $\alpha$ and $\beta$ are given in Eq.~(\ref{eq:shorthands}).

This coincides with the result we obtained for infinite complete networks, Eq.~(\ref{eq supp: complete network steady-state magnetisation}), using the pair approximation [see Sec.~\ref{appendix: pair approximation magnetisation}], and using the annealed approximation, Eq.~(\ref{eq supp: annealed steady mag}). We note that the derivation in the current section did not require any approximation, and is therefore valid for any finite undirected network. 

\subsection{Generalisation to `weighted magnetisation'} \label{appendix: random walks weighted mag}
It is possible to generalise the procedure in the previous subsection. Consider the following `weighted' magnetisation,
\begin{equation}
    \avg{m_{z}} = \frac{\frac{1}{N}\sum_{\bx}z_{\bx}S(\bx)}{\frac{1}{N}\sum_{\bf x}z_{\bf x}},
\end{equation}
where the $z_{\bx}\geq 0$ are site dependent weightings. We define
\be
    \avg{m_{z, n}} = \frac{\frac{1}{N}\sum_{\bx}z_{\bx, n}S(\bx)}{\frac{1}{N}\sum_{\bf x}z_{\bf x, n}},
\ee
with
\begin{equation}
    \begin{aligned}
        z_{\bx, n} = 
        \begin{cases}
            \sum_{\bx_{1} \in \bx^{(1)}}\frac{1}{\nu^{(1)}(\bx_{1})}z_{\bx_{1}, n-1} \hspace{5mm} &n = 1,2,3,... \\
            z_{\bx} \hspace{5mm} &n=0
        \end{cases}.
    \end{aligned}
\end{equation}

Similar to Eq.~(\ref{eq:Lambda1eq1}) we then have
\be
    \zeta_{n} \equiv \frac{1}{N}\sum_{\bx}z_{\bx, n} = 1 \hspace{5mm} \forall n.
    \ee
 Thus the differential equation for $\avg{m_{z, n}}$ follows the same recurrence relation as Eq.~(\ref{eq supp: general network dmn/dt}),
\begin{equation}
    \frac{\dd \avg{m_{z, n}}}{\dd t} = \alpha -\eta \avg{m_{z, n}} + \gamma \avg{m_{z, n+1}}.
\end{equation}
We can solve this in the same way as before, leading to
\begin{align}
    \avg{m_{z}}_{\rm st} &= \avg{m}_{\rm st}. \label{eq supp: random walk weighted steady-state mag}
\end{align}
Thus the steady-state `weighted' magnetisation is always the same, regardless of the weighting. This means, for example, that the magnetisation of a group of specific degree nodes will have the same steady-state independent of the degree chosen. This will become relevant in Sec.~\ref{appendix: SPA}.

\FloatBarrier
\subsection{Interface density} \label{appendix: random walks interface density}
It turns out to be convenient to introduce the following object,
\begin{equation}
    S^{(\ell)}(\bx) = \frac{1}{\nu^{(\ell)}(\bx)}\sum_{\by\in \bx^{(\ell)}}W^{(\ell)}(\bx,\by)\left<s(\bx)s(\by)\right>.\label{eq supp: random walk correlation}
\end{equation}
Broadly speaking, this is the correlation function of $s(\bx)$ with spins at sites $\by$ that can be reached from $\bx$ in walks of length $\ell$. Each $\by\in\bx^{(\ell)}$ is weighted by the number of walks of length $\ell$ connecting $\bx$ and $\by$. Noting the definition of $\nu^{(\ell)}(\bx)$ from Eq.~(\ref{eq supp: nuW}), the pre-factor $1/\nu^{(\ell)}(\bx)$ ensures overall normalisation, such that $-1\leq S^{(\ell)}(\bx)\leq 1$.  

As in Sec.~\ref{appendix: square lattice interface density}, we use again the fact that $s(\bx)s(\by)$ changes by an amount $-2s(\bx)s(\by)$ when either $s(\bx)$ or $s(\by)$ flip ($\bx\neq \by$). Noting that the node $\bx$ itself may be contained in the set $\bx^{(\ell)}$ for some $\ell$, and recognising that $s(\bx)^2$ always takes the value one (and hence does not change if the spin at $\bx$ flips), we find in the continuous-time limit
\begin{align}
    \frac{\dd S^{(\ell)}(x)}{\dd t} &=  \frac{4W^{(\ell)}(\bx,\bx)}{\nu^{(\ell)}(\bx)}\left<w(\bx)\right> -\frac{2}{\nu^{(\ell)}(\bx)}\sum_{\by\in \bx^{(\ell)}}W^{(\ell)}(\bx,\by) \left<s(\bx)s(\by)\left[w(\bx)+w(\by)\right]\right>. \label{eq supp: random walk dS/dt initial}
\end{align}
The first term corrects for any contribution from $\by=\bx$ in the second term (in the event that $\bx\in\bx^{(\ell)}$).

We will now use Eq.~(\ref{eq supp: random walk spin flip full}), which we repeat here for convenience,
\begin{align}w(\bx) 
    &= \frac{\eta}{2} - \frac{\alpha}{2}s(\bx)-\frac{\gamma}{2}\frac{1}{\nu^{(1)}(\bx)}\sum_{\by\in\bx^{(1)}}s(\bx)s(\by), \label{eq supp: random walk spin flip full-2}
\end{align}
and the definition in Eq.~(\ref{eq supp: random walk correlation}). We find 
\begin{equation}
    \left<w(\bx)\right> = \frac{1}{2}\left[\eta - \alpha S(\bx)-\gamma S^{(1)}(\bx)\right]. \label{eq supp: random walk w expectation}
\end{equation}
We also have
\begin{align}
    \sum_{\by\in \bx^{(\ell)}}&W^{(\ell)}(\bx,\by)\left<s(\bx)s(\by)\left[w(\bx)+w(\by)\right]\right>  \nonumber \\
    & = \eta\nu^{(\ell)}(\bx) S^{(\ell)}(\bx) -\frac{\alpha}{2}\Bigg[\nu^{(\ell)}(\bx)S(\bx)+ \sum_{\by\in \bx^{(\ell)}}W^{(\ell)}(\bx,\by)S(\by)\Bigg] \nonumber \\
    &\qquad -\frac{\gamma}{2}\sum_{\by\in \bx^{(\ell)}}W^{(\ell)}(\bx,\by)
    \left[\frac{1}{\nu^{(1)}(\by)}\sum_{\bz\in \by^{(1)}}\left<s(\bx)s(\bz)\right> + \frac{1}{\nu^{(1)}(\bx)}\sum_{\bz\in \bx^{(1)}}\left<s(\bz)s(\by)\right>\right]. \label{eq supp: random walk more complicated expectation}
\end{align}
Eq.~(\ref{eq supp: random walk dS/dt initial}) then becomes 
\begin{align}
    \frac{\dd S^{(\ell)}(\bx)}{\dd t} &= \frac{2W^{(\ell)}(\bx,\bx)}{\nu^{(\ell)}(\bx)}\left[\eta-\alpha S(\bx)-\gamma S^{(1)}(\bx)\right] \nonumber \\
    &\qquad -2\eta S^{(\ell)}(\bx) +\alpha\Bigg[S(\bx)+ \frac{1}{\nu^{(\ell)}(\bx)}\sum_{\by\in \bx^{(\ell)}}W^{(\ell)}(\bx,\by)S(\by)\Bigg] \nonumber \\
    &\qquad + \frac{\gamma}{\nu^{(\ell)}(\bx)}\sum_{\by\in \bx^{(\ell)}}W^{(\ell)}(\bx,\by)
    \left[\frac{1}{\nu^{(1)}(\by)}\sum_{\bz\in \by^{(1)}}\left<s(\bx)s(\bz)\right> + \frac{1}{\nu^{(1)}(\bx)}\sum_{\bz\in \bx^{(1)}}\left<s(\bz)s(\by)\right>\right].
        \label{eq supp: random walk dS/dt}
\end{align}
In the steady-state, the derivative on the left-hand side vanishes. We can then divide through by $\gamma$ so that the model parameters only appear in the following combinations,
\begin{subequations}
\begin{align}
    \frac{\eta}{\gamma} &= 1+\tau,  \label{eq supp: eta over gamma relation}\\
    \frac{\alpha}{\gamma} &= \tau\avg{m}_{\rm st}, \label{eq supp: alpha over gamma relation}
\end{align}
\end{subequations}
where $\avg{m}_{\rm st}$ and $\tau$ are as in Eqs.~(\ref{eq supp: complete network steady-state magnetisation}) and (\ref{eq supp: tau}) respectively. As we will show below, this means that the steady-state interface density can also be determined solely from $\avg{m}_{\rm st}$ and $\tau$. \par
We now sum both sides of Eq.~(\ref{eq supp: random walk dS/dt}) over all nodes $\bx$ in the network to derive an equation for the overall correlation $C^{(\ell)}$, defined as
\begin{equation}
    C^{(\ell)}\equiv\frac{1}{N}\sum_{\bx}S^{(\ell)}(\bx). \label{eq supp: random walk average correlation}
\end{equation}
We find,
\begin{align}
    \frac{\dd C^{(\ell)}}{\dd t} &= 2\frac{1}{N}\sum_{\bx}\left\{\frac{W^{(\ell)}(\bx,\bx)}{\nu^{(\ell)}(\bx)}\left[\eta-\alpha S(\bx) -\gamma S^{(1)}(\bx)\right]\right\} - 2\eta C^{(\ell)} \nonumber \\
    &\qquad +\alpha\Bigg[\avg{m}+\frac{1}{N}\sum_{\bx}\frac{1}{\nu^{(\ell)}(\bx)} \sum_{\by\in \bx^{(\ell)}}W^{(\ell)}(\bx,\by)S(\by)\Bigg] \nonumber \\
    &\qquad +\gamma\frac{1}{N}\sum_{\bx}\frac{1}{\nu^{(\ell)}(\bx)}\sum_{\by\in \bx^{(\ell)}} W^{(\ell)}(\bx,\by)\Bigg[\frac{1}{\nu^{(1)}(\by)}\sum_{\bz\in \by^{(1)}}\left<s(\bx)s(\bz)\right> \nonumber \\
    &\qquad\qquad\qquad\qquad\qquad\qquad\qquad\qquad\qquad +\frac{1}{\nu^{(1)}(\bx)}\sum_{\bz \in \bx^{(1)}}\left<s(\bz)s(\by)\right>\Bigg]. \label{eq supp: random walk dC/dt}
\end{align}
\par
To proceed we introduce a number of shorthands. We write  
\begin{subequations}
\begin{align}
    \Omega^{(\ell)}_{\bx} &\equiv \frac{W^{(\ell)}(\bx,\bx)}{\nu^{(\ell)}(\bx)}, \label{eq supp: def greeks 1}\\
    \chi^{(\ell)}_{\bx} &\equiv \sum_{\by\in\bx^{(\ell)}}\frac{W^{(\ell)}(\by, \bx)}{\nu^{(\ell)}(\by)}, \\
    \psi^{(\ell+1)}_{\bx\by} &\equiv \sum_{\bz\in\bx^{(\ell)}\cap\by^{(1)}}\left[\frac{W^{(\ell)}(\bx,\bz)W^{(1)}(\bz, \by)}{\nu^{(\ell)}(\bx)\nu^{(1)}(\bz)}\right], \label{eq supp: def greeks 3} \\
    \phi^{(\ell+1)}_{\bx\by} &\equiv \sum_{\bz\in\by^{(\ell)}\cap\,\bx^{(1)}}\left[\frac{W^{(\ell)}(\bz, \by)W^{(1)}(\bz, \bx)}{\nu^{(\ell)}(\bz)\nu^{(1)}(\bz)}\right]. \label{eq supp: def greeks 4}
\end{align}
\end{subequations}
In practice these coefficients can be evaluated from the $N\times N$ adjacency matrix $\bA$ of the network. We here recall that the $\bx\by$ element of the $\ell$\textsuperscript{th} power of the adjacency gives the number of possible length $\ell$ walks starting at $\bx$ and ending at $\by$, denoted $W^{(\ell)}(\bx, \by)$, see Eq.~(\ref{eq supp: aux_id1}). The sum of such elements over all $\by$ gives the total number of walks of length $\ell$ starting from $\bx$, denoted $\nu^{(\ell)}(\bx)$, see Eq.~(\ref{eq supp: aux_id2}). Using these, we have
\begin{subequations}
\begin{align}
    \Omega^{(\ell)}_{\bx} &= \frac{(A^{\ell})_{\bx\bx}}{\sum_{\by}(A^{\ell})_{\bx\by}}, \\
    \chi^{(\ell)}_{\bx} &= \sum_{\by}\frac{(A^{\ell})_{\by\bx}}{\sum_{\bz}(A^{\ell})_{\by\bz}},   \\
    \psi^{(\ell+1)}_{\bx\by} &= \frac{1}{\sum_{\bz}(A^{\ell})_{\bx\bz}}\left[\sum_{\bz}(A^{\ell})_{\bx\bz}\frac{1}{\sum_{\bu}A_{\bz\bu}}A_{\bz\by}\right],   \\
    \phi^{(\ell+1)}_{\bx\by} &= \sum_{\bz}\left[\frac{(A^{\ell})_{\bz\by}A_{\bz\bx}}{\sum_{\bu}(A^{k})_{\bz\bu}\sum_{\bu }A_{\bz\bu}}\right]. \label{eq:defs_via_adjacency}
\end{align}
\end{subequations}
We now proceed to write Eq.~(\ref{eq supp: random walk dC/dt}) in more compact form. Using Eq.~(\ref{eq supp: random walk identity 1 special case 2}) we observe for the term on the second line,
\begin{align}
    \frac{1}{N}\sum_{\bx}\frac{1}{\nu^{(\ell)}(\bx)}\sum_{\by\in\bx^{(\ell)}}W^{(\ell)}(\bx, \by)S(\by) &= \frac{1}{N}\sum_{\bx}\sum_{\by\in\bx^{(\ell)}}\left(\frac{W^{(\ell)}(\bx, \by)}{\nu^{(\ell)}(\bx)}S(\by)\right) \nonumber \\
    &= \frac{1}{N}\sum_{\bx}\left(\sum_{\by\in\bx^{(\ell)}}\frac{W^{(\ell)}(\by, \bx)}{\nu^{(\ell)}(\by)}\right)S(\bx) \nonumber \\
    &= \frac{1}{N}\sum_{\bx}\chi_{\bx}^{(\ell)}S(\bx).
\end{align}
For the first term on the third line,
\begin{align}
    & \frac{1}{N}\sum_{\bx}\frac{1}{\nu^{(\ell)}(\bx)}\sum_{\by\in \bx^{(\ell)}}W^{(\ell)}(\bx,\by)\frac{1}{\nu^{(1)}(\by)} \sum_{\bz\in \by^{(1)}}\left<s(\bx)s(\bz)\right> \nonumber \\
    &\qquad = \frac{1}{N}\sum_{\bx}\sum_{\by\in\bx^{(\ell)}}\sum_{\bz\in\by^{(1)}}\frac{W^{(\ell)}(\bx,\by)}{\nu^{(\ell)}(\bx)\nu^{(1)}(\by)}\avg{s(\bx)s(\bz)} \nonumber \\
   &\qquad = \frac{1}{N}\sum_{\bx}\sum_{\by\in\bx^{(\ell+1)}}\left(\sum_{\bz\in\bx^{(\ell)}\cap\,\by^{(1)}}\frac{W^{(\ell)}(\bx,\bz)}{\nu^{(\ell)}(\bx)\nu^{(1)}(\bz)}\right)\left<s(\bx)s(\by)\right> \nonumber \\
   &\qquad = \frac{1}{N}\sum_{\bx}\sum_{\by\in\bx^{(\ell+1)}}\left(\sum_{\bz\in\bx^{(\ell)}\cap\,\by^{(1)}}\frac{W^{(\ell)}(\bx,\bz)W^{(1)}(\bz, \by)}{\nu^{(\ell)}(\bx)\nu^{(1)}(\bz)}\right)\left<s(\bx)s(\by)\right> \nonumber \\
    &\qquad = \frac{1}{N}\sum_{\bx}\sum_{\by\in \bx^{(\ell+1)}}\psi^{(\ell+1)}_{\bx\by}\left<s(\bx)s(\by)\right>,
\end{align}
where we have first used Eq.~(\ref{eq supp: aux_id3}) in going from the second to third line, and then introduced $W^{(1)}(\bz,\by)=1$ for $\bz\in\by^{(1)}$, before finally using the definition of $\psi^{(\ell)}_{\bx, \by}$ from Eq.~(\ref{eq supp: def greeks 3}). In a similar manner one shows that
\begin{align}
    &\frac{1}{N}\sum_{\bx}\frac{1}{\nu^{(\ell)}(\bx)}\sum_{\by\in\bx^{(\ell)}}W^{(\ell)}(\bx,\by)\frac{1}{\nu^{(1)}(\bx)}\sum_{\bz\in\bx^{(1)}}\left<s(\bz)s(\by)\right> \nonumber \\
    &\qquad = \frac{1}{N}\sum_{\bx}\sum_{\by\in\bx^{(\ell)}}\sum_{\bz\in\by^{(1)}}\frac{W^{(\ell)}(\bx,\by)}{\nu^{(\ell)}(\bx)\nu^{(1)}(\bx)}\left<s(\bz)s(\by)\right> \nonumber \\
    &\qquad =\frac{1}{N}\sum_{\bx}\sum_{\by\in\bx^{(\ell+1)}}\left(\sum_{\bz\in\by^{(\ell)}\cap\,\bx^{(1)}}\frac{W^{(\ell)}(\bz,\by)}{\nu^{(\ell)}(\bz)\nu^{(1)}(\bz)}\right)\left<s(\bx)s(\by)\right> \nonumber \\
    &\qquad =\frac{1}{N}\sum_{\bx}\sum_{\by\in\bx^{(\ell+1)}}\left(\sum_{\bz\in\by^{(\ell)}\cap\,\bx^{(1)}}\frac{W^{(1)}(\bx,\bz)W^{(\ell)}(\bz,\by)}{\nu^{(\ell)}(\bz)\nu^{(1)}(\bz)}\right)\left<s(\bx)s(\by)\right>\nonumber \\
    &\qquad = \frac{1}{N}\sum_{\bx}\sum_{\by\in\bx^{(\ell+1)}}\phi^{(\ell+1)}_{\bx\by}\left<s(\bx)s(\by)\right>,
\end{align}
where we have first used Eq.~(\ref{eq supp: aux_id4}) in going from the second to third line, then introduced $W^{(1)}(\bx,\bz)=1$ for $\bx\in\bz^{(1)}$, and finally used the definition of $\phi^{(\ell)}_{\bx,\by}$ from Eq.~(\ref{eq supp: def greeks 4}). Eq.~(\ref{eq supp: random walk dC/dt}) then becomes
\begin{align}
    \frac{\dd C^{(\ell)}}{\dd t} =& 2\eta\frac{1}{N}\sum_{\bx}\Omega^{(\ell)}_{\bx} -\alpha\frac{1}{N}\sum_{\bx}\left[2\Omega^{(\ell)}_{\bx}-\chi^{(\ell)}_{\bx}\right]S(\bx) +\alpha \avg{m} - 2\eta C^{(\ell)} \nonumber \\
    &\qquad - 2\gamma\frac{1}{N}\sum_{\bx}\Omega^{(\ell)}_{\bx}S^{(1)}(\bx) +\gamma\frac{1}{N}\sum_{\bx}\sum_{\by\in \bx^{(\ell+1)}}    \left[\psi^{(\ell+1)}_{\bx\by}+\phi^{(\ell+1)}_{\bx\by}\right]\left<s(\bx)s(\by)\right>. \label{eq supp: minor simplified dC/dt no approx}
\end{align}
Several of the terms in the last expression take the form of weighted magnetisations, whose steady-state we can determine from Eq.~(\ref{eq supp: random walk weighted steady-state mag}). However the two final terms are problematic and we have not attempted to solve this equation in full generality. Progress is possible for what we will call `homogeneous' networks, as discussed in the next section.

Alternatively we can choose model parameters such that $\gamma = 0$. However, this would mean that $\tau \to \pm \infty$ and the steady-state configuration would always be random which is not of particular interest (see also Sec.~\ref{appendix: H and tau}).

\subsection{Homogeneous networks} \label{appendix: homogeneous networks}
\subsubsection{Homogeneity assumptions}
We define networks that fulfill the following two properties as \textit{homogeneous}:
\begin{enumerate}
    \item[(1)] the total number walks of length $\ell$ starting and ending at any node $\bx$ is the same for all $\bx$, $W^{(\ell)}(\bx, \bx)\equiv W^{(\ell)}$;
    \item[(2)] the total number of walks of length $\ell$ starting at $\bx$ (ending at any point) is also the same for all $\bx$, $\nu^{(\ell)}(\bx) \equiv \nu^{(\ell)}$.
\end{enumerate}

There are several types of network that fall under this category. For example, infinite regular lattices in any dimension or Bethe lattices (infinite regular trees) have these properties. Individual realisations of degree-regular networks do not fulfill (1) and (2), but these conditions hold as averages over the ensemble of degree-regular networks, i.e. the expected value of $W^{(\ell)}(\bx,\bx)$ is the same for all nodes $\bx$, and similarly for the expected value of $\nu^{(\ell)}(\bx)$. \par

We now proceed making these assumptions, writing $W^{(\ell)}$ and $\nu^{(\ell)}$ for the common values of all $W^{(\ell)}(\bx, \bx)$ and $\nu^{(\ell)}(\bx)$ respectively. The assumptions imply in particular that all nodes in the network have the same degree, $\nu^{(1)}$. We note that we then have $\nu^{(\ell+1)}=\nu^{(\ell)}\nu^{(1)}$, and from this we find $\nu^{(\ell)}=(\nu^{(1)})^\ell$. \par
Noting again that $W^{(1)}(\bz,\by)=A_{\bz\by}$ is equal to one if and only if $\bz$ and $\by$ are nearest neighbours, and zero otherwise, Eqs.~(\ref{eq supp: def greeks 1})-(\ref{eq supp: def greeks 4}) can then be simplified as follows:
\begin{subequations}
\begin{align}
    \Omega^{(\ell)}_{\bx} &= \frac{W^{(\ell)}(\bx,\bx)}{\nu^{(\ell)}(\bx)} = \frac{W^{(\ell)}}{\nu^{(\ell)}} \equiv \Omega^{(\ell)}, \hspace{10mm} \forall \bx \label{eq supp: homogeneous network omega} \\
    \chi^{(\ell)}_{\bx} &= \sum_{\by\in\bx^{(\ell)}}\frac{W^{(\ell)}(\by, \bx)}{\nu^{(\ell)}(\by)} = \frac{1}{\nu^{(\ell)}}\sum_{\by\in\bx^{(\ell)}}W^{(\ell)}(\by, \bx) = 1, \hspace{10mm} \forall \bx \\  
    \psi^{(\ell+1)}_{\bx\by} &= \frac{1}{\nu^{(\ell)}(\bx)}\left[\sum_{\bz\in\bx^{(\ell)}\cap\by^{(1)}}W^{(\ell)}(\bx,\bz)\frac{1}{\nu^{(1)}(\bz)}W^{(1)}(\bz, \by)\right] \nonumber \\
    &= \frac{1}{\nu^{(\ell)}\nu^{(1)}}\sum_{\bz\in\by^{(1)}}W^{(\ell)}(\bx,\bz)W^{(1)}(\bz, \by)\nonumber \\
    &= \frac{W^{(\ell+1)}(\bx,\by)}{\nu^{(\ell+1)}}, \hspace{10mm} \forall \bx,\by \\
    \phi^{(\ell+1)}_{\bx\by} &= \sum_{\bz\in\by^{(\ell)}\cap\bx^{(1)}}\left[\frac{1}{\nu^{(\ell)}(\bz)}W^{(\ell)}(\bz, \by)\frac{1}{\nu^{(1)}(\bz)}W^{(1)}(\bz, \bx)\right] \nonumber \\
    &=  \frac{1}{\nu^{(\ell+1)}}\sum_{\bz\in\bx^{(1)}}W^{(\ell)}(\bz, \by) \nonumber \\
    &= \frac{W^{(\ell+1)}(\bx, \by)}{\nu^{(\ell+1)}}. \hspace{10mm} \forall \bx,\by 
\end{align}
\end{subequations}

\subsubsection{Interface density for homogeneous networks, and resulting simplifications} \label{appendix: homogeneous networks interface density}
Assuming that the underlying network is homogeneous (in the sense as defined above), Eq.~(\ref{eq supp: minor simplified dC/dt no approx}) reduces to
\begin{equation}
    \frac{\dd C^{(\ell)}}{\dd t} = 2\Omega^{(\ell)}\Big[\eta -\alpha \avg{m} -\gamma C^{(1)}\Big]-2\Big[\eta C^{(\ell)} -\alpha \avg{m} - \gamma C^{(\ell+1)}\Big]. \label{eq supp: homogeneous network dC/dt}
\end{equation}
In the steady-state we then have the following relation:
\begin{equation}
    \eta C^{(\ell)}_{\rm st}-\gamma C^{(\ell+1)}_{\rm st} = \Omega^{(\ell)}\Big[\eta -\alpha \avg{m}_{\rm st} - \gamma C^{(1)}_{\rm st}\Big] +\alpha \avg{m}_{\rm st}. \label{eq supp: homogeneous network steady-state C}
\end{equation}
The interface density, the quantity we are aiming to calculate, is given by 
\begin{equation}
    \avg{\sigma} = \frac{1-C^{(1)}}{2}, \label{eq supp: interface from correlation function}
\end{equation}
thus we need to determine $C^{(1)}_{\rm st}$. However setting $\ell=0$ in Eq.~(\ref{eq supp: homogeneous network steady-state C}) is not useful, as all terms involving $C^{(1)}_{\rm st}$ then cancel out because of $\Omega^{(\ell=0)} = 1$. To see the latter we note that there is one walk of length zero steps starting at $\bx$, and this walk trivially also ends at $\bx$. Thus Eq.~(\ref{eq supp: homogeneous network omega}) evaluates to one.

Instead, to proceed we multiply both sides of Eq.~(\ref{eq supp: homogeneous network steady-state C}) by $\gamma^{\ell}/\eta^{\ell}$, and then sum over $\ell$ from zero to infinity. We find
\begin{align}   
    \sum_{\ell=0}^{\infty}\Bigg(\frac{\gamma^{\ell}}{\eta^{\ell-1}}C^{(\ell)}_{\rm st}&-\frac{\gamma^{\ell+1}}{\eta^{\ell}}C^{(\ell+1)}_{\rm st}\Bigg)-\alpha \avg{m}_{\rm st} \sum_{\ell=0}^{\infty}\frac{\gamma^{\ell}}{\eta^{\ell}} = \Big[\eta -\alpha \avg{m}_{\rm st} - \gamma C ^{(1)}_{\rm st}\Big]\sum_{\ell=0}^{\infty}\Omega^{(\ell)}\frac{\gamma^{\ell}}{\eta^{\ell}}. \label{eq supp: homogeneous network C multiplied by fraction with sum}
\end{align}
The first term on the left-hand side is a telescopic sum and simplifies to
\begin{equation}
    \sum_{\ell=0}^{\infty}\Bigg(\frac{\gamma^{\ell}}{\eta^{\ell-1}}C^{(\ell)}_{\rm st}-\frac{\gamma^{\ell+1}}{\eta^{\ell}}C^{(\ell+1)}_{\rm st}\Bigg)= \eta C^{(0)}_{\rm st}-\lim_{\ell\to\infty}\frac{\gamma^{\ell+1}}{\eta^{\ell}}C^{(\ell+1)}_{\rm st}. \label{eq supp: LHS limit}
\end{equation}
We can use analogous arguments to that at the end of Sec.~\ref{appendix: random walks magnetisation}, except considering $C^{(\ell)}_{\rm st}$ instead of $\avg{m_{n}}_{\rm st}$, to perform the geometric sum in Eq.~(\ref{eq supp: homogeneous network C multiplied by fraction with sum}) and show that the limit in Eq.~(\ref{eq supp: LHS limit}) vanishes. Eq.~(\ref{eq supp: homogeneous network C multiplied by fraction with sum}) then becomes
\begin{equation}
    \eta C^{(0)}_{\rm st} - \frac{\alpha \avg{m}_{\rm st}}{1-\frac{\gamma}{\eta}}=\Big[\eta-\alpha \avg{m}_{\rm st} -\gamma C^{(1)}_{\rm st}\Big]\sum_{\ell=0}^{\infty}\Omega^{(\ell)}\left(\frac{1}{1+\tau}\right)^{\ell}.
\end{equation}
Using the fact that $C^{(0)}_{\rm st}=\avg{s(\bx)s(\bx)}_{\rm st}=1$, and the identities from Eqs.~(\ref{eq supp: eta over gamma relation}) and (\ref{eq supp: alpha over gamma relation}), and Eq.~(\ref{eq supp: interface from correlation function}), we find
\begin{align}
    \avg{\sigma}_{\rm st} 
    &= \frac{1}{2}H_{\rm HG}(\tau, \mathbf{A})\left(1-\avg{m}^{2}_{\rm st}\right), \label{eq supp: homogenous network steady-state interface}
\end{align}
with
\be
    H_{\rm HG}(\tau)= \frac{1+\tau}{\sum_{\ell=0}^{\infty}\Omega^{(\ell)}\left(\frac{1}{1+\tau}\right)^{\ell}}-\tau. \label{eq supp: H_HG}
\ee
We note that calculation of $\Omega^{(\ell)}$ requires the adjacency matrix of the network $\mathbf{A}$. The subscript `HG' stands for `homogeneous graph', and indicates that the relation holds only when the homogeneity assumptions apply. See Sec.~\ref{appendix: H and tau} for detailed discussion on $\tau$ and $H(\tau)$.

Eq.~(\ref{eq supp: H_HG}) is not closed form but the coefficients $\Omega^{(\ell)}$ can, in principle, be obtained through direct enumeration of walks on a given network. We note that this only needs to be performed once for a given network, to obtain the function $H_{\rm HG}(\tau)$. We discuss this further in Sec.~\ref{appendix: omega calc and non-homogeneous networks}.

We now move on to use Eq.~(\ref{eq supp: H_HG}) for a number of special cases (finite complete networks, hyper-cubic lattices, and Bethe lattices) where closed form solutions can be found.

\FloatBarrier
\subsubsection{First special case: Finite complete networks} \label{appendix: homogeneous to complete}
We consider a complete network with $N$ nodes. For any fixed node, we write $a_N(\ell)$ for the number of walks starting and ending at that node (this number will be the same for all nodes, given the all-to-all connectivity).
\begin{figure}[htbp]
    \centering
	\includegraphics[scale=0.25]{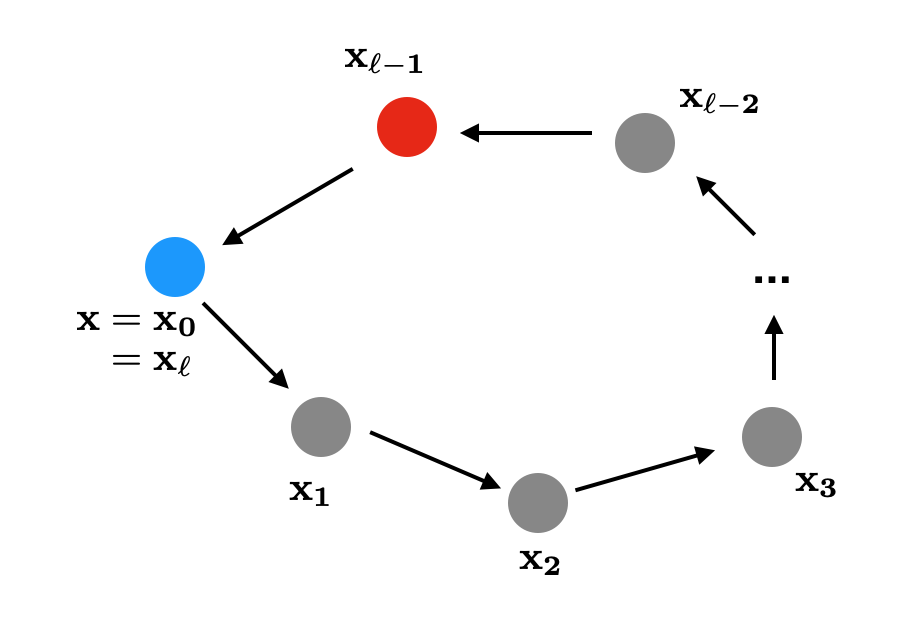}
	\caption{Illustration of closed walks on a complete network (see text). }
	\label{fig supp:closed_walks_CG}
    \end{figure}
One then has the following recursion
\be\label{eq supp: iteration}
a_N(\ell)=(N-1)^{\ell-1}-a_N(\ell-1),
\ee
with $a_N(0)\equiv 1$. \par
This recursion can be derived as follows (see Fig.~\ref{fig supp:closed_walks_CG}), keeping in mind that the walker cannot stay put in any step of the walk. Consider a fixed node $\bx$, and a closed walk of length $\ell$ starting and ending at $\bx$. The walk consists of a sequence of $\ell+1$ nodes in total, which we label $\bx_0, \bx_1,\dots,\bx_\ell$. The first and the last of these are given by $\bx$ ($\bx_0=\bx_\ell=\bx$), shown in blue in Fig.~\ref{fig supp:closed_walks_CG}. Now focus on node $\bx_1$. This node can be any node in the network other than node $\bx_0$ (the walker must make a hop in the first step). Similarly, $\bx_2$ can be any node other than $\bx_1$, and so on. If we follow this though until (and including) node $\bx_{\ell-1}$, we find a total of $(N-1)^{\ell-1}$ possible sequences for the nodes $\bx_1,\bx_2,\dots,\bx_{\ell-1}$ such that no node is the same as the previous one in the sequence. This produces the term $(N-1)^{\ell-1}$ in Eq.~(\ref{eq supp: iteration}). There is however another constraint, and not all of these sequences produce valid walks of length $\ell$ starting and ending at $\bx$. In a valid walk, node $\bx_{\ell-1}$ (shown in red in the figure) must not be node $\bx$, as the walker must make a hop in the $\ell$\textsuperscript{th} step. We therefore need to subtract the number of all sequences for which this is the case. This number is $a_N(\ell-1)$, as the sequence $\bx_0=\bx, \bx_1,\dots,\bx_{\ell-1}=\bx$ is then a closed walk of length $\ell-1$ starting and ending at $\bx$. This produces the second term in Eq.~(\ref{eq supp: iteration}). Thus, we have demonstrated the validity of the recursion. \par
One can directly show that (see e.g. sequences A109499, A109500, A109501 in \cite{sloane2007line}) 
\be
a_N(\ell)=\frac{(N-1)^\ell+(N-1)(-1)^\ell}{N}
\ee
fulfills the recursion in Eq.~(\ref{eq supp: iteration}), as well as the condition $a_N(0)=1$. \par
Given that further the number of all walks of length $\ell$ starting at $\bx$ on a complete network of size $N$ is given by $\nu^{(\ell)}(\bx)=(N-1)^\ell$, we have
\be
    \Omega^{(\ell)}=\frac{W^{(\ell)}}{\nu^{(\ell)}(\bx)}=\frac{a_N(\ell)}{(N-1)^\ell}=\frac{(N-1)^{\ell}+(N-1)(-1)^{\ell}}{N (N-1)^\ell}.
\ee
The generating function for the sequence $\Omega^{(\ell)}$ can be directly evaluated to be
\begin{equation}
    \sum_{\ell=0}^{\infty}\Omega^{(\ell)}z^{\ell} = \frac{1-\left(\frac{N-2}{N-1}\right)z}{1-\left(\frac{N-2}{N-1}\right)z-\left(\frac{1}{N-1}\right)z^{2}}.
\end{equation}
Thus Eq.~(\ref{eq supp: homogenous network steady-state interface}) for the finite complete network is
\begin{align}
    \avg{\sigma}_{\rm st} &= \frac{1}{2}\left[\frac{(N-1)\tau}{1+(N-1)\tau}\right]\left(1-\avg{m}^{2}_{\rm st}\right) \nonumber \\
    &\equiv \frac{1}{2}H_{\rm CG}(\tau, N)\left(1-\avg{m}^{2}_{\rm st}\right),
\end{align}
(the subscript `CG' stands for `complete graph'). We note that in the limit $N\rightarrow\infty$ we recover Eq. (\ref{eq supp: complete network steady-state interface}).

\FloatBarrier
\subsubsection{Second special case: Infinite hyper-cubic lattice in \texorpdfstring{$d$}{}-dimensions} \label{appendix: homogeneous to d-dim square lattice}
We next consider the case of a hyper-cubic lattice in $d$-dimensions. We then have $\Omega^{(\ell)}=0$ for odd values of $\ell$, as it is impossible to return to a given point in an odd number of steps on a hyper-cubic lattice. Thus we focus on even $\ell=2k$. So looking at Eq.~(\ref{eq supp: H_HG}) we need to calculate $\Omega^{(\ell)}\to\Omega^{(2k)}_{d}$, where the $d$ subscript denotes the dimension of the hyper-cubic lattice. From Eq.~(\ref{eq supp: homogeneous network omega}) we have
\begin{equation}
    \Omega^{(2k)}_{d} = \frac{W^{(2k)}_{d}}{\nu^{(2k)}_{d}}, \label{eq supp: d-dim lattice omega 2k}
\end{equation}
On a $d$-dimensional hyper-cubic lattice, the total number of walks of length $2k$ starting at a given point is
\begin{equation}
    \nu^{(2k)}_{d} = (2d)^{2k}, \label{eq supp: d-dim lattice nu 2k}
\end{equation}
given that the walker can choose from one of $2d$ directions at each step. The number of walks starting and ending at the same point is
\begin{equation}
    W_{d}^{(2k)} = \sum_{n_{1}+...+n_{d}=k}\frac{(2k)!}{\prod_{i=1}^{d}(n_{i}!)^{2}}. \label{eq supp: d-Dim lattice multiplicy of origin}
\end{equation}
In this expression the $n_i$ ($i=1,\dots,d$) are integer numbers; $n_i$ indicates how many steps in total the walker takes in the positive $i$ direction along the walk. Given that the walker returns to the starting point, $n_i$ is also the number of steps taken in the negative $i$ direction. Since a total of $2k$ steps must be made, we have $\sum_i 2n_i=2k$, i.e. $\sum_i n_i=k$. The term inside the sum in Eq.~(\ref{eq supp: d-Dim lattice multiplicy of origin}) is the number of distinct walks that can be generated for a given choice of $n_1,\dots,n_d$. The multinomial coefficient is the number of permutations of the $n_1$ forward steps in direction $1$, the $n_1$ backward steps in direction $1$, the $n_2$ forward steps in direction $2$, the $n_2$ backward steps in direction $2$ and so on, keeping in mind that the total number of steps is $2k$. \par
Using this we can construct the following so-called `exponential generating function' [note the factors of $1/(2k)!$ which is the difference between an exponential generating function and an ordinary generating function],
\begin{equation}
    \sum_{k=0}^{\infty}W_{d}^{(2k)}\frac{x^{2k}}{(2k)!} = I_{0}(2x)^{d}, \label{eq supp: d-dim lattice exp generating function}
\end{equation}
where $I_{0}(\cdot)$ is the modified Bessel function of the first kind. A proof of this can be found in Sec.~\ref{appendix: bessel generating function}. A Laplace-Borel transform, which we define as
\be
    \mathcal{L}_{B}[f](x) = \int_{0}^{\infty}e^{-t}f(tx)dt, \label{eq supp: laplace-borel transform}
\ee
can be used to convert an exponential generating function of the type in Eq.~(\ref{eq supp: d-dim lattice exp generating function}) to an ordinary generating function \cite[pp.~374, 566]{graham1989concrete}, \cite[p.~750]{flajolet2009analytic} or \cite[p.~2]{schmidt2019short}. Therefore we have
\begin{equation}
    \sum_{k=0}^{\infty}W_{d}^{(2k)}x^{2k}= \int_{0}^{\infty}e^{-t}[I_{0}(2xt)]^{d} \dd t. \label{eq supp: bessel exponential generating function}
\end{equation}
What we want is a summation of the form in Eq.~(\ref{eq supp: H_HG}), so we need to incorporate the $\nu^{(2k)}_{d}=(2d)^{2k}$ factor which can be done by applying the transform $x\to\frac{x}{2d}$ to Eq.~(\ref{eq supp: bessel exponential generating function}),
\begin{equation}
    \sum_{k=0}^{\infty}W_{d}^{(2k)}\frac{x^{2k}}{(2d)^{2k}}= \int_{0}^{\infty}e^{-t}\left[I_{0}\left(\frac{xt}{d}\right)\right]^{d} \dd t. \label{eq supp: laplace of product of bessel}
\end{equation}
With $x=\frac{1}{1+\tau}$, the summation on the LHS is exactly that in Eq.~(\ref{eq supp: H_HG}) for an infinite hyper-cubic $d$-dimensional lattice. So we are left with solving the integral on the RHS.

To proceed, it is now useful to make the substitution $xt=z$,
\begin{align}
    \int_{0}^{\infty}e^{-t}\left[I_{0}\left(\frac{xt}{d}\right)\right]^{d} \dd t &= \frac{1}{x}\int_{0}^{\infty}e^{-\frac{z}{x}}\left[I_{0}\left(\frac{z}{d}\right)\right]^{d} \dd z \nonumber \\
    &= \frac{1}{x}\mathcal{L}\left(\left[I_{0}\left(\frac{z}{d}\right)\right]^{d}\right).
\label{eq supp: laplace transform to solve}
\end{align}
where $\mathcal{L}$ is the standard Laplace transform defined as
\be 
 \mathcal{L}[f](s) = \int_{0}^{\infty} e^{-sz}f(z) \dd z,
\ee
where in this case $s=\frac{1}{x}$. We can now make use of the following general identity \cite[p. 346]{prudnikov2018integrals},
\begin{align}
    \mathcal{L}\left(z^{\lambda}\prod_{i=1}^{d}I_{\nu_{i}}(a_{i}z)\right)=&\frac{\Gamma(\lambda+\nu+1)}{2^{\nu}p^{\lambda+\nu+1}}\left[\prod_{i=1}^{d}\frac{a_{i}^{\nu_{i}}}{\Gamma(\nu_{i}+1)}\right]\nonumber \\  &\times F_{C}^{(d)}\left(\frac{\lambda+\nu+1}{2},\frac{\lambda+\nu+2}{2};\nu_{1},...,\nu_{d}+1;\frac{a_{1}^{2}}{p^{2}},...,\frac{a_{d}^{2}}{p^{2}}\right), \label{eq supp: laplace transform of products of modified bessel functions}
\end{align}
where $F_{C}^{(d)}$ is a Lauricella hypergeometric series \cite[p. 113]{lauricella1893sulle}\cite{koornwinder2020encyclopedia}, 
\begin{equation}
    F_{C}^{(d)}(a,b;c_{1},...,c_{d};y_{1},...,y_{d})=\sum_{i_{1},...,i_{d}=0}^{\infty}\frac{(a)_{i_{1}+...+i_{d}}(b)_{i_{1}+...+i_{d}}}{(c_{1})_{i_{1}}...(c_{d})_{i_{d}}i_{1}!...i_{d}!}y_{1}^{i_{1}}...y_{d}^{i_{d}},
\end{equation}
with 
\begin{equation}
    (q)_{i}\equiv q(q+1)...(q+i-1),
\end{equation}
for $i=0,1,2,3,\dots$, referred to as the Pochhammer symbol or rising factorial. 

We note that the identity in Eq.~(\ref{eq supp: laplace transform of products of modified bessel functions}) as printed in  \cite{prudnikov2018integrals} contains a typographical error, for further details see also \cite{jordanabbott_1969}. This error has been corrected in Eq.~(\ref{eq supp: laplace transform of products of modified bessel functions}). \par 
We now substitute Eq.~(\ref{eq supp: laplace transform of products of modified bessel functions}) in the special case $\lambda=0$, $\nu_{i}=0$ and $a_{i}=\frac{1}{d}$ into Eq.~(\ref{eq supp: laplace transform to solve}) and find
\begin{align}
    \int_{0}^{\infty}e^{-t}\left[I_{0}\left(\frac{zt}{d}\right)\right]^{d} \dd t &= F_{C}^{(d)}\left(\frac{1}{2},1;1,...,1;\frac{1}{\left[d(1+\tau)\right]^{2}},...,\frac{1}{\left[d(1+\tau)\right]^{2}}\right),
\end{align}
where $z=\frac{1}{1+\tau}$. With this, we find the following expression for the interface density on a $d$-dimensional hyper-cubic lattice,
\begin{equation}
    \left<\sigma\right>_{\rm st}=\frac{1}{2}\left[\frac{1+\tau}{F_{C}^{(d)}\left(\frac{1}{2},1;1,...,1;\frac{1}{\left[d(1+\tau)\right]^{2}},...,\frac{1}{\left[d(1+\tau)\right]^{2}}\right)}-\tau\right]\left(1-\avg{m}^{2}_{\rm st}\right). \label{eq supp: d-dim lattice steady-state interface}
\end{equation}

\subsubsection{Application to infinite two-dimensional square lattice} \label{appendix: d-dim to 2D lattice}
We have already derived an analytical expression for $H(\tau)$ for infinite 2D square lattice in Sec.~\ref{appendix: square lattice} of this Supplement [see Eq.~(\ref{eq supp: H 2d square lattice})]. However as a test of our results for the homogeneous network and hyper-cubic lattice in $d$-dimensions we will re-derive it, with reference to several points of the derivation.

\noindent{\em First method: Closed form expression for $\Omega^{(2k)}$}\\
We can directly evaluate the summation in Eq.~(\ref{eq supp: H_HG}) by first determining a closed form expression for $\Omega^{(\ell)}$. On an infinite square lattice only even length walks can return us to the initial position, so it is really $\Omega^{(2k)}$ we wish to calculate.

The number of closed walks of length $2k$ can be calculated with the following combinatorial argument. If we take $p$ steps North, we must then also take $p$ steps South. There are $\binom{2k}{2p}$ ways to choose these $2p$ steps from $2k$ and $\binom{2p}{p}$ ways to organise the North and South steps. From the remaining steps, half of them must be East and the other half must be West, there are $\binom{2k-2p}{k-p}$ ways of arranging these steps. Thus the total number of closed walks of length $2k$ is
\begin{align}
    \sum_{p=0}^{k}\binom{2k}{2p}\binom{2p}{p}\binom{2(k-p)}{k-p} &= \sum_{p=0}^{k}\frac{(2k)!}{(2p)!(2(k-p))!}\cdot\frac{(2p)!}{p!^{2}}\cdot\frac{(2(k-p))!}{(k-p)!^{2}} \nonumber \\
    &= \frac{(2k)!}{k!^{2}}\sum_{p=0}^{k}\frac{k!^{2}}{p!^{2}(k-p)!^{2}} \nonumber \\
    &= \binom{2k}{k}\sum_{p=0}^{k}\binom{k}{p}\binom{k}{k-p} \nonumber \\
    &= \binom{2k}{k}^{2},
\end{align}
where in the last step we use Vandermonde's identity. We also note that the total number of length $2k$ walks on a 2D square lattice is $\nu^{(2k)}=4^{2k}$. Thus, letting $x=\frac{1}{1+\tau}$, the summation in Eq.~(\ref{eq supp: H_HG}) becomes
\begin{equation}
    \sum_{k=0}^{\infty}\Omega^{(2k)}x^{2k} = \sum_{k=0}^{\infty}\binom{2k}{k}^{2}\frac{x^{2k}}{4^{2k}} = \frac{2}{\pi}K(x),
\end{equation}
where we have used the power series for the complete elliptical integral of the first kind \cite[(1.3.6)]{borwein1987pi}. With this, Eq.~(\ref{eq supp: H_HG}) reduces exactly to Eq.~(\ref{eq supp: H 2d square lattice}).


\noindent{\em Second method: Relation of elliptic integral to modified Bessel function}\\
The second way of verifying the special case $d=2$ is by evaluation of Eq.~(\ref{eq supp: laplace of product of bessel}) for $d=2$. \par
We are looking to prove that the Laplace-Borel transformation [see Eq.~(\ref{eq supp: laplace-borel transform})] of the square of modified Bessel functions of the first kind is proportional to the complete elliptical integral of the first kind i.e.
\begin{equation}
    \int_{0}^{\infty}e^{-t}I_{0}\left(\frac{xt}{2}\right)^{2}\dd t = \frac{2}{\pi}K(x).
\end{equation}
First, make the change of variables $xt=z$ to convert the Laplace-Borel transform into a regular Laplace transform,
\begin{align}
    \int_{0}^{\infty}e^{-t}I_{0}\left(\frac{xt}{2}\right)^{2}\dd t &= \frac{1}{x}\int_{0}^{\infty}e^{-pz}I_{0}\left(\frac{z}{2}\right)^{2}\dd z, \label{eq supp: laplace transform proof}
\end{align}
where $p=\frac{1}{x}$. From Sec.~\ref{appendix: bessel generating function} we find the exponential generating function for the square of modified Bessel functions of the first kind is
\begin{align}
    I_{0}(2y)^{2} &= \sum_{k=0}^{\infty}W_{2}^{(2k)}\frac{y^{2k}}{(2k)!} \nonumber \\
    &= \sum_{k=0}^{\infty}{\binom{2k}{k}}^{2}\frac{y^{2k}}{(2k)!}, \label{eq supp: exponential generating function}
\end{align}
where $W_{2}^{(2k)}$ can be found from Eq.~(\ref{eq supp: d-Dim lattice multiplicy of origin}) with $d=2$. Substituting Eq.~(\ref{eq supp: exponential generating function}) into Eq.~(\ref{eq supp: laplace transform proof}) gives,
\begin{align}
    \frac{1}{x}\int_{0}^{\infty}e^{-pz}I_{0}\left(\frac{z}{2}\right)^{2}\dd z &= \frac{1}{x}\int_{0}^{\infty}e^{-pz}\sum_{k=0}^{\infty}{\binom{2k}{k}}^{2}\left(\frac{z}{4}\right)^{2k}\frac{1}{(2k)!}\dd z \nonumber \\
    &= \frac{1}{x}\sum_{k=0}^{\infty}{\binom{2k}{k}}^{2}\frac{1}{(2k)!}\frac{1}{4^{2k}}\int_{0}^{\infty}e^{-pz}{z}^{2k}\dd z. \label{eq supp: 2nd proof 2D lattice laplace}
\end{align}
The integral here is simply the Laplace transform of ${z}^{2k}$ given by \cite[p.~15]{prudnikov2018integrals},
\begin{align}
    \int_{0}^{\infty}e^{-pz}{z}^{2k}\dd z &= \frac{\Gamma(2k+1)}{p^{2k+1}} \nonumber \\
    &= x^{2k+1}(2k)!,
\end{align}
where we recall $p=\frac{1}{x}$ and where we have used the definition of the Gamma function for integer $n$ \cite[p.~255]{abramowitz1988handbook},
\begin{equation}
    \Gamma(n) = (n-1)!.
\end{equation}
Putting this back into Eq.~(\ref{eq supp: 2nd proof 2D lattice laplace}) we finally have
\begin{align}
    \int_{0}^{\infty}e^{-t}I_{0}\left(\frac{xt}{2}\right)^{2}\dd t &= \sum_{k=0}^{\infty}{\binom{2k}{k}}^{2}\frac{x^{2k}}{4^{2k}} \nonumber \\
    &= \frac{2}{\pi}K(x),
\end{align}
where we have used the power series for the complete elliptical integral of the first kind \cite[(1.3.6)]{borwein1987pi}. Thus we have shown that Eq.~(\ref{eq supp: laplace of product of bessel}) evaluated at $d=2$, which is exactly the relevant summation appearing in Eq.~(\ref{eq supp: H_HG}) for the 2D infinite square lattice, is proportional to $K(x)$, thus Eq.~(\ref{eq supp: H_HG}) reduces to Eq.~(\ref{eq supp: H 2d square lattice}).

\noindent{\em Third proof: Relation of elliptic integral to Lauricella hyper-geometric series}\\
The final method is to evaluate Eq.~(\ref{eq supp: d-dim lattice steady-state interface}), for the $d$-dimensional lattice, for the case $d=2$.

For $d=2$, the following Lauricella series corresponds to the Appell hypergeometric series \cite{koornwinder2020encyclopedia},
\begin{align}
    F_{C}^{(2)}\Bigg(\frac{1}{2},1;1,1;\frac{1}{4(1+\tau)^{2}}&,\frac{1}{4(1+\tau)^{2}}\Bigg) = F_{4}\left(\frac{1}{2},1;1,1;\frac{1}{4(1+\tau)^{2}},\frac{1}{4(1+\tau)^{2}}\right).
\end{align}
An Appell hypergeometric series of this form can be written as a generalised hypergeometric series \cite[Eq.~(37)]{burchnall1942differential}
\begin{equation}
    F_{4}\left(\frac{1}{2},1;1,1;\frac{1}{4(1+\tau)^{2}},\frac{1}{4(1+\tau)^{2}}\right) = {}_{4}F_{3}\left(\frac{1}{2},1,1,\frac{1}{2};1,1,1;\frac{1}{(1+\tau)^{2}}\right).
\end{equation}
The generalised hypergeometric function $_{p}F_{q}$ is defined as \cite[p.~439]{prudnikov1990more},
\begin{equation}
    _{p}F_{q}(a_{1},...,a_{p};b_{1},...,b_{q};x)=\sum_{k=0}^{\infty}\frac{(a_{1})_{k},...,(a_{p})_{k}}{(b_{1})_{k},...,(b_{q})_{k}}\frac{x^{k}}{k!}.
\end{equation}
 Using this we can show that
\begin{equation}
    _{4}F_{3}\left(\frac{1}{2},1,1,\frac{1}{2};1,1,1;x^{2}\right) = {}_{2}F_{1}\left(\frac{1}{2},\frac{1}{2},1;x^{2}\right).
\end{equation}
Finally it is known that \cite[p.~51]{anderson1992hypergeometric}
\begin{equation}
    {}_{2}F_{1}\left(\frac{1}{2},\frac{1}{2},1;x^{2}\right)\equiv\frac{2}{\pi}K(x).
\end{equation}
Thus we have shown that Eq.~(\ref{eq supp: d-dim lattice steady-state interface}) reduces to Eq.~(\ref{eq supp: sigma steady-state 2d square lattice}) for the 2D infinite square lattice.

\subsubsection{Third special case: Infinite Bethe lattice} 
\label{appendix: homogeneous to bethe lattice}
The infinite regular tree, or Bethe lattice, is a network in which every node has degree $n$ and which has a cascading tree-like structure such that there are no loops or leaf nodes.

On such trees, only walks of even length can return to the starting point. So we have $W^{(2k+1)}=0$ for all $k\in\mathbb{Z}_{\geq 0}$, and thus only need to perform the summation in Eq.~(\ref{eq supp: H_HG}) over even $k$. Additionally, it can be seen by inspection that $\nu^{(\ell)}=n^{\ell}$. Ultimately, the summation in Eq.~(\ref{eq supp: H_HG}) reduces to
\begin{equation}
    T\left(z\right) = \sum_{k=0}^{\infty}\mathcal{T}^{(k)}z^{k}, \label{eq supp: generating function to evaluate}
\end{equation}
for
\be
    z=\frac{1}{n^{2}(1+\tau)^{2}}, \label{eq supp: bethe lattice z}
\ee
where $\mathcal{T}^{(k)} \equiv W^{(2k)}$ is the number of self returning walks of length $2k$ from any point on the network.

This means that we need the generating function for the sequence $\mathcal{T}^{(k)}$. To calculate this we follow the approach in \cite{wanless2010counting} and first consider an infinite rooted tree, in which the root has degree $r$ and all other nodes  have degree $c+1$. We write $\mathbb T$ for this tree.

A closed rooted walk is defined as a walk which starts and ends at the root node. We will refer to the nearest neighbours of the root as the `children'.
\par
Any closed rooted walk of length $2k$ in $\mathbb T$ can be broken down as follows:
\begin{enumerate}
    \item[1.] The walker starts at the root, and takes one step from the root to one of its $r$ nearest neighbours (i.e., a child).
    \item[2.] A closed walk of length $2i$, avoiding the root, is undertaken, starting at the child node in step 1, and bringing the walker back to the same child.
    \item[3.] The walker returns to the root node, having taken $2(i+1)$ steps in total.
    \item[4.] A closed rooted walk of length $2k-2(i+1)$ can now be undertaken.
\end{enumerate}
We define ${\cal R}_{r}^{(k)}$ as the number of self-returning walks of length $2k$ from the root node, and ${\cal C}_{c+1}^{(k)}$ as the number of self-returning root-avoiding walks of length $2k$ from one of the children. The subscripts indicate the degree of the root ($r$) and that of all other nodes ($c+1$) respectively. These quantities are specific to the tree $\mathbb{T}$. However, the number of self-returning root-avoiding walks starting at a child will not depend on the degree $r$ of the root.

From the above breakdown on rooted walks of length $2k$ we can obtain the following recursive relation:
\begin{equation}
    {\cal R}_{r}^{(n+1)}=r\sum_{i=0}^{n}{\cal R}_{r}^{(i)}{\cal C}_{c+1}^{(n-i)}.
\end{equation}
We next define 
\be\label{eq supp: bethe lattice R}
    R_r(z)\equiv \sum_{k=0}^\infty {\cal R}^{(k)}_{r} z^k
\ee
as the generating function for self-returning walks in the tree from the root node which has degree $r$. Similarly, 
\be\label{eq supp: bethe lattice C}
    C_{c+1}(z)=\sum_{k=0}^\infty {\cal C}^{(k)}_{c+1} z^k
\ee
is the generating function for self-returning walks in the tree from a child of the root node and avoiding the root. 

With these definitions we have
\BE
    R_{r}(z) &\equiv&\sum_{k=0}^\infty {\cal R}^{(k)}_{r} z^k \nonumber \\
    &=& 1+ \sum_{k=1}^\infty {\cal R}^{(k)}_{r} z^k \nonumber \\
    &=& 1+r\sum_{k=1}^\infty \sum_{i=0}^{k-1} {\cal R}_r^{(i)}{\cal C}_{c+1}^{(k-1-i)} z^k \nonumber \\
    &=& 1+rz\sum_{k=0}^\infty \sum_{i=0}^k {\cal R}_r^{(i)} {\cal C}_{c+1}^{(k-i)} z^k \nonumber \\
    &=& 1+rz R_r(z) C_{c+1}(z).\label{eq supp: generating function recurrance relation}
\EE
The last step can be seen by writing out the first few terms of the series for $R_{r}(z)$ and $C_{c+1}(z)$ in Eqs.~(\ref{eq supp: bethe lattice R}) and (\ref{eq supp: bethe lattice C}).

We can determine $R_{r}(z)$ from Eq.~(\ref{eq supp: generating function recurrance relation}) if we can calculate $C_{c+1}(z)$. To do this, we note again that ${\cal C}_{c+1}^{(k)}$ does not depend on the degree of the root. Hence $C_{c+1}(z)$ is not a function of the root-degree either. With this in mind, we now consider a tree $\tilde{\mathbb{T}}$ with root of degree $c$ and where, as before, all other nodes have degree $c+1$. Setting $r=c$ in Eq.~(\ref{eq supp: generating function recurrance relation}) we then have
\be
    \tilde R_c(z) = 1+cz\tilde R_c(z)\tilde C_{c+1}(z),\label{eq supp: generating function inter recurrance}
\ee
where the tildes indicate that these are generating functions for the tree $\tilde{\mathbb T}$. The only difference between the trees $\mathbb T$ and $\tilde{\mathbb T}$ is the degree of the root. Hence $\tilde C_{c+1}(z)=C_{c+1}(z)$.

Next we realise that $\tilde R_c(z)=\tilde C_{c+1}(z)$. This is because $\tilde C_{c+1}(z)$ is the generating function of self-returning walks starting from a child and avoiding the root on the tree $\tilde{\mathbb T}$. For the purpose of counting such walks the child effectively has degree $c$ (the same degree as the root of $\tilde{\mathbb{T}}$). Hence on the tree $\tilde{\mathbb{T}}$, the number of self-returning root-avoiding walks of a given length and starting from a child is the same as the number of self-returning walks of this length starting at the root. 

Using $\tilde R_c(z)=\tilde C_{c+1}(z)$ and $\tilde C_{c+1}(z)=C_{c+1}(z)$ Eq.~(\ref{eq supp: generating function inter recurrance}) thus becomes
\be
    C_{c+1}(z) = 1+czC^2_{c+1}(z).
\ee
We can then solve for $C_{c+1}(z)$ and find the physically relevant solution to be
\begin{equation}\label{eq supp: T_c}
    C_{c+1}(z) = \frac{1-\sqrt{1-4zc}}{2zc}
\end{equation}
(this is the solution ensuring that $\lim_{z\to 0}C_{c+1}(z) = 1$, i.e. there is exactly one walk of length zero). It is interesting to note that this generating function is closely related to the generating function for the Catalan numbers \cite{larcombe2001generating,stanley2015catalan}.

Having now determined $C_{c+1}(z)$ we can substitute this back into Eq.~(\ref{eq supp: generating function recurrance relation}) to determine $R_r(z)$,
\begin{equation}
    R_{r}(z)=\frac{2c}{2c-r+r\sqrt{1-4zc}}. \label{eq supp: T_r soln}
\end{equation}

For the special case of the $n$-regular tree (i.e. the tree where all nodes have the same degree $n$) we simply set $n=r=c+1$ in Eq.~(\ref{eq supp: T_r soln}). This gives the generating function for self-returning walks on an infinite $n$-regular tree, which we define as $T(z)$ in Eq.~(\ref{eq supp: generating function to evaluate}),
\begin{equation}
    T(z) = \frac{n-1}{n-2+n\sqrt{1-4(n-1)z}}.
\end{equation}
Then with $z$ as in Eq.~(\ref{eq supp: bethe lattice z}) we have a closed-form expression for the summation in Eq.~(\ref{eq supp: generating function to evaluate}):
\begin{equation}
    T(z) = \frac{2(n-1)}{n-2+n\sqrt{1-4(n-1)\left(\frac{1}{n(1+\tau)}\right)^{2}}}. \label{eq supp: bethe lattice final sum}
\end{equation}
With the replacement $n\to \mu$, Eqs.~(\ref{eq supp: homogenous network steady-state interface}) and (\ref{eq supp: bethe lattice final sum}) combine to give
\begin{align}
    \avg{\sigma}_{\rm st} &= \frac{1}{2}\left[\frac{\mu-2 - \mu\tau+\sqrt{(\tau+1)^{2}\mu^{2}-4(\mu-1)}}{2(\mu-1)}\right]\left(1-\avg{m}^{2}_{\rm st}\right) \nonumber \\
    &\equiv  \frac{1}{2}H_{\text{BL}}(\tau)\left(1-\avg{m}^{2}_{\rm st}\right), 
\end{align}
(where `BL' stands for `Bethe lattice'). We note that this is the same result that we derived in Sec.~\ref{appendix: pair approximation} using the pair approximation for infinite uncorrelated networks. In the current section we did not have to make any approximations about spin correlations as we did for the pair approximation. The only assumption that we made was that the dynamics was being taking place on a particular topology, hence this result is exact for Bethe lattices. This is not surprising as the pair approximation is known to be exact on Bethe lattices \cite[p; 346]{dieckmann2000geometry} as there are no loops.

\subsubsection{Calculation of \texorpdfstring{$\Omega^{(\ell)}$}{} and validity of the method} \label{appendix: omega calc and non-homogeneous networks}
For networks that are not homogeneous in the above sense, we can still apply the approach based on network walks [Eq.~(\ref{eq supp: H_HG})] as an approximation. We replace $\Omega^{(\ell)}$ with its average over all nodes,
\begin{equation}
    \Omega^{(\ell)} \rightarrow \frac{1}{N}\sum_{\bx} \Omega^{(\ell)}_{\bx} \equiv \overline{\Omega^{(\ell)}}.
\end{equation}
It is difficult to find this object in closed form. However, for many networks we find that $\overline{\Omega^{(\ell)}} \rightarrow \frac{1}{N}$ for large $\ell$ (where $N$ is the size of the graph). Thus we compute $\overline{\Omega^{(\ell)}}$ directly from the adjacency matrix for $\ell$ below some cutoff $L$, and use  $\overline{\Omega^{(\ell)}}\approx \frac{1}{N}$ for $\ell > L$. This approximation should be tested for each network on an individual basis, from which an appropriate choice for $L$ should be made. Based on this then we can write, using the geometric series,
\begin{equation}
    \sum_{\ell = 0}^{\infty} \overline{\Omega^{(\ell)}}x^{\ell} \approx \frac{1}{N}\frac{x^{L+1}}{1-x} + \sum_{\ell = 0}^{L}\overline{\Omega^{\ell}}x^{\ell}.
\end{equation}

\begin{figure}[htbp]
    \includegraphics[scale = 0.9]{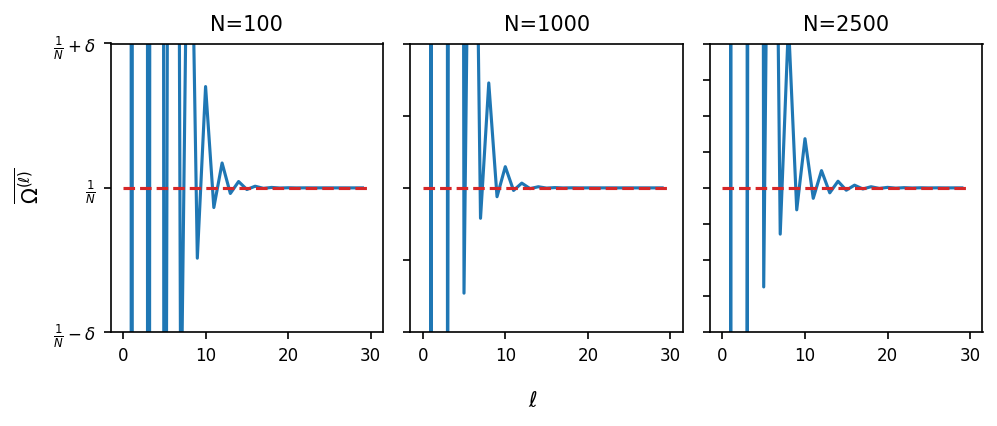}
    \caption{Demonstration of $\overline{\Omega^{(\ell)}}$ converging to $\frac{1}{N}$ for increasing $\ell$ for difference size Barab\'asi--Albert networks with $4$ new links established upon adding a new node. The extent of the vertical axis shown is from $1/N-\delta$ to $1/N+\delta$, with $\delta = 1\times 10^{-4}$.}
    \label{fig supp: omega convergence}
    \end{figure}

In Fig.~\ref{fig supp: omega convergence} we show the convergence of $\overline{\Omega^{(\ell)}}$ to $\frac{1}{N}$ for Barab\'asi--Albert networks of differen size. These networks are not homogeneous. In this case it would be suitable to take $L\approx 30$.

\begin{figure}[htbp]
    \includegraphics{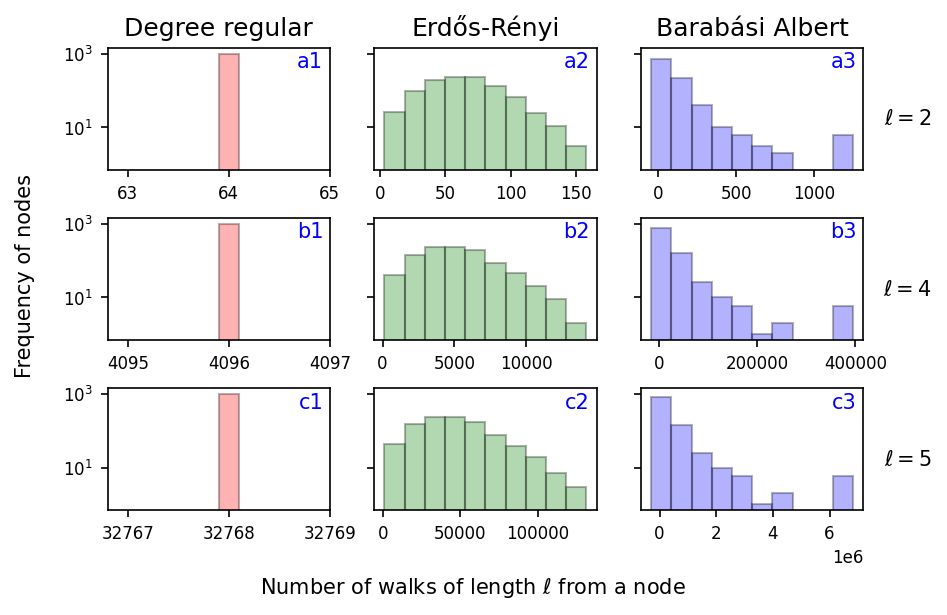}
    \includegraphics{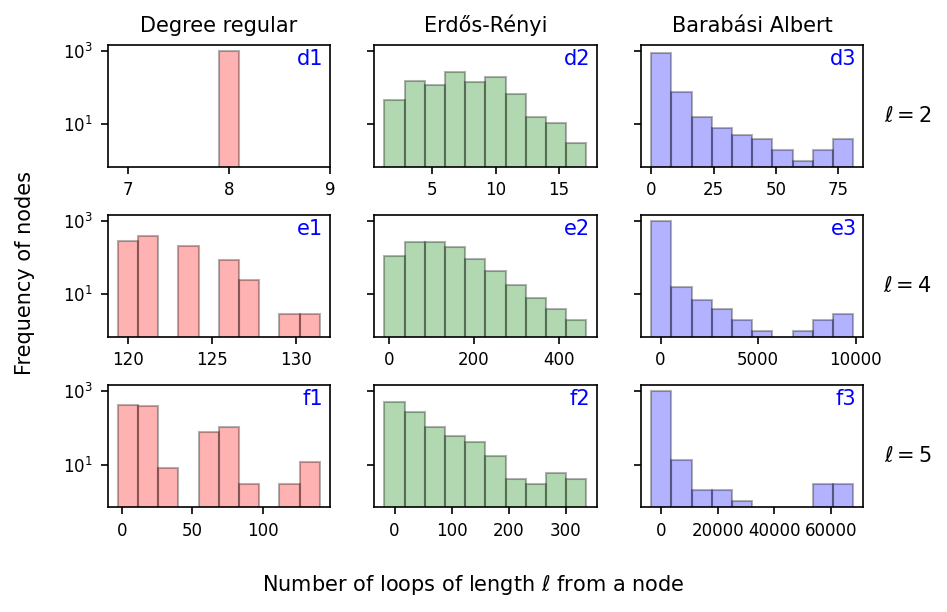}
    \caption{Histograms of the frequency of length $\ell$ walks (top) and loops (bottom) from each node in three different networks: degree-regular (left column) with $\mu=8$, Erd\"os--R\'enyi (middle column) with $p=8/N$ and Barab\'asi--Albert (right column) with with $4$ new links established upon adding a new node. All networks have $N=2,500$.}
    \label{fig supp: histograms}
\end{figure}

Although the approaches based on network walks is only exact on homogeneous networks, we find that it serves as a satisfactory approximation for networks in which the variation of $\Omega^{\ell}_{\bx}$ across nodes is small. In Fig.~\ref{fig supp: histograms} we show the frequencies at which length $\ell$ walks (top) and length $\ell$ loops (bottom) appear as we vary $\ell$ for degree-regular, Erd\"os--R\'enyi, and Barab\'asi--Albert networks.

On degree-regular networks with degree $k$, the number of walks of a given length $\ell$ starting from any node is always $k^\ell$, as confirmed in Fig.~\ref{fig supp: histograms}(a1), (a2) and (a3). The number of loops can vary slightly from node to node (see panels d1-f1) as the network is not exactly homogeneous, but the distributions are still relatively tight, compared to the two other types of network shown.  

For Erd\"os--R\'enyi networks there is more variance in the distributions, and for Barab\'asi--Albert networks there is significantly more. Thus we can anticipate the approximation to be somewhat accurate on Erd\"os--R\'enyi networks, and less so on  Barab\'asi--Albert networks. 

\begin{figure}[htbp]
    \includegraphics[scale = 0.9]{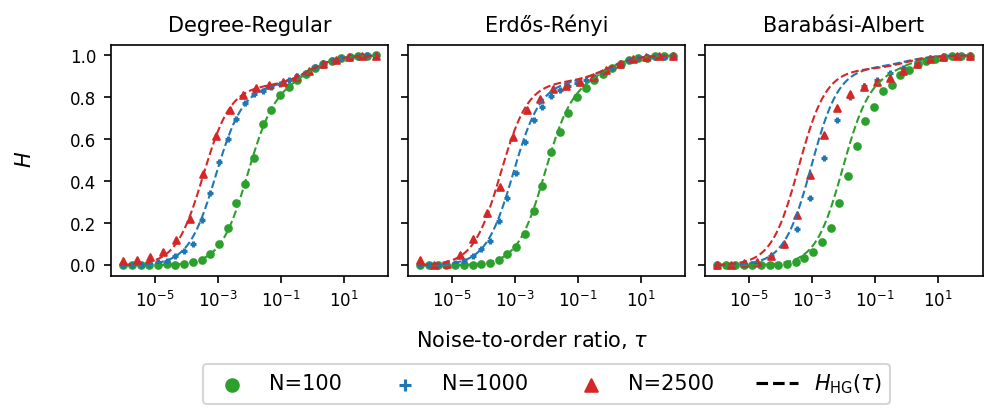}
    \caption{Plots of $H(\tau)=\avg{\sigma}_{\rm st}/[\frac{1}{2}(1-\avg{m}_{\rm st}^{2})]$, which characterises the scattering, as a function of the noise-to-order ratio, $\tau$, for networks of size $N$. Degree-regular with $\mu=8$ (left), Erd\"os--R\'enyi with $p=8/N$ (middle), Barab\'asi--Albert with $4$ new links established upon adding a new node (right). For all three types of network we have $\mu\approx8$ (deviations are due to the finite size of the graphs). The dashed lines are the results from our theory, Eq.~(\ref{eq supp: H_HG}). Each marker is the result from averaging 1000 independent simulations in the steady-state at a particular $\tau$ value. Model parameters for a particular $\tau$ can be generated via the algorithm detailed in Sec.~\ref{appendix: determining model parameters tau and m}.}
    \label{fig supp: H omega method}
    \end{figure}

In Fig.~\ref{fig supp: H omega method} we show how $H_{\rm HG}(\tau)$ from Eq.~(\ref{eq supp: H_HG}) varies as a function of $\tau$ for different degree-regular (left), Erd\"os--R\'enyi (middle) and Barab\'asi--Albert (right) networks. The theory works very well for degree-regular networks and is a good approximation for Erd\"os-–R\'enyi networks as expected because the variance of their degree distributions is small. For Barab\'asi--Albert there is greater deviation.

\FloatBarrier
\subsubsection{Small disconnected networks} \label{appendix: exotic networks}
It is interesting to note that the result $H_{\rm HG}(\tau)$ applies to homogeneous networks of any size, for example see Fig.~\ref{fig main: theory comparison} where we show the result working for an $N=10$ degree-regular network. Further, the theory does not assume that the network is connected, in fact many low-degree Erd\"os--R\'enyi networks are disconnected. 

\begin{figure}[htbp]
    \includegraphics{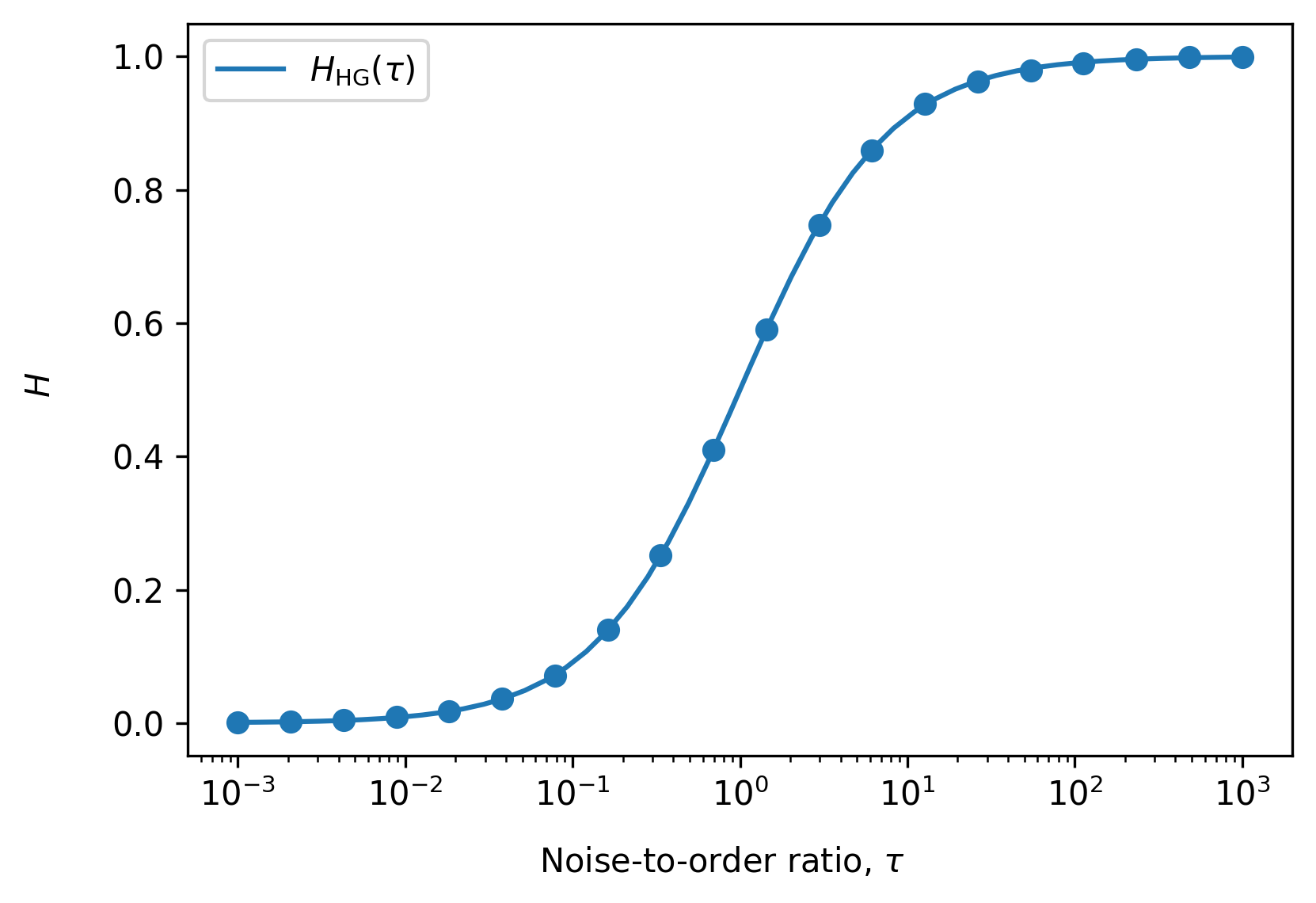}
    \caption{Plot of $H(\tau)=\avg{\sigma}_{\rm st}/[\frac{1}{2}(1-\avg{m}_{\rm st}^{2})]$, which characterises the scattering, against the noise-to-order ratio $\tau$ for a network of size $N=4$ containing two connected pairs. Solid line is $H_{\rm HG}(\tau)$ from Eq.~(\ref{eq supp: H_HG}). Markers are the result from averaging over $1\times 10^{5}$ independent simulations in the steady-state at a specific value of $\tau$. Model parameters for specific values of $\tau$ are generated via the algorithm detailed in Sec.~\ref{appendix: determining model parameters tau and m}.}
    \label{fig supp: small disconnected}
    \end{figure}

Fig.~\ref{fig supp: small disconnected} demonstrates that the theory works well on a network of $N=4$ with only two connected pairs of nodes. 

The pathological case of a system in which all nodes are isolated from one another is not handled by the theory. In this case there are no links, and thus the fraction of active links is undefined. 

\FloatBarrier
\clearpage
\newpage
\section{Stochastic pair approximation} \label{appendix: SPA}
The stochastic pair approximation (SPA) is an extension of the regular pair approximation discussed in Sec.~\ref{appendix: pair approximation}. The SPA targets large but finite uncorrelated networks. This was originally proposed in \cite{peralta2018stochastic} (pp 7-16 in particular), which studied the noisy voter model. Our model is more general, and reduces to the noisy voter model in the limit of no copying errors in the horizontal process, $h_{+}'=h_{-}'=0$, and when the vertical process is symmetric, $v_{+}=v_{-}$. We here construct the SPA for the more general setup of our model. More precisely we develop what is called `S2PA' in \cite{peralta2018stochastic} but here we refer to it as simply `SPA'.

Unlike in the main paper and in the remaining sections of this supplement we do not use $-1$ and $+1$ for the node states. Instead we denote these states by $0$ and $1$. This is to keep the notation as close to that of \cite{peralta2018stochastic}.

\subsection{Master equation for reduced set of degrees of freedom}
To begin we consider the reduced set of variables $\left\{\mathbf{N}_{1}, \mathcal{L}\right\}$ where $\mathbf{N}_{1}$ is the vector
\begin{equation}
    \mathbf{N}_{1} = \left(\mathbf{N}_{1,k_{\text{min}}}, ..., \mathbf{N}_{1,k_{\text{max}}}\right),
\end{equation}
whose components $N_{1,k}$ are the number of spin-up nodes with degree $k$. The notation $N_{0,k}$ will be used to represent the number of spin-down nodes with degree $k$, and thus the total number of nodes of degree $k$ is $N_{k}=N_{0,k}+N_{1,k}$. $\mathcal{L}$ is the number of active links. We also define $n_{i}\in[0,k_{i}]$ as the number of neighbours of node $i$ in the up-state.

When a node $i$ flips from the down to the up-state, we have the process $N_{1,k_{i}}\rightarrow N_{1,k_{i}}+1$. The number of active links connected to $i$ before flipping is $n_{i}$, after the flip it will be $k_{i}-n_{i}$, thus a change of $k_{i}-2n_{i}$ in the number of active links connected to $i$. Analogously, when a node flips from up to down, we have $N_{1,k_{i}}\rightarrow N_{1,k_{i}}-1$ and the total number of active links changes as $\mathcal{L}\rightarrow \mathcal{L} 
- k_{i}+2n_{i}$. 

We then define the probability distribution of $\left\{\mathbf{N}_{1}, \mathcal{L}\right\}$ at time $t$ as $P(\mathbf{N}_{1}, \mathcal{L};t)$ and we can construct the following master equation:
\begin{align}
    \frac{\partial}{\partial t}P(\mathbf{N}_{1}, \mathcal{L};t) = \sum_{k}\sum_{n=0}^{k}\Big\{&\big(E_{N_{1,k}}^{+}E_{\mathcal{L}}^{k-2n}-1\big)\big[W^{(-,k,n)}P(\mathbf{N}_{1},\mathcal{L};t)\big] \nonumber \\
    &\qquad +\big(E_{N_{1,k}}^{-}E_{\mathcal{L}}^{-k+2n}-1\big)\big[W^{(+,k,n)}P(\mathbf{N}_{1}, \mathcal{L};t)\big]\Big\}, \label{eq supp: SPA master eqn}
\end{align}
where $W^{(+,k,n)}$ and $W^{(-,k,n)}$ are the effective rates of the two processes described above. The $E$'s are step operators which act on functions of $N_{1,k}$ or ${\cal L}$ as follows 
\begin{subequations}
\begin{align}
    E_{N_{1,k}}^{+}f(N_{1,k}) &\rightarrow f(N_{1,k}+1), \\
    E_{\mathcal{L}}^{k-2n}f(\mathcal{L}) &\rightarrow f(\mathcal{L}+k-2n), \\
    E_{N_{1,k}}^{-}f(N_{1,k}) &\rightarrow f(N_{1,k}+1), \\
    E_{\mathcal{L}}^{-k+2n}f(\mathcal{L}) &\rightarrow f(\mathcal{L}-k+2n).
\end{align}
\end{subequations}
Next we define the degree specific magnetisation $m_{k}$, i.e. the magnetisation among the nodes of degree $k$. The interface density is denoted by $\sigma$. These two quantities can be written in terms of the variables $\left\{\mathbf{N}_{1}, \mathcal{L}\right\}$ as follows:
\begin{subequations}
\begin{align}
    m_{k} &= 2\frac{N_{1,k}}{N_{k}}-1, \\
    \sigma &= \frac{\mathcal{L}}{\frac{1}{2}\mu N},
\end{align}
\end{subequations}
where $\mu$ is the mean degree of the network, and $N$ is the total number of nodes, thus $\frac{1}{2}\mu N$ gives the total number of links. 

\subsection{Dynamics of first and second moments}
Eq.~(\ref{eq supp: SPA master eqn}) can be used to derive differential equations for the average degree specific magnetisation, $\avg{m_{k}(t)}$, the average density of active links, $\avg{\sigma(t)}$, and correlations such as $\avg{m_{k}(t)m_{k'}(t)}$. Angle brackets, as always, represent averages over independent realisations of the dynamics i.e.
\begin{equation}
    \avg{F[\mathbf{s}(t)]} = \sum_{\mathbf{s}}F[\mathbf{s}(t)]P(\mathbf{s}; t),
\end{equation}
where the sum is over all $2^N$ spin configurations $\mathbf{s}=(s_1,\dots,s_N)$, and where $P(\mathbf{s};t)$ is the probability of finding the spins in the configuration $\mathbf{s}$ at time $t$. Specifically one finds (after some algebra),
\begin{subequations}
\begin{gather} 
    \frac{\dd \avg{m_{k}}}{\dd t} = \avg{F_{k}}, \label{eq supp: F and G diff eqns 1} \\
    \frac{\dd \avg{\sigma}}{\dd t} = \avg{F_{\sigma}},  \label{eq supp: F and G diff eqns 2} \\
    \frac{\dd \avg{m_{k}m_{k'}}}{\dd t} = \avg{m_{k'}F_{k}} + \avg{m_{k}F_{k'}} + \delta_{k,k'}\frac{\avg{G_{k}}}{N},  \label{eq supp: F and G diff eqns 3}
\end{gather}
\end{subequations}
where
\begin{subequations}
\begin{gather} 
    F_{k} = \frac{2}{N_{k}}\sum_{n=0}^{k}\left[W^{(+,k,n)}-W^{(-,k,n)}\right], \label{eq supp: F and G raw 1} \\
    F_{\sigma} = \frac{2}{\mu N}\sum_{k}\sum_{n=0}^{k}(k-2n)\left[W^{(+,k,n)}-W^{(-,k,n)}\right], \label{eq supp: F and G raw 2}  \\
    G_{k} = \frac{4}{P_{k}N_{k}}\sum_{n=0}^{k}\left[W^{(+,k,n)}+W^{(-,k,n)}\right],  \label{eq supp: F and G raw 3}
\end{gather}
\end{subequations}
and where we have written $P_{k}=\frac{N_{k}}{N}$ for the degree distribution of the network.

We now proceed to calculate the effective rates. $W^{(+,k,n)}$ is the product of the number of degree $k$ nodes in the down-state, the probability of a node in the down-state with degree $k$ having $n$ neighbours in the up-state, $P_{0}(k,n)$, and the rate at which such a down-state node flips to an up-state node, $R^{+}_{k,n}$. The latter is determined by the dynamics of the model. Therefore,
\begin{equation}
    W^{(+,k,n)} = (N_{k}-N_{1,k})P_{0}(k,n)R^{+}_{k,n}.
\end{equation}
In a similar manner,
\begin{equation}
    W^{(-,k,n)} = N_{1,k}P_{1}(k,n)R^{-}_{k,n},
\end{equation}
where $P_{1}(k,n)$ is the probability of a node in the up-state with degree $k$ having $n$ neighbours in the up-state, and $R^{-}_{k,n}$ is the rate at which such a node in the up-state flips to a down-state node.

The spin flip rates in our model are different from the ones in \cite{peralta2018stochastic}. We have
\begin{subequations}
\begin{align}
    R^{+}_{k,n} &= v_{+}'+h_{+}\frac{n}{k}+h_{+}'\frac{k-n}{k}, \\
    R^{-}_{k,n} &= v_{-}'+h_{-}\frac{k-n}{k}+h_{-}'\frac{n}{k}.
\end{align}
\end{subequations}
The probabilities $P_{0}(k,n)$ and $P_{1}(k,n)$ are intricate objects, determined by the distribution of the full spin configuration on the network. In order to make analytical progress we use the pair approximation and describe these probabilities in terms of $\left\{\mathbf{N}_{1}, \mathcal{L}\right\}$. This is the same procedure as in Sec.~\ref{appendix: pair approximation}, specifically Eq.~(\ref{eq supp: pair approx}), where we assumed that the probability of a degree $k$ node having $n$ neighbours in the opposite state is binomial. Using this approximation we have
\begin{subequations}
\begin{align}
    P_{0}(k,n) &= \binom{k}{n}c_{0}^{n}(1-c_{0})^{k-n}, \\
    P_{1}(k,n) &= \binom{k}{n}c_{1}^{k-n}(1-c_{1})^{n}.
\end{align}
\end{subequations}
In these expressions $c_{0/1}$ is the probability that a given neighbour of a node in the down/up-state is in the up/down-state (i.e. in both cases, the probability that a link, connected to the focal node, is active). Formally, 

\begin{equation}
    c_{0/1} = \frac{\mathcal{L}}{\sum_{k}kN_{0/1, k}},
\end{equation}
which is the ratio between the total number of active links and the number of links connected emanating from nodes in the up/down-state respectively. Evaluating the above expression gives
\begin{equation}
    c_{0/1} = \frac{\sigma}{1\mp m_{L}},
\end{equation}
where $m_{L}$ is the degree averaged magnetisation defined by
\begin{equation}
    m_{L} = \frac{1}{\mu}\sum_{k}kP_{k}m_{k}. \label{eq supp: SPA1 mL defn}
\end{equation}
With these ingredients we can now evaluate the objects on the right-hand sides of Eqs.~(\ref{eq supp: F and G raw 1})-(\ref{eq supp: F and G raw 3}) further. Due to the complexity this is done using Mathematica. An accompanying Mathematica notebook can be found at the following GitHub repository \cite{github}. 

The resulting expressions for the objects in Eqs.~(\ref{eq supp: F and G raw 1})-(\ref{eq supp: F and G raw 3}) are long, so we do not report them here in full. We include one expression as an example,  
\begin{equation}
    F_{k} = \alpha - \beta m_{k}+\frac{\sigma}{1-m_{L}^{2}}\Big[2\gamma(m_{L}-m_{k})+ (\gamma_{+}-\gamma_{-})(1-m_{k}m_{L})\Big], \label{eq supp: SPA1 Fk}
\end{equation}
where we have used the shorthands from Eq.~(\ref{eq:shorthands}). 

As before we now make the restriction from Eq.~(\ref{eq supp: gamma restriction}), $\gamma_{+}=\gamma_{-}$, which helps maintain analytic tractability. The different quantities in Eqs.~(\ref{eq supp: F and G raw 1})-(\ref{eq supp: F and G raw 3}) can be expressed in terms of $m_{k}$, $m_{L}$, $m$, $\mu$, $\mu_{2}$, and the shorthand parameter combinations defined in Eq.~(\ref{eq:shorthands}). The quantity $m$ is the global magnetisation used throughout this paper, and can be calculated from the degree specific magnetisation $m_k$, 
\begin{equation}
    m = \sum_{k}P_{k}m_{k}. \label{eq supp: SPA1 global mag defn}
\end{equation}
We use the notation
\begin{equation}
    \mu_{n} = \sum_{k}P_{k}\mu^{n}
\end{equation}
for the $n$\textsuperscript{th} moment of the degree distribution.

We also point out a typo in the second moment of $P_{1}(k,n)$ given on p.~9 of \cite{peralta2018stochastic}. This should in fact be $\sum_{n=0}^{k}n^{2}P_{1}(k,n)=k\left[k-(2k-1)c_{1}+(k-1)c_{1}^{2}\right]$.
\par
We find that Eqs.~(\ref{eq supp: F and G diff eqns 1})-(\ref{eq supp: F and G diff eqns 3}) are not  closed. However we can use Eq.~(\ref{eq supp: F and G diff eqns 1}) to obtain a differential equation for $\avg{m_{L}}$. To do this, we evaluate $\frac{1}{\mu}\sum_{k}kP_{k}\avg{F_{k}}$, and  find
\begin{equation}
    \frac{\dd \avg{m_{L}}}{\dd t} = \alpha-\beta\avg{m_{L}}. \label{eq supp: SPA1 diff m_L}
\end{equation}
As a consequence we have
\begin{equation}
    \avg{m_{L}} = \avg{m_{L}}_{\rm st}-\big[\avg{m_{L}}_{\rm st}-\avg{m_{L}(0)}\big]e^{-\beta t}, \label{eq supp: SPA mL time solution}
\end{equation}
and the steady-state value of $\avg{m_L}$ is
\begin{equation}
    \avg{m_{L}}_{\rm st} = \frac{\alpha}{\beta} = \frac{(h_{+}'-h_{-}')+(v_{+}'-v_{-}')}{(h_{+}'+h_{-}')+(v_{+}'+v_{-}')}. \label{eq supp: SPA1 mL steady}
\end{equation}
This is the same solution that we have found in all other cases, for instance see Sec.~\ref{appendix: pair approximation magnetisation} where we analysed the standard pair approximation. This is not surprising, as all we have done in this section is perform the pair approximation again but in a style better suited to the extension we will make in Sec.~\ref{appendix: SPA system size expansions}. In fact, we have already shown in Sec.~\ref{appendix: random walks weighted mag} that Eq.~(\ref{eq supp: SPA1 mL steady}) is a very general solution, a fact we will use in the next section.

\subsection{System-size expansion} \label{appendix: SPA system size expansions}
We now use what is called `S2PA' (stochastic pair approximation 2) in \cite{peralta2018stochastic} to address Eqs.~(\ref{eq supp: F and G diff eqns 1})-(\ref{eq supp: F and G diff eqns 3}). This method focuses on expanding around the deterministic solution for the variable $m_{L}$.

Motivated by the fact that $\avg{m_{L}}$ is a self governed variable, as we saw from Eq.~(\ref{eq supp: SPA1 diff m_L}), and that the steady-state interface density under the standard pair approximation is as in Eq.~(\ref{eq supp: pair approx interface}), we purpose the following expansion:
\begin{subequations}
\begin{align}
    m_{k} &= m_{L} + \frac{\lambda_{k}}{\sqrt{N}} + \frac{\nu_{k}}{N}+..., \label{eq supp: SPA2 mk expansion} \\
    \sigma &= \frac{1}{2}H_{\text{PA}}(\tau)(1-m_{L}^2) + \frac{\lambda_{\sigma}}{\sqrt{N}}+..., \label{eq supp: SPA2 interface expansion}
\end{align}
\end{subequations}
where $\lambda_{k}$, $\nu_{k}$ and $\lambda_{\sigma}$ are stochastic variables. All objects in these equations, excluding $H_{\rm PA}(\tau)$ which is given in Eq.~(\ref{eq supp: H pair}), are understood to be time-dependent. Ultimately however, we are only interested in the stationary state values. We neglect higher-order terms in the expansion in powers of $N^{-1/2}$. The goal is to calculate $\avg{\sigma}_{\rm st}$ to leading and sub-leading order. Any expansions we perform in the remainder of this section can be found in a Mathematica notebook at the following GitHub repository \cite{github}.

This expansion is further justified by the fact that individual simulated trajectories in $(m,\sigma)$-space, performed on finite size uncorrelated networks, fluctuate around a parabola with height $H_{\rm PA}(\tau)$. This is demonstrated in Fig.~\ref{fig main: parabolas} in the main paper. Simulated trajectories on a truly infinite uncorrelated network would move exactly along this parabola, in fact they would follow the path described by Eqs.~(\ref{eq supp: complete network analytic magnetisation}) and (\ref{eq supp: pair approximation interface differential}) which is also shown in Fig.~\ref{fig main: parabolas}. Time averaging such analytical trajectories on infinite networks in the steady-state (equivalently we can average over many such independent trajectories at a single time in the steady state) will result in the point given by Eqs.~(\ref{eq supp: SPA1 mL steady}) and (\ref{eq supp: pair approx interface}). However,  stochasticity means that the steady-state averages in finite networks sit on a different parabola. This is what we wish to determine and what motivates the system-size expansion in Eqs.~(\ref{eq supp: SPA2 mk expansion}) and (\ref{eq supp: SPA2 interface expansion}).

To continue, we note that based on the results from Sec.~\ref{appendix: random walks weighted mag}, that any normalised weighted magnetisation in the steady-state is equal to the steady-state global magnetisation, Eq.~(\ref{eq supp: random walk weighted steady-state mag}). The degree-specific average magnetisation $\avg{m_{k}}_{\rm st}$ is effectively a weighted magnetisation with weights $\delta_{k,k_{i}}$, i.e. we have
\begin{equation}
    \avg{m_{k}} = \frac{\sum_{i}\delta_{k,k_{i}}S_{i}}{\sum_{i}\delta_{k,k_{i}}},
\end{equation}
where $S_{i}$ is the average spin state at site $i$. Thus we have $\avg{m_{k}}_{\rm st} = \avg{m}_{\rm st}$. Given the definition of $m_L$ in Eq.~(\ref{eq supp: SPA1 mL defn}) this then also means that $\avg{m_k}_{\rm st}=\avg{m_L}_{\rm st}$ for all $k$. With this we see immediately from Eq.~(\ref{eq supp: SPA2 mk expansion}) that $\avg{\lambda_{k}}_{\rm st}=0$ and $\avg{\nu_{k}}_{\rm st}=0$. 

Next we substitute Eqs.~(\ref{eq supp: SPA2 mk expansion}) and (\ref{eq supp: SPA2 interface expansion}) into Eqs.~(\ref{eq supp: F and G diff eqns 2}) and (\ref{eq supp: F and G diff eqns 3}), and equate powers of $N^{-1/2}$. Many model parameters that appear, i.e. $\alpha$, $\beta$, $\gamma$ etc. from Eq.~(\ref{eq:shorthands}), can be expressed in terms of the noise-to-order ratio $\tau=\frac{\beta}{\gamma}$ from Eq.~(\ref{eq supp: tau}) and the steady-state global magnetisation $\avg{m}_{\rm st}=\frac{\alpha}{\beta}$ from Eq.~(\ref{eq supp: complete network steady-state magnetisation}). We find
\begin{align}
    \frac{\dd \avg{\lambda_{\sigma}}}{\dd t} &= -2\left\{\tau-1+2H_{\rm PA}(\tau)+\frac{2}{\mu}\big[1-H_{\rm PA}(\tau)\big]\right\}\avg{\lambda_{\sigma}} \nonumber \\
    &\qquad + \frac{2\gamma}{\mu}H_{\rm PA}(\tau)\big[1-H_{\rm PA}(\tau)\big]\sum_{k}P_{k}\avg{\lambda_{k}m_{L}} \label{eq supp: SPA2 lambda_sigma}
\end{align}
and
\be
    \frac{\dd \avg{\lambda_{k}m_{L}}}{\dd t} = -\gamma\big[H_{\text{PA}}(\tau)+2\tau\big]\avg{\lambda_{k}m_{L}}+\gamma\tau \avg{m}_{\rm st}\avg{\lambda_{k}} \label{eq supp: SPA2 lambda_k mL diff}
\ee
respectively. We note that to get Eq.~(\ref{eq supp: SPA2 lambda_k mL diff}) we set $k=k'$ in Eq.~(\ref{eq supp: F and G diff eqns 3}). 

Taking both of these equations to the steady-state, and given that we already know that $\avg{\lambda_{k}}_{\rm st}=0$, we also find that $\avg{\lambda_{\sigma}}_{\rm st}=0$. With this $\avg{\sigma}_{\rm st}$ reduces to
\begin{equation}
    \avg{\sigma}_{\rm st} = \frac{1}{2}H_{\rm PA}(\tau)\Big(1-\avg{m_{L}^{2}}_{\rm st}\Big), \label{eq supp: SPA2 interface steady raw}
\end{equation}
and we are left with determining $\avg{m_{L}^{2}}_{\rm st}$. Comparing with the result from the conventional pair approximation in Eq.~(\ref{eq supp: pair approx interface}), we note that the final result contains the object $\avg{m_L^2}_{\rm st}$ instead of $\avg{m}_{\rm st}^2$. In particular the square is now inside the average.

To calculate $\avg{m_L^2}_{\rm st}$, we start from Eq.~(\ref{eq supp: F and G diff eqns 3}),
\be
    \frac{\dd \avg{m_{k}m_{k'}}}{\dd t} = \avg{m_{k'}F_{k}} + \avg{m_{k}F_{k'}} + \delta_{k,k'}\frac{\avg{G_{k}}}{N}. \label{eq supp: SPA1 diff m_k m_k' in SPA2}
\ee
We expand this equation in powers of $\frac{1}{\sqrt{N}}$ using Eqs.~(\ref{eq supp: SPA2 mk expansion}). The LHS of Eq.~(\ref{eq supp: SPA1 diff m_k m_k' in SPA2}) then becomes
\begin{align}
    \frac{\dd \avg{m_{k}m_{k'}}}{\dd t} &= \frac{\dd \avg{m_{L}^{2}}}{\dd t} + \frac{1}{\sqrt{N}}\left(\frac{\dd \avg{\lambda_{k}m_{L}}}{\dd t} + \frac{\dd \avg{\lambda_{k'}m_{L}}}{\dd t}\right) \nonumber \\
    &\qquad + \frac{1}{N}\left(\frac{\dd \avg{\lambda_{k}\lambda_{k'}}}{\dd t} + \frac{\avg{\nu_{k}m_{L}}}{\dd t} + \frac{\avg{\nu_{k'}m_{L}}}{\dd t}\right)+{\cal O}(N^{-3/2}).
\end{align}
Next we consider the term $\avg{m_{k'}F_{k}}$ on the RHS of Eq.~(\ref{eq supp: SPA1 diff m_k m_k' in SPA2}). This expands as \cite{github}
\begin{align}
    \avg{m_{k'}F_{k}} &= \gamma\tau\left[\avg{m}_{\rm st}\avg{m_{L}}-\avg{m_{L}^{2}}\right] \nonumber \\
    &\qquad + \frac{\gamma}{\sqrt{N}}\Big[-\Big(H_{\rm PA}(\tau) + \tau\Big)\avg{\lambda_{k}m_{L}}-\tau\avg{\lambda_{k'}m_{L}}+\tau\avg{\lambda_{k'}}\avg{m}_{\rm st}\Big] \nonumber \\
    &\qquad +\frac{\gamma}{N}\Bigg[-\Big(H_{\rm PA}(\tau)+\tau\Big)\avg{\lambda_{k}\lambda_{k'}}+\tau\avg{\nu_{k}}\avg{m}_{\rm st} - 2\avg{\frac{m_{L}\lambda_{\sigma}\lambda_{k'}}{1-m_{L}^{2}}} \nonumber \\
    &\qquad\qquad -\tau\avg{\nu_{k}m_{L}}-\Big(H_{\rm PA}(\tau)+\tau\Big)\avg{m_{L}\nu_{k'}}\Bigg] \nonumber \\
    &\qquad + {\cal O}(N^{-3/2}).
\end{align}
The $\avg{m_{k}F_{k'}}$ term gives a similar expression, but with $k$ and $k'$ interchanged. 

Since $\avg{G_{k}}$ carries a pre-factor $1/N$ in Eq.~(\ref{eq supp: SPA1 diff m_k m_k' in SPA2}) we only consider the leading order term of $\avg{G_{k}}$. We have \cite{github}
\begin{equation}
    \avg{G_{k}} = \frac{2\gamma}{P_{k}}\Big[H_{\rm PA}(\tau) + \tau - H_{\rm PA}(\tau)\avg{m_{L}^{2}}-\tau\avg{m}_{\rm st}\avg{m_{L}}\Big]+{\cal O}(N^{-1/2}).
\end{equation}
Collecting terms in the expansions of the left-hand and right-hand sides of Eq.~(\ref{eq supp: SPA1 diff m_k m_k' in SPA2}), and keeping only expressions up to and including order $1/N$, we have  
\begin{align}
    \frac{\dd \avg{m_{L}^{2}}}{\dd t} &+ \frac{1}{\sqrt{N}}\left(\frac{\dd \avg{\lambda_{k}m_{L}}}{\dd t} + \frac{\dd \avg{\lambda_{k'}m_{L}}}{\dd t}\right) + \frac{1}{N}\left(\frac{\dd \avg{\lambda_{k}\lambda_{k'}}}{\dd t} + \frac{\avg{\nu_{k}m_{L}}}{\dd t} + \frac{\avg{\nu_{k'}m_{L}}}{\dd t}\right) =  \nonumber \\
    &\qquad 2\gamma\tau\left[\avg{m}_{\rm st}\avg{m_{L}}-\avg{m_{L}^{2}}\right] \nonumber \\
    &\qquad +\frac{\gamma}{\sqrt{N}}\Bigg\{-\Big(H_{\rm PA}(\tau)+2\tau\Big)\Big[\avg{\lambda_{k}m_{L}}+\avg{\lambda_{k'}m_{L}}\Big]+\tau\avg{m}_{\rm st}\Big[\avg{\lambda_{k}} + \avg{\lambda_{k'}}\Big]\Bigg\} \nonumber \\
    &\qquad + \frac{\gamma}{N}\Bigg\{-2\Big(H_{\rm PA}(\tau)+\tau\Big)\avg{\lambda_{k}\lambda_{k'}}+\tau\avg{m}_{\rm st}\Big[\avg{\nu_{k}}+\avg{\nu_{k'}}\Big] - 2\avg{\frac{(\lambda_{k}+\lambda_{k'})m_{L}\lambda_{\sigma}}{1-m_{L}^{2}}} \nonumber \\
    &\qquad\qquad\qquad -\Big(H_{\rm PA}(\tau)+2\tau\Big)\Big[\avg{\nu_{k}m_{L}} + \avg{\nu_{k'}m_{L}}\Big] \nonumber \\
    &\qquad\qquad\qquad + \frac{2}{P_{k}}\Big[H_{\rm PA}(\tau) + \tau - H_{\rm PA}(\tau)\avg{m_{L}^{2}}-\tau\avg{m}_{\rm st}\avg{m_{L}}\Big]\delta_{k,k'}\Bigg\}. \label{eq supp: SPA2 full expanded mkmk' diff}
\end{align}
Performing the operation $\sum_{kk'}kk'P_{k}P_{k'}$ on both sides, several of the contributions can be eliminated. To see this, we perform the sum $\sum_{k}kP_{k}$ on the expansion of $m_{k}$ from Eq.~(\ref{eq supp: SPA2 mk expansion}), and find  
\be
    \sum_{k}kP_{k}m_{k} = \sum_{k}kP_{k}m_{L} + \frac{1}{\sqrt{N}}\sum_{k}kP_{k}\lambda_{k}+\frac{1}{N}\sum_{k}kP_{k}\nu_{k}, 
\ee
resulting in
\be
    \mu m_{L} = \mu m_{L} + \frac{1}{\sqrt{N}}\sum_{k}kP_{k}\lambda_{k}+\frac{1}{N}\sum_{k}kP_{k}\nu_{k}.
\ee
Equating powers of $N^{-1/2}$ we therefore conclude
\begin{subequations}
\begin{align}
    \sum_{k}kP_{k}\lambda_{k} &= 0,  \label{eq supp: SPA2 sum identity 1} \\
    \sum_{k}kP_{k}\nu_{k} &= 0. \label{eq supp: SPA2 sum identity 2}
\end{align}
\end{subequations}
We now use the identities in Eqs.~(\ref{eq supp: SPA2 sum identity 1}) and (\ref{eq supp: SPA2 sum identity 2}) when performing the operation \\ $\sum_{kk'}kk'P_{k}P_{k'}$ in Eq.~(\ref{eq supp: SPA2 full expanded mkmk' diff}). This results in all stochastic terms vanishing and we are left with
\begin{align}
    \frac{\dd \avg{m_{L}^{2}}}{\dd t} &= -2\gamma\tau\Big[\avg{m_{L}^{2}}-\avg{m}_{\rm st}\avg{m_{L}}\Big] \nonumber \\
    &\qquad + \frac{2\gamma\mu_{2}}{N\mu^{2}}\Big[H_{\rm PA}(\tau)+ \tau - H_{\rm PA}(\tau)\avg{m_{L}^{2}}-\tau\avg{m}_{\rm st}\avg{m_{L}}\Big], \label{eq supp: SPA2 diff m_L^2}
\end{align}
where we note the introduction of $\mu_{2}$ and $\mu^{2}$ arising from terms of the form $\sum_{k}k^{2}P_{k}$ and $(\sum_{k}kP_{k})^2$. It is possible to solve this equation analytically and obtain a full time dependent solution for $\avg{m_{L}^{2}(t)}$ and thus $\avg{\sigma(t)}$ using Eq.~(\ref{eq supp: SPA mL time solution}).

However, we are mostly interested in the steady-state, so we move to that regime. Eq.~(\ref{eq supp: SPA2 diff m_L^2}) then becomes
\begin{equation}
    0 = -\tau N\Big[\avg{m_{L}^{2}}_{\rm st} - \avg{m}^{2}_{\rm st}\Big] + \frac{\mu_{2}}{\mu^{2}}\Big[H_{\rm PA}(\tau) + \tau - H_{\rm PA}(\tau)\avg{m_{L}^{2}}_{\rm st} - \tau\avg{m}^{2}_{\rm st}\Big]
\end{equation}
and we can solve for $\avg{m_{L}^{2}}_{\rm st}$:
\begin{equation}
    \avg{m^{2}_{L}}_{\rm st} = \frac{(H_{\rm PA}(\tau)+\tau)\mu_{2}-\tau(\mu_{2}-N\mu^{2})\avg{m}_{\rm st}^{2}}{N\mu^{2}\tau+H_{\rm PA}(\tau)\mu_{2}}. 
\end{equation}
Finally, we substitute this back into Eq.~(\ref{eq supp: SPA2 interface steady raw}) and obtain
\begin{equation}
    \avg{\sigma}_{\rm st} = \frac{1}{2}H_{\text{SPA}}(\tau)\Big(1-\avg{m}_{\rm st}^{2}\Big),
\end{equation}
where
\begin{equation}
    H_{\text{SPA}}(\tau) = \frac{\tau H_{\text{PA}}(\tau)(N\mu^{2}-\mu_{2})}{N\mu^{2}\tau+H_{\text{PA}}(\tau)\mu_{2}}, \label{eq supp: H SPA2}
\end{equation}
(the subscript `SPA' of course stands for `stochastic pair approximation'). Expanding in $1/N$ Eq.~(\ref{eq supp: H SPA2}) can be written as
\be
    H_{\text{SPA}}(\tau)=H_{\text{PA}}(\tau)\left[1-\frac{\mu_{2}}{N\mu^{2}}\frac{\tau+H_{\text{PA}}(\tau)}{\tau}\right]+{\cal O}(N^{-2}), \label{eq supp: H_SPA2 expanded}
\ee
making it transparent that the result from the stochastic pair approximation is a finite-size correction to the expression obtained from the conventional pair approximation.

Due to the terms involving $H_{\rm PA}(\tau)$ this solution applies in the same regions as that of the conventional pair approximation, see Sec.~\ref{appendix: uncorrelated network - interface density} for a detailed discussion of this, as well as Secs.~\ref{appendix: H and tau} and \ref{appendix: determining model parameters tau and m} for discussions on the $\tau$ limits in general. 

Focusing on positive $\tau$, we substitute the solution for $H_{\rm PA}(\tau)$ from Eq.~(\ref{eq supp: H pair}) into Eq.~(\ref{eq supp: H SPA2}) and then expand in powers of $\tau$. To lowest order we find
\be
    H_{\rm SPA}(\tau) = \left(\frac{N\mu^{2}}{\mu_{2}}-1\right)\tau + \order{\tau^2}.
\ee 
Thus it is clear that in the limit $\tau\to 0$, $H_{\rm SPA}(\tau)\to 0$.  This matches what we see in simulations on finite size networks. 

In the limit $\tau\to\infty$, Eq.~(\ref{eq supp: H_SPA2 expanded}) becomes
\be
    \lim_{\tau\to\infty}H_{\rm SPA}(\tau) = 1-\frac{\mu_{2}}{N\mu^2}. \label{eq supp: SPA2 tau to inf}
\ee
This is of similar form to Eq.~(\ref{eq supp: annealed tau to inf}) when we studied the annealed approximation. We know the large $\tau$ regime is dominated by random state changes so there are no correlations between neighbours, hence we expect $H(\tau)\to 1$, as this corresponds to random scattering. Eq.~(\ref{eq supp: SPA2 tau to inf}) reproduces this, up to corrections in $1/N$.

\begin{figure}[htbp]
    \includegraphics[scale = 0.95]{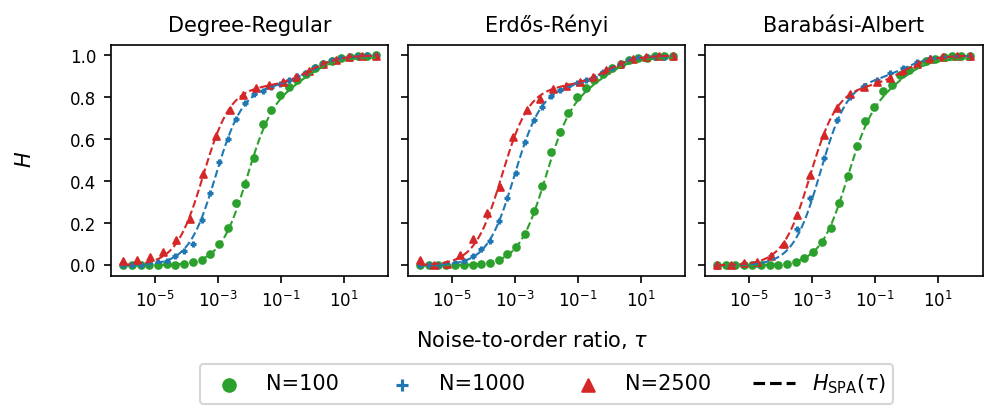}
    \caption{Plots of $H(\tau)=\avg{\sigma}_{\rm st}/[\frac{1}{2}(1-\avg{m}_{\rm st}^{2})]$, which characterises the scattering, as a function of the noise-to-order ratio, $\tau$, for networks of size $N$. Degree-regular with $\mu=8$ (left), Erd\"os--R\'enyi with $p=8/N$ (middle), Barab\'asi--Albert with $4$ new links established upon adding a new node (right). For all three types of network we have $\mu\approx8$ (deviations are due to the finite size of the graphs). The dashed lines are the results from the SPA approximation, Eq.~(\ref{eq supp: H SPA2}). Each marker is the result from averaging 1000 independent simulations in the steady-state at a particular value of $\tau$. Model parameters for a particular $\tau$ value can be generated via the algorithm detailed in Sec.~\ref{appendix: determining model parameters tau and m}.}
    \label{fig supp: H SPA2}
    \end{figure}

In Fig.~\ref{fig supp: H SPA2} we demonstrate how Eq.~(\ref{eq supp: H SPA2}) compares to simulation for different size degree-regular (left), Erd\"os-–R\'enyi (middle), and Barab\'asi--Albert (right) networks. We find convincing agreement across all ranges of $\tau$. We note that the approximation becomes inaccurate for very small networks [see Fig.~\ref{fig main: theory comparison}(b) in the main paper].

\FloatBarrier
\clearpage
\newpage
\section{Scattering and the noise-to-order ratio} \label{appendix: H and tau}

\subsection{Relation between \texorpdfstring{$\mathbb{H}$}{} and \texorpdfstring{$H(\tau)$}{}}

\subsubsection{Degree of scatter for individual realisations}
For individual configurations of the system, the parameter $\mathbb{H}$ is defined as [Eq.(\ref{eq main: H})]
\be
    \mathbb{H} \equiv \frac{\sigma}{\frac{1}{2}(1-m^{2})}. \label{eq supp: H}
\ee
$\mathbb{H}$ quantifies the `scatter' of a spin configuration, relative to a random configuration with the same magnetisation (same fraction of up-spins). More specifically, the denominator in Eq.~(\ref{eq supp: H}) is the (expected) fraction of links that are active if spins of total magnetisation $m$ are placed randomly on the network. The quantity $\mathbb{H}$ is not defined if $m=\pm1$, i.e. all spins are in the same state. From an ensemble of independent simulation runs we can measure the quantity $\avg{\mathbb{H}(t)}$. In computing this average, we discard any configurations where $\mathbb{H}$ is not defined. Such configurations are very rare provided $\tau$ is not too small. The exclusion is always implied when we write $\avg{\mathbb{H}}$. 

\begin{figure}[t]
    \includegraphics[scale = 0.8]{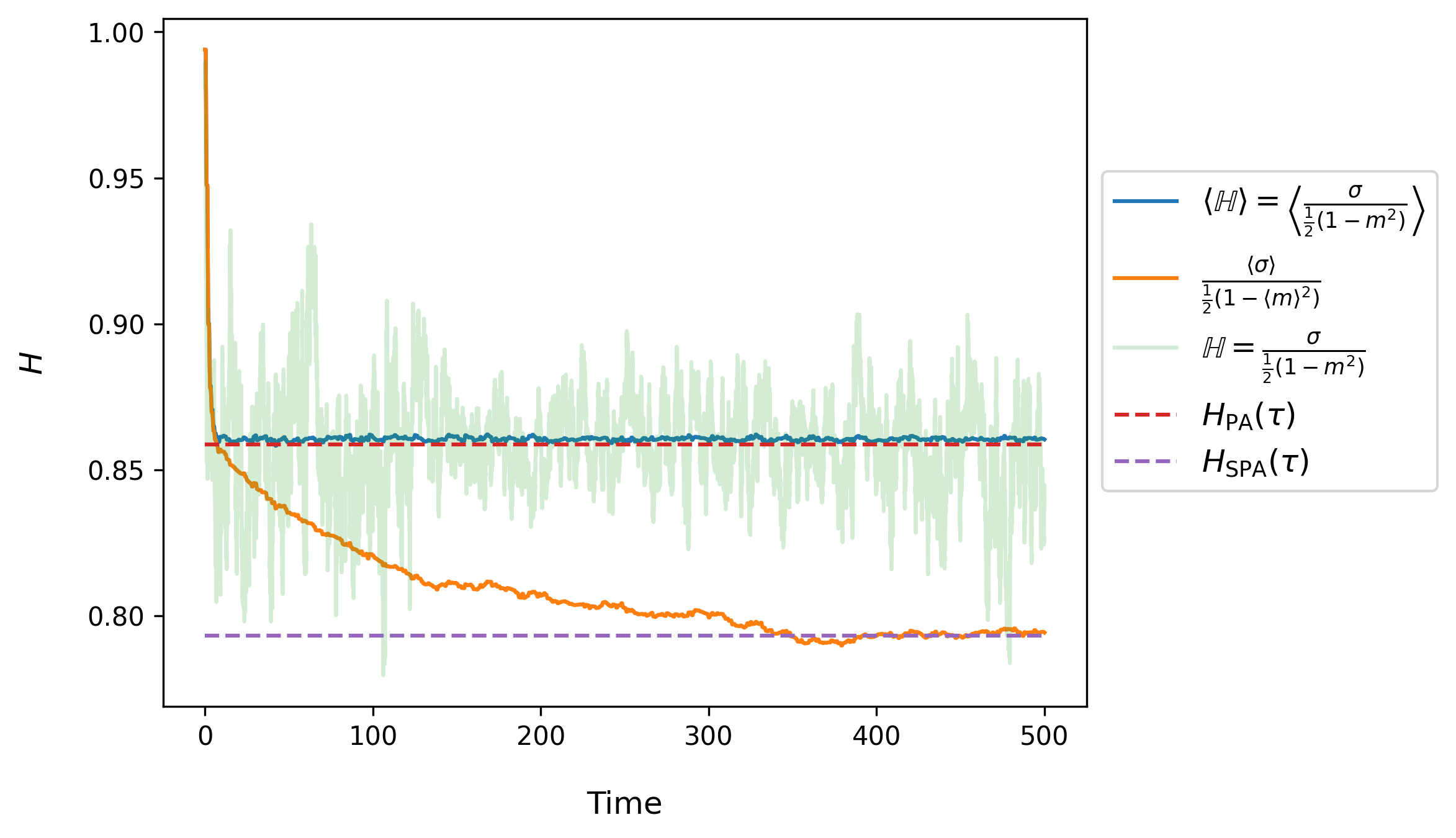}
    \caption{Time evolution of various order parameters for a Barab\'asi-Albert network with $\mu\approx 8$ and $N=2500$. The red and purple dashed lines are theory lines from $H_{\rm PA}(\tau)$ [Eq.~(\ref{eq supp: H pair})] and $H_{\rm SPA}(\tau)$ [Eq.~(\ref{eq supp: H SPA2})] respectively. The solid light green line is the measured value of $\mathbb{H}(t)$ in a single simulation. The solid blue line is $\avg{\mathbb{H}(t)}$ over many such simulations. The solid orange line is $H(\tau,t)$, as defined in the text and in the legend, and obtained as an average over many simulations. Data in this example is for $\tau=10^{-2}$.}
    \label{fig supp: comparing H time}
    \end{figure}

In Fig.~\ref{fig supp: comparing H time} we show $\mathbb{H}(t)$ (green line) for a single simulation on a Barab\'asi-Albert network of size $N=2500$, and $\avg{\mathbb{H}(t)}$ obtained from many independent simulations (blue line). The value of $\mathbb{H}(t)$ fluctuates around $\avg{\mathbb{H}}_{\rm st}$. The amplitude of the fluctuations is about one tenth of the mean in this example. Therefore, the value of $\mathbb H$ measured from a single configuration in the stationary state will be well approximated by $\avg{\mathbb{H}}_{\rm st}$.

\subsubsection{Relation between average density of active interfaces and average magnetisation}
We define a second order parameter
\be
    H(\tau) \equiv \frac{\avg{\sigma}_{\rm st}}{\frac{1}{2}(1-\avg{m}_{\rm st}^{2})}. \label{eq supp: H(tau)}
\ee
Even if individual configurations are in `consensus' states ($m=\pm 1$), time averages will lead to $\avg{m}_{\rm st} \neq \pm 1$ (except for pathological choices of the model parameters). We can also define $H(\tau,t) \equiv \frac{\avg{\sigma(t)}}{\frac{1}{2}(1-\avg{m(t)}^{2})}$. This quantity is shown as an orange line in Fig.~\ref{fig supp: comparing H time} (for $\tau=10^{-2}$). The data demonstrates that $H(\tau)$ will in general not take the same value as $\avg{\mathbb{H}}_{\rm st}$ (blue line). 

We note in Fig.~\ref{fig supp: comparing H time} that, in the stationary state, $\mathbb{H}$ for a single realisation (green line) fluctuates around $H_{\rm PA}(\tau)$ while $H(\tau, t)$ (orange line) approaches $H_{\rm SPA}(\tau)$. This is analogous to Fig.~\ref{fig main: parabolas} in the main paper, where individual trajectories fluctuate around a parabola whose height is set by $H_{\rm PA}(\tau)$ but where the points defined by $\avg{\sigma}_{\rm st}$ and $\avg{m}_{\rm st}$ sit on a parabola with a different height $H_{\rm SPA}(\tau)$.

\subsubsection{\texorpdfstring{$H(\tau)$}{} can be inferred from measurement \texorpdfstring{$\mathbb{H}$}{} in individual configurations, for large \texorpdfstring{$\tau$}{}}
\begin{figure}[t]
    \includegraphics[scale = 0.72]{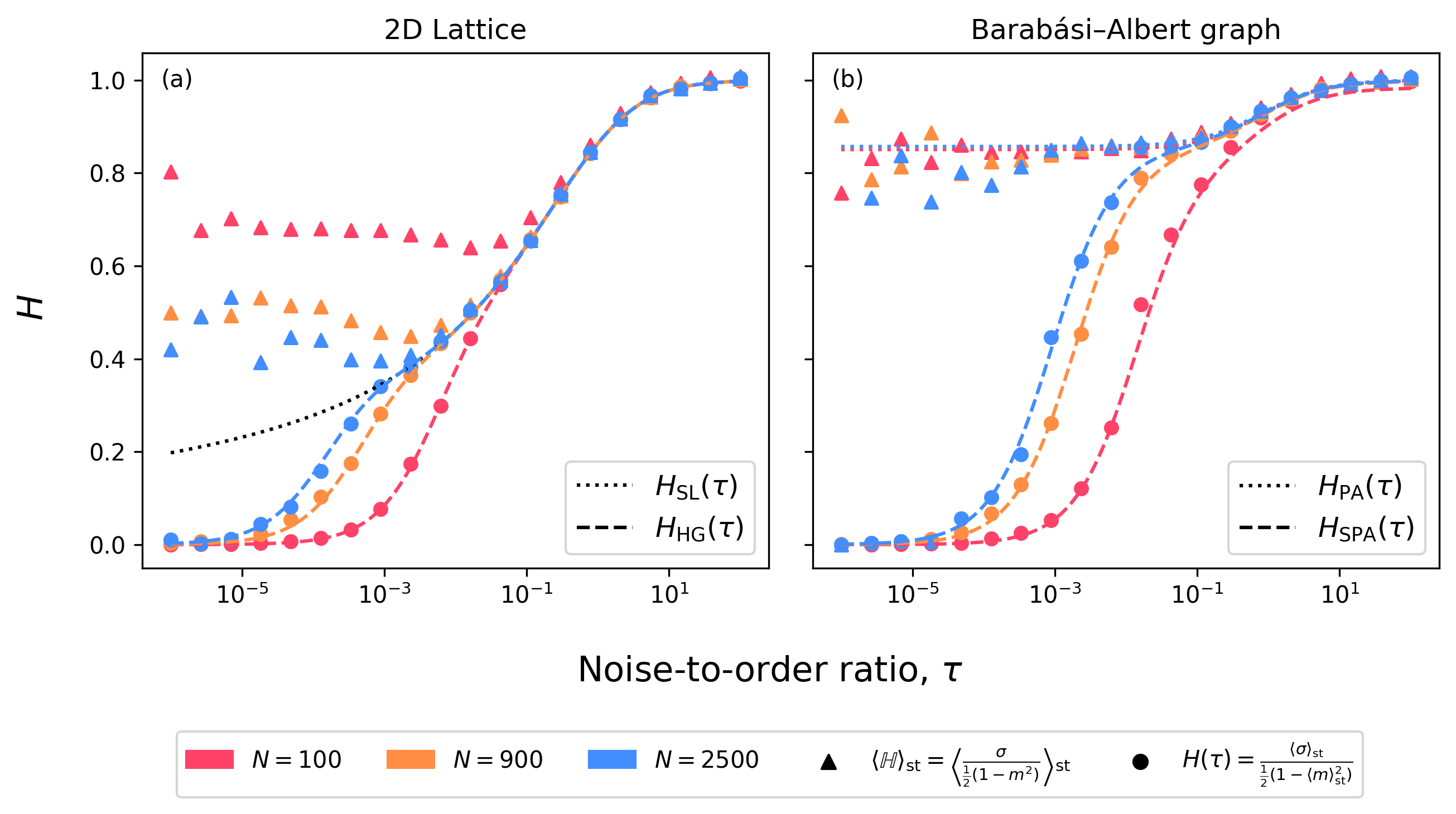}
    \caption{Plots comparing $\avg{\mathbb{H}}_{\rm st}$ (triangles) and $H(\tau)$ (circles), from Eqs.~(\ref{eq supp: H}) and (\ref{eq supp: H(tau)}), in simulations on square lattices with periodic boundary conditions (left) and Barab\'asi-Albert networks with $\mu\approx 8$ (right), of varying size $N$. Left panel: dotted line is the $H_{\rm SL}(\tau)$ theory line for an infinite square lattice [Eq.~(\ref{eq supp: H 2d square lattice})], dashed lines are the $H_{\rm HG}(\tau)$ theory lines for homogeneous networks [Eq.~(\ref{eq supp: H_HG})]. Right panel: dotted lines are $H_{\rm PA}(\tau)$ theory lines for infinite uncorrelated networks [Eq.~(\ref{eq supp: H pair})], dashed lines are the $H_{\rm SPA}(\tau)$ theory lines for large but finite uncorrelated networks [Eq.~(\ref{eq supp: H SPA2})]. In both panels we find that there is much more stochasticity in $\avg{\mathbb{H}}_{\rm st}$ (triangles) than in $H(\tau)$ (circles). This is because if the system reaches consensus (all spins up or down), $\mathbb{H}_{\rm st}$ cannot be measured and the sample is discarded. This happens the majority of the time at low $\tau$ values meaning many more simulations are needed to measure $\avg{\mathbb{H}}_{\rm st}$ than $H(\tau)$}
    \label{fig supp: comparing H}
    \end{figure}

In Fig.~\ref{fig supp: comparing H} we compare $\avg{\mathbb{H}}_{\rm st}= \avg{\frac{\sigma}{\frac{1}{2}(1-m^{2})}}_{\rm st}$ and $H(\tau)=\frac{\avg{\sigma}_{\rm st}}{\frac{1}{2}(1-\avg{m}_{\rm st}^{2})}$ over a range of values of $\tau$ for square lattices and Barab\'asi--Albert networks of different sizes. We find that $H(\tau)$ measured in simulations follows the analytical predictions $H_{\rm HG}(\tau)$ in the case of finite lattices, and $H_{\rm SPA}(\tau)$ for the Barab\'asi--Albert network. 

We note in Fig.~\ref{fig supp: comparing H} that $\avg{\mathbb{H}}_{\rm st}$ measured in simulations agrees with the analytical predictions for $H(\tau)$ (and with the measurements of this quantity in simulation) above some $\tau_C$. The value of $\tau_C$ depends on the type and size of the network. Deviations are seen for $\tau<\tau_{C}$. The value of $\tau_{C}$ decreases as $N$ becomes larger, see e.g. the triangle markers in Fig.~\ref{fig supp: comparing H}(a).

Analytical functions for the order parameter $H(\tau)$ can be derived for many networks. These are the functions $H_{X}(\tau)$ [Eqs.~(\ref{eq supp: H pair}), (\ref{eq supp: H annealed}), (\ref{eq supp: H_HG}) and (\ref{eq supp: H SPA2})]. These functions depend only on properties of the network (e.g. the average degree $\mu$), and their argument is $\tau$ [see Sec.~\ref{appendix: tau}]. This, together with the above, provides a method to infer $\tau$ from snap shots of spin configurations in the steady-state.

\subsubsection{Range of possible \texorpdfstring{$H(\tau)$}{} values}
Since $m$ is bound on $[-1, 1]$, and $\sigma$ on $[0, 1]$, $H(\tau)$ can, in principle, take any real value greater than or equal to 0. However, not all combinations of $\avg{m}_{\rm st}$ and $\avg{\sigma}_{\rm st}$ are realised by the model dynamics. In practice the maximum value of $H(\tau)$ that can be achieved depends on the topology of the network. For general networks, finding the assignment of states $\pm 1$ for each node leading to the largest possible number of active links is closely related to the maximum-cut problem, which is known to NP-hard \cite{maxcutNPhard}.

\subsection{Noise-to-order ratio, \texorpdfstring{$\tau$}{}} \label{appendix: tau}

\subsubsection{Definition and interpretation}
The noise-to-order ratio, $\tau$, is given by [see Eq.~(\ref{eq supp: tau})]
\be
    \tau = 2\frac{h_{+}'+h_{-}'+v_{+}'+v_{-}'}{(h_{+}+h_{-})-(h_{+}'+h_{-}')}. \label{eq supp: negative tau tau defn}
\ee

{\em Interpretation of the numerator}\\ 
To interpret the numerator we imagine a fully ordered spin configuration, that is, all spins up or all spins down. For example, if all spins are up, then the only possible events are spontaneous down-flips or down-flips due to incorrect copying. As seen from Eqs.~(\ref{eq supp: dyn1}) and (\ref{eq supp: dyn2}) the total event rate for the former is $v_{-}' N$ and for the latter $h_{-}'N$. So the total event rate, starting from the all-up-state is $(v_-'+h_-')N$. Similarly, the total event rate starting from the all down configuration is $(v_+'+h_+')N$. Up to the factor of $N$, we can thus think of the numerator in Eq.~(\ref{eq supp: negative tau tau defn}) as the average event rate in a fully ordered state. Any event in such a state moves the system towards disorder.

{\em Interpretation of the denominator}\\
We now show that the denominator describes the initial rate of the build-up of nearest-neighbour correlations if the system is started from a fully uncorrelated state with $m=0$. In other words, we assume that, in the initial state, each spin is up or down with equal probability, and independently of its neighbours. 
Then we calculate $\frac{\dd}{\dd t} \left[\avg{ss'}-\avg{s}\avg{s'}\right]$.

 If a give node is the state $+1$ and if $n$ of its $k$ nearest neighbours are in state $-1$, then the rate at which it flips to $-1$ is
\begin{align}\label{eq:aux_corr1}
    T_{n,k,t}^- &=  h_{-}\frac{n}{k}  +  h_{-}'\frac{k-n}{k}  +  v_-'.
\end{align}

Spontaneous down flips in one node (described by $v_-'$) cannot affect correlations between neighbours. In fact, in the fully random state with $m=0$ such a spontaneous flip will, on average, convert as many active interfaces into inactive interfaces as the other way round. Therefore, we do not consider this term any further for the purposes of calculating the build-up of correlations.

The first two terms in Eq.~(\ref{eq:aux_corr1}) describe the following: A random neighbour of the node is chosen. If that neighbour is in the down-state (this happens with probability $1/2$ in the fully uncorrelated state) the focal node flips with rate $h_-$. The pair of focal node and neighbour change from $+-$ to $--$. The change in $ss'$ is thus $+2$. If instead the neighbour is in the up-state (which also happens with probability $1/2$), the focal node flips with rate $h_-'$. The pair of focal node and neighbour change from $++$ to $-+$, and the change in $ss'$ is $-2$.

 We note that interfaces of the focal node with any other nearest neighbours (other than the designated interaction partner) are equally likely to change from active to inactive or from inactive to active, given that we are starting from the fully uncorrelated state. These interfaces therefore do not contribute to any build-up of correlations.

The total expected rate of change of $\avg{ss'}$ due to events described by Eq.~(\ref{eq:aux_corr1})  is thus $\frac{1}{2}h_-\cdot(+2)+\frac{1}{2}h_-'\cdot(-2)=h_--h_-'$.

Similarly, if the focal node is in state $-1$
and $n$ is the number of neighbours in the up-state, then the rate at which it flips to state $+1$ is
\begin{align}\label{eq:aux_corr2}
    T_{n,k,t}^+ &=  h_{+}\frac{n}{k}  +  h_{+}'\frac{k-n}{k}  +  v_+'.
\end{align}
We can make an analogous argument as above to show that the total expected rate of change of $\avg{ss'}$ due to events described by Eq.~(\ref{eq:aux_corr2}) is thus $h_+-h_+'$.

Finally, we note that 
\be
    \frac{\dd}{\dd t} \big(\avg{s}\avg{s'}\big)=\avg{s}\frac{\dd}{\dd t}\avg{s'} + \avg{s'}\frac{\dd}{\dd t}\avg{s}=0
\ee
in the initial state in which $m=0$, i.e. $\avg{s}=0$.

Putting everything together we therefore find that
\be
    \frac{\dd}{\dd t}\avg{ss'}\propto h_--h_-' + h_+-h_+'.
\ee
Thus, we have shown that the denominator of the expression in Eq.~(\ref{eq supp: negative tau tau defn}) describes  (up to constant pre-factors) the initial rate of the build-up of nearest neighbour correlations if the dynamics is started from a fully disordered state in which each spin is up or down with probability $1/2$, and independently of its neighbours.

\subsubsection{Range of possible values for \texorpdfstring{$\tau$}{}}\label{sec:range_tau}
The definition in Eq.~(\ref{eq supp: negative tau tau defn}) is of the form
\be
    \tau = 2 \frac{a+b}{c-a}, 
\ee
with $a,b,c,d\geq 0$. 

It is clear from this definition that all values $\tau\geq 0$ can be realised through a suitable choice of $a,b,c,d$.

The quantity $\tau$ takes negative values, if $a>c$. In this case we have
\be
-\tau^{-1}=\frac{1}{2}\frac{a-c}{a+b},
\ee
and $0<a-c\leq a \leq a+b$. Thus -$\tau^{-1}\leq 1/2$, and hence $\tau\leq -2$.

We conclude that $\tau$ can only take values in the ranges $\tau\geq 0$ or $\tau\leq -2$.

\subsection{Behaviour in different regimes} \label{appendix: behaviour in tau regimes}
\begin{figure}[t]
    \includegraphics[scale=0.16]{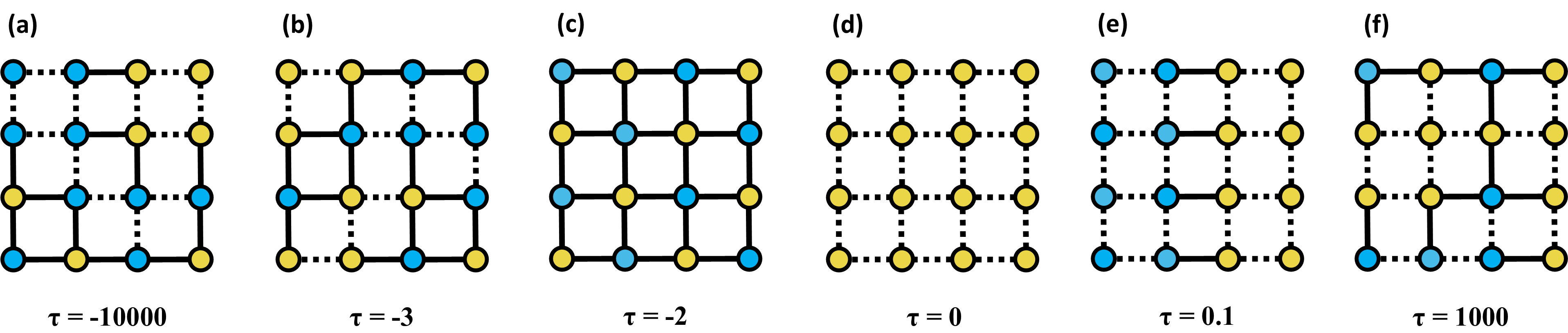}
    \caption{Snapshots of the spin distribution of a single realisation on a grid network after running the model dynamics for a long time, for different values of $\tau$. Nodes in state $+1$ are blue. Nodes in state $-1$ are yellow. Links connecting different state nodes are active (solid). Links connecting same state nodes are inactive (dashed).}
    \label{fig supp: H grid}
\end{figure}

In Fig.~\ref{fig supp: H grid} we show snapshots of the spin configuration in the stationary state, obtained from simulations of the model on a $4 \times 4$ square grid.

\subsubsection{The limit \texorpdfstring{$\tau \to -\infty$}{}}
The limit $\tau\to-\infty$ emerges in two different ways:

(i) One or both of $v_\pm'$ tend to infinity, while $h_{+}'+h_{-}'>h_{+}+h_{-}$; or 

(ii) The denominator of Eq.~(\ref{eq supp: negative tau tau defn}) approaches zero from below. 

Typical configurations of the system will be mostly random, but there will be some small anti-correlation between nodes so the number of active links is slightly greater than random. We find $H(\tau)\to 1^{-}$ in this case. This case is illustrated by the configuration in Fig.~\ref{fig supp: H grid}(a), for which we find $\mathbb{H}=\sigma/[\frac{1}{2}(1-m^{2})]=1.02$.

\subsubsection{Finite negative values of \texorpdfstring{$\tau<-2$}{}}
The copying process is mostly unfaithful, but there can also be faithful copying and/or spontaneous spin flips. There will be some anti-correlation between neighbouring spins, and thus, for a given trait frequency $m$, one expects more active interfaces than a configuration in which these spins are placed randomly on the network. Hence, we expect values of $H(\tau)>1$. 

A case in which $\tau$ is negative and finite is illustrated by the configuration in Fig.~\ref{fig supp: H grid}(b), for which $\mathbb{H}=\sigma/[\frac{1}{2}(1-m^{2})]=1.27$.

\subsubsection{The limit \texorpdfstring{$\tau \to -2^-$}{}}
From the definition of $\tau$ in Eq.~(\ref{eq supp: negative tau tau defn}) we have
\be
\tau = -2 \frac{h_+'+h_-' + v_+' + v_-'}{(h_+'+h_-') - (h_++h_-)}.
\ee
Hence $\tau=-2$ when either $h_+'+h_-'$ diverges and the $v_\pm', h_\pm$ remain finite, or if the $v'_{\pm}$ and $h_\pm$ all vanish. In either case, the dynamics is dominated by unfaithful copying from neighbours, i.e. the model is type of `contrarian voter model', in which a spin copies the state opposite to that of a neighbour. This creates strong anti-correlations between the states of neighbouring spins, and $H(\tau)$ will take its maximum value which is specific to the network. 

This scenario is illustrated in Fig.~\ref{fig supp: H grid}(c), where we show a perfect `checkerboard' pattern seen in simulations on a small square grid. For this pattern we have $\mathbb{H}=\sigma/[\frac{1}{2}(1-m^{2})]=2$.

\subsubsection{The limit \texorpdfstring{$\tau\to 0^{+}$}{}}
The limit $\tau\to 0^{+}$ can occur in one of two ways:

(i) The numerator in Eq.~(\ref{eq supp: negative tau tau defn}) goes to zero (from above) and, at the same time, $h_{+}+h_{-} > h_{+}'+h_{-}'$, so that denominator is positive; or 

(ii) One has $h_{+}+h_{-}\to\infty$. 

In the former case we have $h_\pm'\to 0$ and $v_\pm'\to 0$. There is no transmission noise. In the latter case faithful copying from neighbouring spins dominates the dynamics. 

In either case the number of active links is low and we expect configurations in which either all spins are up or all spins are down, as illustrated in Fig.~\ref{fig supp: H grid}(d). Due to the convergence to a fully ordered state, the ratio $\mathbb{H}=\sigma/[\frac{1}{2}(1-m^{2})]$ cannot be defined for single realisations. However all realisations have $\sigma=0$ eventually, and we measure $H(\tau)=\avg{\sigma}_{\rm st}/[\frac{1}{2}(1-\avg{m}_{\rm st}^{2})]=0$.

\subsubsection{Finite \texorpdfstring{$\tau>0$}{}}
For $\tau > 0$ and finite, there are some unfaithful events occurring, and the rate of faithful horizontal events is greater than the total rate of unfaithful horizontal events, i.e. $h_{+}+h_{-} > h_{+}'+h_{-}'$. This leads to steady-state situations where $0<H(\tau)<1$. The case is illustrated in Fig.~\ref{fig supp: H grid}(e), where we measure $\mathbb{H}=\sigma/[\frac{1}{2}(1-m)^{2}]=0.33$.

\subsubsection{The limit \texorpdfstring{$\tau\to\infty$}{}}
This limit emerges in two different ways: 

(i) The denominator in Eq.~(\ref{eq supp: negative tau tau defn}) remains positive, and either $v_+'$ or $v_-'$ (or both) tend to infinity; or

(ii) The $v_\pm'$ remain finite, but the denominator in Eq.~(\ref{eq supp: negative tau tau defn}) diverges, i.e., $h_++h_-=h_+'+h_-'$. 

In the former case, the system is dominated by random spin flips. In the second scenario, correct and incorrect copying from neighbouring nodes occur at the same rates. This means that the copying spin takes a state that is effectively unrelated to that of the state it is trying to copy. 

In neither of these two situations correlations between neighbouring states can build up, and so the resulting configurations are effectively random. In general $H(\tau)\to 1^{-}$. This case is illustrated in Fig.~\ref{fig supp: H grid}(f), where we find $\mathbb{H}=\sigma/[\frac{1}{2}(1-m^{2})]=0.98$.

\subsubsection{Illustration and test against simulations}
\begin{figure}[htbp]
    \includegraphics[scale = 0.8]
    {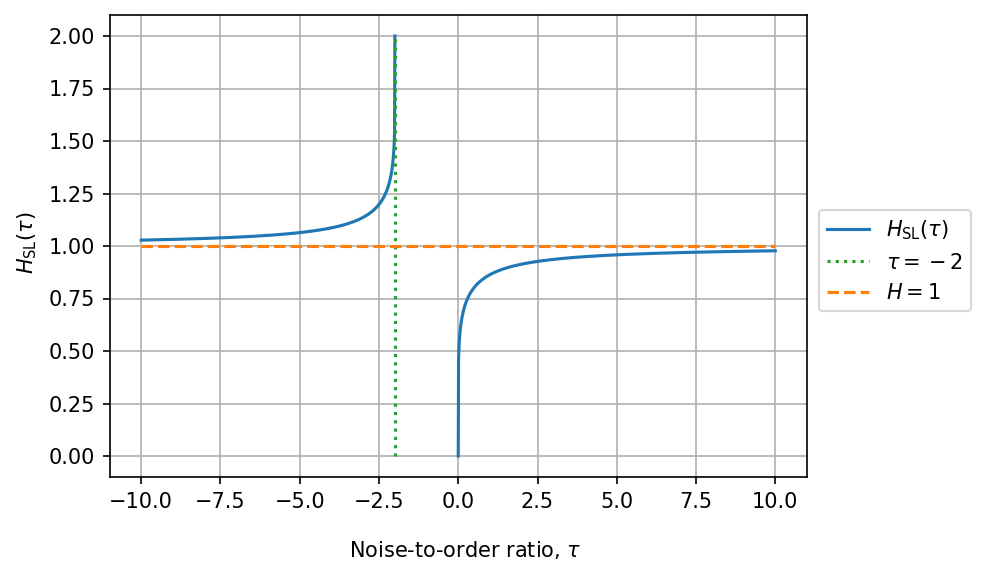}
    \caption{Plot of $H_{\rm SL}(\tau)$ from Eq.~(\ref{eq supp: H 2d square lattice}) over a range of positive and negative values of $\tau$. The horizontal orange dashed line shows $H=1$, and the vertical green dotted line is at $\tau=-2$ for reference. }
    \label{fig supp: negative tau two panel analytic}
\end{figure}

Fig.~\ref{fig supp: negative tau two panel analytic} shows $H_{\rm SL}(\tau)$, from Eq.~(\ref{eq supp: H 2d square lattice}), as an example, for both positive and negative values of $\tau$. 

\begin{figure}[htbp]
    \includegraphics[width=0.9\textwidth]{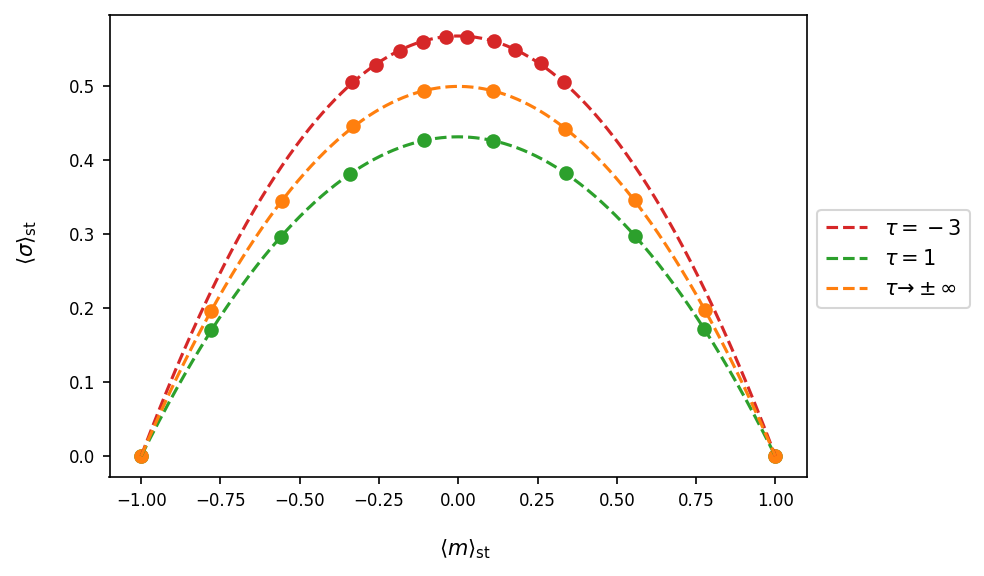}
    \caption{Parabolas $\avg{\sigma}_{\rm st}=\frac{1}{2}H_{\rm SL}(\tau)(1-\avg{m}_{\rm st}^{2})$ for the model on an infinite square lattice, and for different values of $\tau$. Dashed lines are from the theory [Eq.~(\ref{eq supp: H 2d square lattice})], markers are the results from averaging the steady-state values of $10$ independent simulations on a lattice with $N=10,000$ nodes (periodic boundary conditions). For $\tau=-3$ the algorithm we use to generate model parameters [see Sec.~\ref{appendix: determining model parameters tau and m}] limits the values of the stationary magnetisation that we can probe. More specifically, for parameters generated by this algorithm, $\avg{m}_{\rm st}$ can only take values in the range given in Eq.~(\ref{eq supp: negative tau bounds}). For the other cases $\avg{m}_{\rm st}$ can take any value in the interval $[-1,1]$. The dashed line labelled $\tau\to\pm \infty$ corresponds to $H(\tau)=1$, the orange markers are from simulations with $\tau=1000$.}
    \label{fig supp: negative tau parabolas}
\end{figure}

In Fig.~\ref{fig supp: negative tau parabolas} we show simulation results to illustrate the effects of negative values of $\tau$, and to demonstrate the validity of our analytical description. Specifically, simulations are on a square lattice and we show plots of $\avg{\sigma}_{\rm st}$ as a function of $\avg{m}_{\rm st}$. As before these curves are obtained by varying the detailed model parameters, while keeping $\tau$ fixed [see Sec.~\ref{appendix: determining model parameters tau and m}]. We also show the analytical predictions for the square lattice, i.e. $\avg{\sigma}_{\rm st}=\frac{1}{2}H_{\rm SL}(\tau)(1-\avg{m}_{\rm st}^{2})$ with $H_{\rm SL}(\tau)$ from Eq.~(\ref{eq supp: H 2d square lattice}).

For $\tau=1$, the model dynamics contains some transmission noise, but copying from neighbours is more likely to be faithful than unfaithful. We have $H(\tau) < 1$. 

For very large positive $\tau$, noise dominates, and the resulting configurations at a given $m$ are effectively random, so $H(\tau)=1$. 

For $\tau=-3$, copying events are mostly unfaithful (that is, spins take the state opposite to the one they are `copying'). There is also spontaneous flipping, and some faithful copying. Overall, some degree of anti-correlation between neighbours ensues, and so $H(\tau)>1$. 

The stationary average density of active interfaces at the apexes of the three parabolas (located at $\avg{m}_{\rm st}=0$) are given by $\avg{\sigma}_{\rm st} = \frac{H(\tau)}{2}$, and are smaller than $0.5$ for $\tau=1$, equal to $0.5$ for $\tau\to\infty$, and larger than $0.5$ for $\tau=-3$.

\FloatBarrier
\newpage
\section{Analysis of linguistic data} \label{appendix: linguistic analysis}
\subsection{Dataset}
In this section we provide further information on the analysis of data from WALS (The World Atlas of Language Structures) \cite{wals} in the main paper. WALS contains information on $2,662 $ human languages, most importantly the co-ordinates of where that language is most densely spoken. We will refer to languages by their WALS-ID, which is a combination of numbers and letters such as `11A', `143F' etc. 

Language possess a number of language traits (known as `typological features of language'). WALS details 192 such traits. Not all traits are recorded for all languages in WALS. We focus on a curated set of $35$ language traits which can be quantified in a sufficient set of languages, and which are `binarizable', that is, if the trait can be quantified in the language it can meaningfully be classified as present or absent. The set of traits and languages is that of \cite{kauhanen2021geospatial}, where further details can be found. Ultimately, we consider $2,248$ languages from WALS, these are those for which at least one trait is quantifiable in binary form. This means that not all nodes (languages) have a well defined binary state for all $35$ traits.

Binary traits can all be naturally mapped to our theoretical set-up. Nodes represent languages, the spin states indicate whether that language posses or lacks a specific language trait, links between nodes indicate interaction.

The relevant data to perform this analysis including details about traits, distances between languages, and information on which languages contain which features can be found at the following GitHub repository \cite{github}. 

\subsection{Empirical noise-to-order ratio} \label{appendix: measuring tau linguistics}
To determine the density of active interfaces for each of the $35$ features from the WALS data we need to decide how the languages are connected. We do not know the true interaction network. We therefore make the following choice: two languages \textit{A} and \textit{B} are connected if \textit{B} is among the $10$ geographically closest neighbours of \textit{A}, or if \textit{A} is among the $10$ geographically closest neighbours of \textit{B}. This creates an undirected network. 

With this method, languages can end up having more than $10$ neighbours in the network. For example, start by considering $\textit{A}$. We want to determine which other languages it connects to. We find the $10$ geographically nearest languages of $\textit{A}$ and connect it to those languages. Assume that $\textit{B}$ is one of these languages. We then consider $\textit{B}$. It already has one connection to $\textit{A}$. We then find its $10$ geographically nearest languages. It is possible that $\textit{A}$ is not within this set. We connect $\textit{B}$ to these $10$ languages. $\textit{B}$ now has $11$ neighbours in the network.

The choice of $10$ nearest neighbours is somewhat arbitrary, however, the results of \cite{kauhanen2021geospatial} for example have been shown to be robust against variation in neighbourhood size. 

\begin{figure}[htbp]
    \centering
    \includegraphics{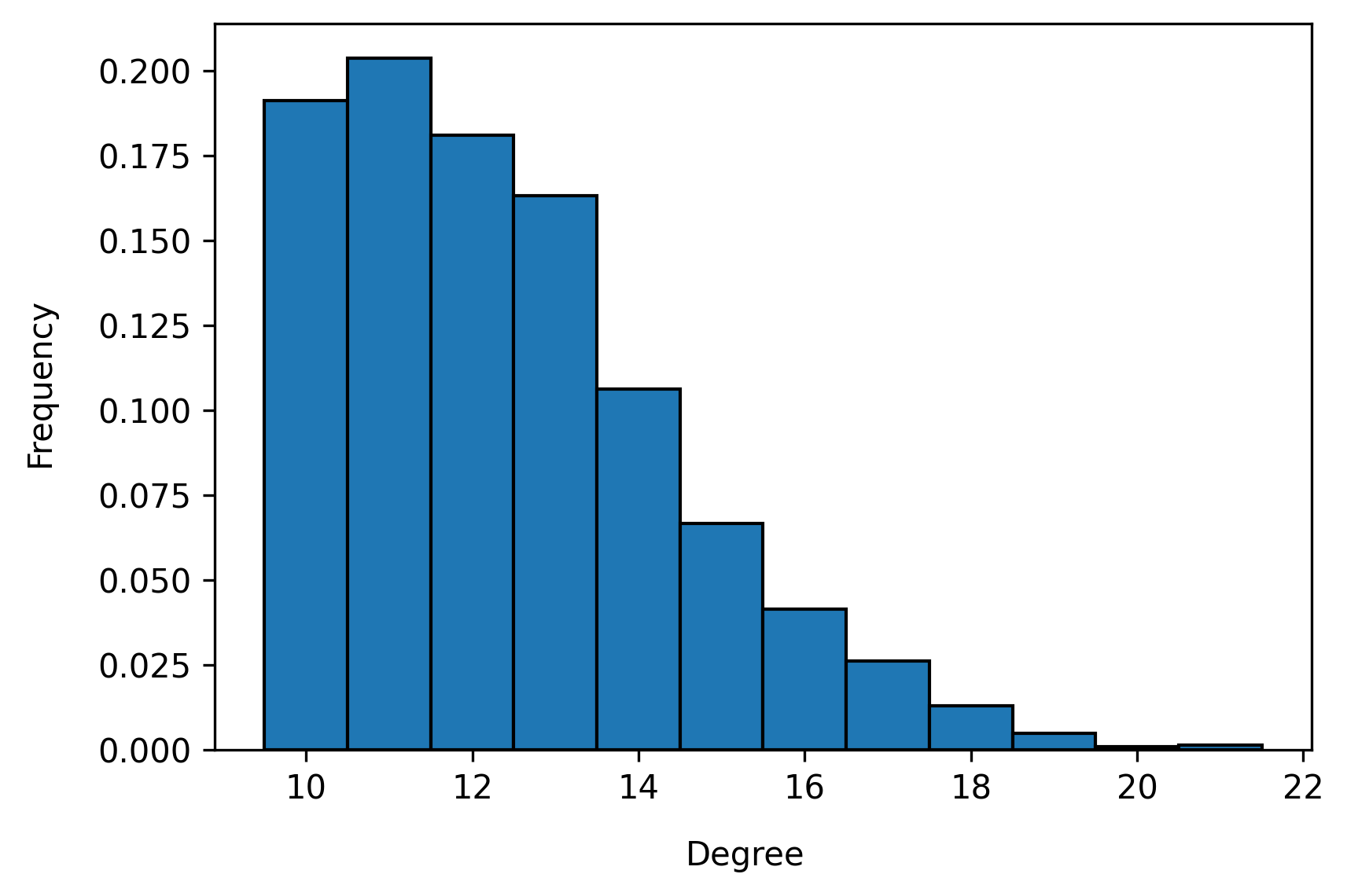}
    \caption{Degree distribution of our approximation to the real-world network of language interactions (see text).}
    \label{fig supp: real network degree dist}
    \end{figure}

The degree distribution for the above network is shown in Fig.~\ref{fig supp: real network degree dist}. The network is not regular, neither is it uncorrelated as it is likely that two neighbours of a given node are connected to one another. However the degree distribution is fairly tight, so we can still apply the various analytical results to good effect. 

To determine an empirical noise-to-order ratio for a particular binarisable trait, we assign states $\pm1$ to those nodes (languages), for which the binary state of the trait is defined. The state $\pm 1$ indicates whether the language represented by the node posses or lacks the trait. If the trait is not quantifiable on a given node, then no spin state is assigned to the node. Taking into account only nodes with states $\pm 1$, we then determine the feature frequency $\rho$, and, using the putative interaction network, the density of active links, $\sigma$.  From this we obtain $\mathbb{H}=\sigma/[2\rho(1-\rho)]$ as a quantifier of the spatial scattering. Finally, the different analytical functions, $H_{\rm AA}(\tau)$ [Eq.~(\ref{eq supp: H annealed})], $H_{\rm SL}(\tau)$ [Eq.~(\ref{eq supp: H 2d square lattice})], $H_{\rm HG}(\tau)$ [Eq.~(\ref{eq supp: H_HG})] and $H_{\rm SPA}(\tau)$ [Eq.~(\ref{eq supp: H SPA2})], can be used to infer values for the noise-to-order ratio, $\tau_{X}$. The $X$ subscript stands for one of AA, SL, HG, or SPA depending on the specific $H_X(\tau)$ function that was used. 

Using this procedure it is sometimes possible to obtain $\mathbb{H}=\sigma/[2\rho(1-\rho)]>1$. As discussed in Sec.~\ref{appendix: H and tau} this would lead to negative values of $\tau$ values. Such occurrences are rare in our dataset (less than $1\%$), so we discard these cases.

\subsection{Comparison against longitudinal stability estimate}
To validate our results we compare to a phylogenetic method which does not take into account any spatial data. This was done in \cite{dediu2011} using a principal component analysis, and in particular the first component PC1. Higher PC1 values corresponding to greater instability. PC1 values were only measured for a subset of the 35 traits that we consider.

\begin{figure}[htb]
    \centering
    \includegraphics[scale = 0.9]{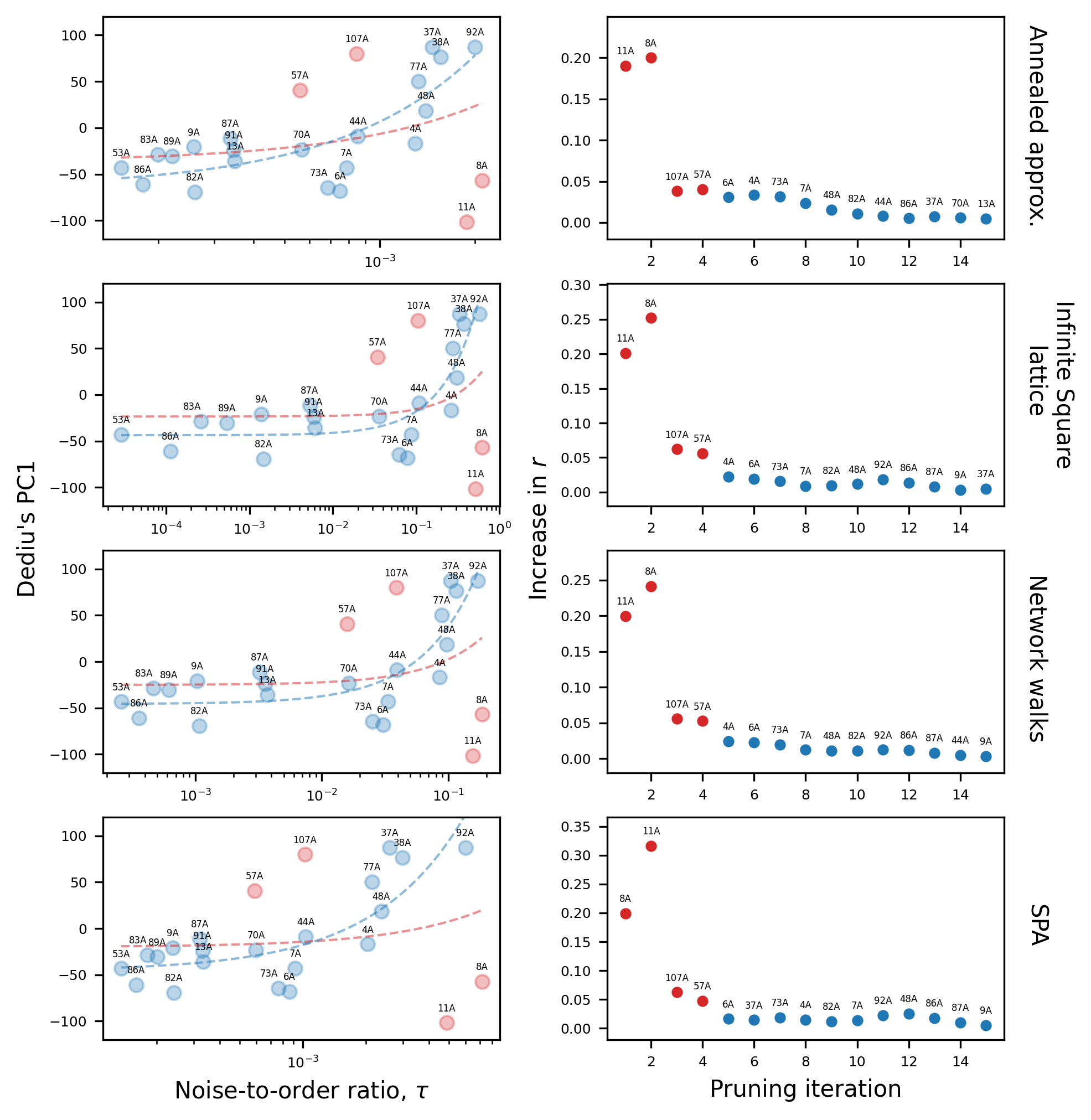}
    \caption{Panels on the left show regression plots of Dediu's PC1 \cite{dediu2011} against the noise-to-order ratio, $\tau$. Red points are deemed outliers (see text). The red and blue dashed lines are the best linear fit before and after outliers are removed (we draw attention to the logarithmic scale on the horizontal axis, making the fits look curved). The different rows correspond to different choices of the analytical form of $H(\tau)$, as indicated on the right-hand side. From top to bottom we have used the annealed approximation [Sec.~\ref{appendix: annealed approximation}], the infinite 2D square lattice [Sec.~\ref{appendix: square lattice}], the  network walk method [Sec.~\ref{appendix: random walks}] and the stochastic pair approximation [Sec.~\ref{appendix: SPA}] 
    After outlier removal the $r$ values are $0.813$, $0.857$, $0.848$, $0.821$ from top to bottom, all with $p$-value$<0.01$. The right column illustrates the outlier removal process. At each pruning iteration the point that gives the largest increase in $r$ is removed. For clarity, we only show the first $15$ iterations.}
    \label{fig supp: dediu all theory}
    \end{figure}

In Fig.~\ref{fig supp: dediu all theory} (left) we show regression plots of PC1 against $\tau_{X}$. The different rows correspond to different $X$ and thus use different analytical forms $H_{X}(\tau)$, as described above and indicated in the figure.  

A similar but restricted analysis was carried out in \cite{kauhanen2021geospatial}. This was solely based on $H_{\rm SL}(\tau)$ from the theory for infinite square lattices. That earlier analysis identified a small set of outlier traits for which the estimates of $\tau$ and Dediu's PC1 gave conflicting indications regarding relative stability. Some of these outliers could be justified based on known properties of some of the WALS features, see the discussion in \cite{kauhanen2021geospatial} for details.

We have carried out a similar identification of outliers to test robustness against the use of different analytical forms $H_{X}(\tau)$. Outliers are determined by recursively pruning the dataset and removing the trait that would provide the largest increase in Pearson correlation coefficient $r$ (measuring linear correlation \cite{boslaugh2012statistics}). Results are shown in the right-hand panels of Fig.~\ref{fig supp: dediu all theory}. Independent of the analytical form used for the function $H_{X}(\tau)$ we find the same set of outliers, namely WALS features 8A, 11A, 107A and 57A. This is the same set as found in \cite{kauhanen2021geospatial}, confirming the robustness against variation of the assumed network structure [reflected by the choice of analytical form of $H_{X}(\tau)$].

After outlier removal we find strong correlation between the inferred values of $\tau_{X}$ and Dediu's PC1, with an $r$ consistently above $0.8$, and $p$-value $<0.01$, for all theories $X$. 

\FloatBarrier
\subsection{Bootstrapping and rankograms} \label{appendix: bootstrapping}
\subsubsection{Construction of rankograms}
The method in Sec.~\ref{appendix: measuring tau linguistics} can be used to rank the $35$ traits in terms of noise-to-order ratio, which in turn can be regarded as a proxy for stability. 

Our analysis is based on the WALS dataset, involving $N=2,248$ languages, each with at least one binarisable feature among our selected set of $35$ traits. In reality, there are other languages for which we do not have the binarisable data. This can be due to lack of observation, or WALS simply is missing the data. We can therefore think of the WALS dataset as an effective random sub-sample of all human languages. To quantify the uncertainty associated with this, we use bootstrap re-sampling  \cite{bootstrap}.

In our context, bootstrapping involves sampling $N=2,248$ languages with replacement from the original WALS dataset. This gives rise to an (assumed) interaction network as described in Sec.~\ref{appendix: measuring tau linguistics}. This network is specific to the bootstrap sample. We note that is possible to select the same language multiple times. This leads to multiple identical languages with geographical distance zero, thus they are essentially guaranteed to be connected. This represents potentially very close neighbouring languages, not included in WALS, which have the same trait values.

For each sample, we can perform the procedure in Sec.~\ref{appendix: measuring tau linguistics} to calculate a noise-to-order ratio $\tau_{X}$ for each trait (discarding nodes in the network for which the status of a given trait cannot be quantified in binary form). Repeating this many times, we generate distributions of $\tau_{X}$ for each trait.

\begin{figure}[htbp]
    \centering
    \includegraphics{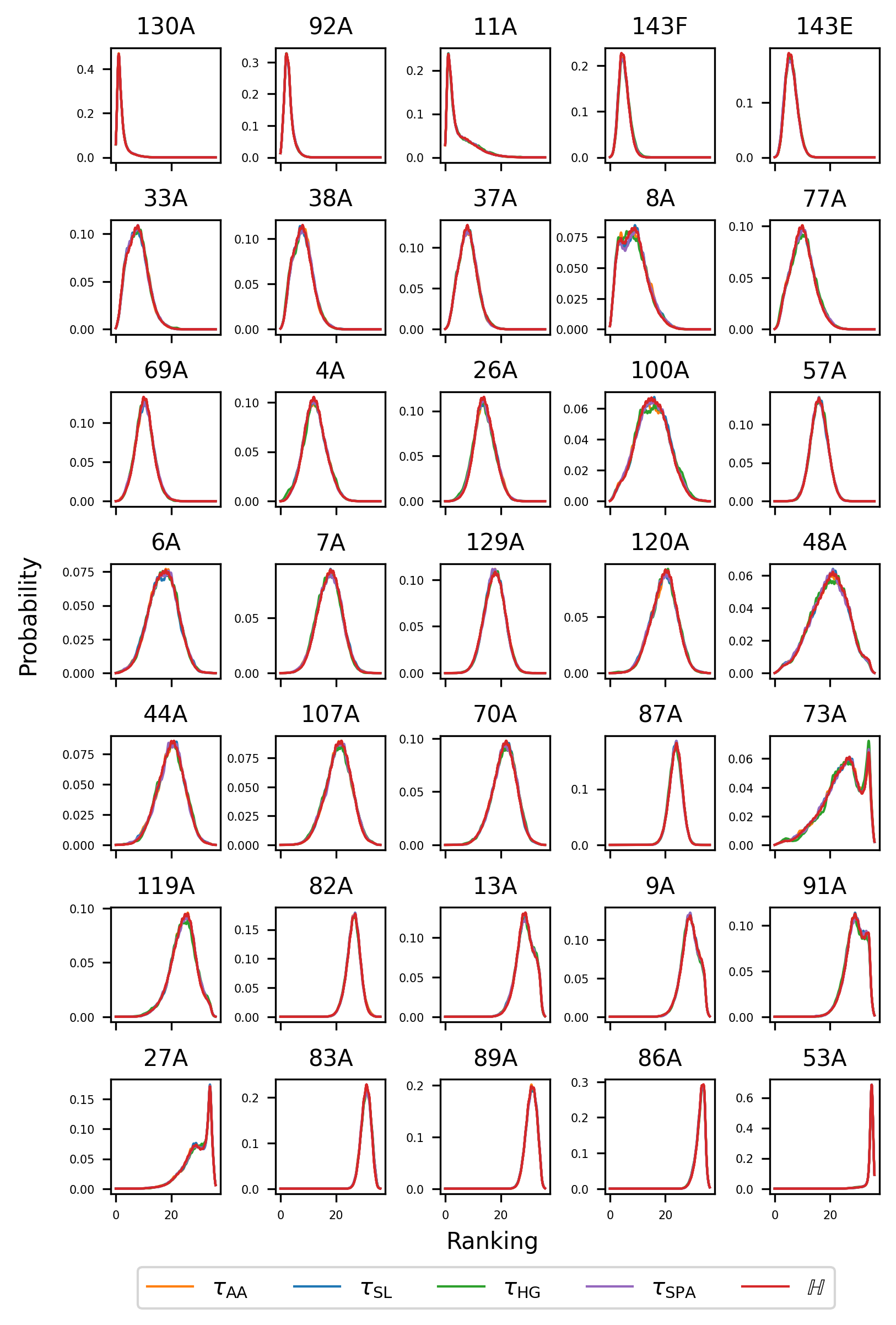}
    \caption{Rankograms for all 35 language traits for 10,000 bootstrap samples. We only show Gaussian kernel density estimates of the distributions for clarity. Each rankogram has five distributions. Four of which are based on inferring values of $\tau_{X}$ from the theories $H_{\rm AA}(\tau)$ [Eq.~(\ref{eq supp: H annealed})], $H_{\rm SL}(\tau)$ [Eq.~(\ref{eq supp: H 2d square lattice})], $H_{\rm HG}(\tau)$ [Eq.~(\ref{eq supp: H_HG})] and $H_{\rm SPA}(\tau)$ [Eq.~(\ref{eq supp: H SPA2})]. The fifth is from ranking raw $\mathbb{H}$ values calculated using Eq.~(\ref{eq main: H}). The height at a given rank is the probability a trait is ranked in that position amongst all traits.}
    \label{fig supp: all rankogram}
    \end{figure}

To produce rankograms for a chosen $X$ we can then draw a value of $\tau_{X}$ for each trait from the respective distributions. The traits can then be ranked based off of these values. This is then repeated many times and thus a distribution of ranks is constructed for each trait. These are the rankograms shown in Figs.~\ref{fig main: rankogram} in the main paper, and in \ref{fig supp: all rankogram} in this supplement. These are histograms where the $i^{\rm th}$ bar indicates the probability that the trait represented by that rankogram ranks in the $i^{\rm th}$ position amongst all traits. 

We observe that the traits on the extreme ends, such as 130A or 53A, have very tight rankograms. This means that we can be confident that those traits will rank in those positions amongst all features. We are less confident about the traits with broad distributions, such as 100A or 48A. 

\FloatBarrier
\subsubsection{Distributions of \texorpdfstring{$\tau_{X}$}{}}

From the rankograms in Fig.~\ref{fig supp: all rankogram} we see that the rankings produced from $\mathbb{H}$ and those produced from $\tau_{X}$ using each of the theories $X$ are very similar. This is a non-trivial result and is not a consequence of the fact that all the $H_{X}(\tau)$ functions are monotonic increasing functions of $\tau$ as explained below.

In a single bootstrap sample, we have a set of languages and construct one interaction network. We then measure 35 values of $\mathbb{H}$, one for each trait, using the procedure in Sec.~\ref{appendix: measuring tau linguistics}. These $\mathbb{H}$ values will have a certain ranking. With the single interaction network we construct the functions $H_{X}(\tau)$, one for each theory $X$. For a chosen $X$ the 35 $\mathbb{H}$ values can be mapped to 35 $\tau_{X}$ values. Since all $H_{X}(\tau)$ are monotonic increasing functions of $\tau$, ordering the traits based on $\tau_{X}$ will be the same as ordering based on $\mathbb{H}$, for all theories $X$.

However, the procedure for constructing rankograms is different in that we do not perform any rankings until distributions of $\tau_X$ have been constructed for each trait. To construct distributions of $\tau_{X}$ for a trait we repeat the procedure in Sec.~\ref{appendix: measuring tau linguistics} many times. Each bootstrap sample consists of a different set of languages, and thus a different interaction network. Focusing on two traits $A$ and $B$, we measure the corresponding values $\mathbb{H}^A=\sigma^A/[2\rho^A (1-\rho^A)]$ and $\mathbb{H}^B=\sigma^B/[2\rho^B (1-\rho^B)]$ in each sample. This gives distributions for $\mathbb{H}^{A}$ and $\mathbb{H}^{B}$. For each bootstrap sample we also map the empirical values of $\mathbb{H}^A$ and $\mathbb{H}^B$ to values $\tau^A_{X}$ and $\tau^B_{X}$, using one of the analytical functions $H_{X}(\tau)$. This gives us distributions of $\tau^{A}_{X}$ and $\tau^{B}_{X}$. 

\begin{figure}[htbp]
    \centering
    \includegraphics[scale = 0.17]{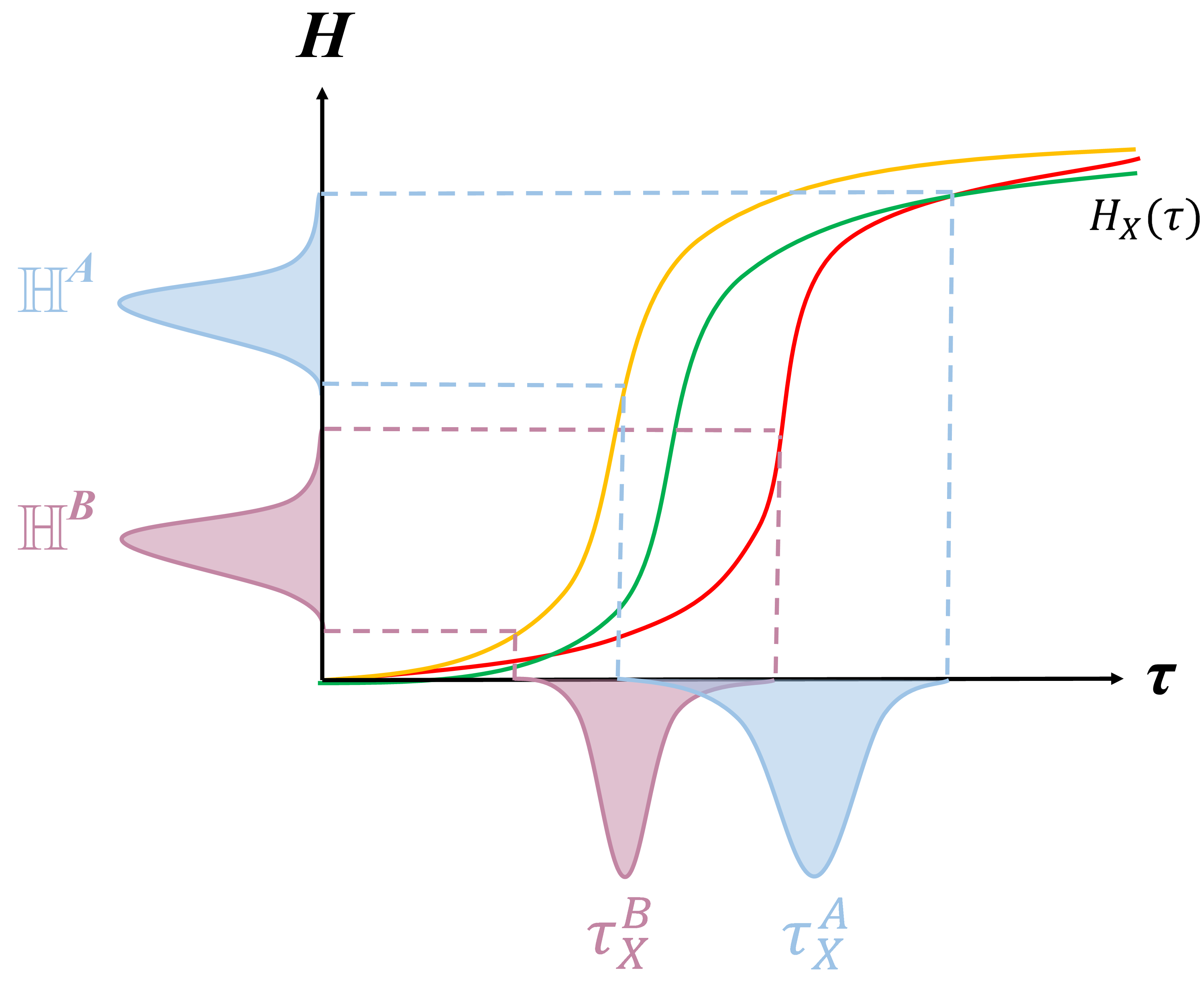}
    \caption{Pictorial demonstration of how non-overlapping $\mathbb{H}$ distributions from two traits A and B can lead to $\tau_{X}$ distributions which do overlap through the bootstrapping process (see text). The yellow, green and red curves are examples of $H_{X}(\tau)$ from different bootstrap samples.}
    \label{fig supp: H and tau dists}
    \end{figure}

Since the interaction network is different in each bootstrap sample, the function $H_{X}(\tau)$ will also change in each sample (for a fixed choice of $X$). This is because the functional form of $H_X(\tau)$ makes use for example of the mean degree or other characteristics of the network. This means, that even if the distributions of $\mathbb{H}^A$ and $\mathbb{H}^B$ do not overlap, for a given $X$ it is possible that the distributions of $\tau^A_{X}$ and $\tau^B_{X}$ do overlap. Fig.~\ref{fig supp: H and tau dists} gives a visual representation of this idea.

\FloatBarrier
\subsection{SUCRA scores and comparison with Dediu} \label{appendix: sucra}
To quantify the rankograms for each of the 35 linguistic traits shown in Fig.~\ref{fig supp: all rankogram} further we use a statistic called SUCRA (surface under the cumulative ranking line) first introduced in \cite{salanti2011graphical} to rank medical treatments. This is a transform of the expected rank,
\be
    \text{SUCRA}_{a} = \frac{N-\mathbb{E}(r)_{a}}{N-1},
\ee
where $\mathbb{E}(r)_{a}$ is the expected rank of item $a$ and where we are ranking $N$ items total. SUCRA values go from 0 to 1 but are typically written as a percentage. If item $a$ ranks first with probability 1 then it will have $\text{SUCRA}_{a}=1$ (or 100\%). Similarly if item $a$ ranks last with probability 1 then it will have $\text{SUCRA}_{a}=0$ (or 0\%). 

In Tab.~\ref{tab supp: SUCRA} we list the SUCRA scores for all $35$ traits for the different choices of the function $H_{X}(\tau)$ as well as the SUCRA derived from a ranking based on $\mathbb{H}=\sigma/[2\rho(1-\rho)]$ itself. We also list Dediu's PC1 values where possible. We can calculate the Spearman rank correlation coefficient, $r_{S}$, between the traits sorted by SUCRA value and PC1. All give $r_{S} > 0.8$ with $p$-value$<0.01$ after removing the outliers.

\begin{table}[htbp]
    \centering
    \begin{tabular}{|p{2cm}|p{2cm}|p{2cm}|p{2cm}|p{2cm}|p{2cm}|p{2cm}|}
        \hline
        \multicolumn{1}{|p{2cm}|}{WALS ID} & \multicolumn{5}{c|}{SUCRA values (\%)} & \multicolumn{1}{|p{2cm}|}{Dediu's PC1} \\
        \cline{2-6}
         & $\mathbb{H}$ & $H_{\rm AA}(\tau)$ & $H_{\rm SL}(\tau)$ & $H_{\rm HG}(\tau)$ & $H_{\rm SPA}(\tau)$ & \\
        \hline
        130A & 96.97 & 96.97 & 97.02 & 96.90 & 97.03 & 97.03 \\
        92A & 94.52 & 94.36 & 94.33 & 94.60 & 94.44 & 87.10 \\
        11A & 88.42 & 88.07 & 88.43 & 87.67 & 88.12 & -101.68 \\
        143F & 88.02 & 87.99 & 87.87 & 87.38 & 88.01 & \\
        143E & 85.30 & 85.61 & 85.42 & 84.97 & 85.54 & \\
        33A & 79.73 & 79.70 & 79.66 & 79.57 & 79.66 & \\
        38A & 79.49 & 79.12 & 79.23 & 79.56 & 79.34 & 76.40 \\
        37A & 78.77 & 78.77 & 78.64 & 78.77 & 78.77 & 87.38 \\
        8A & 77.89 & 77.25 & 77.42 & 77.93 & 77.04 & -57.03 \\
        77A & 73.38 & 73.26 & 73.10 & 73.01 & 73.01 & 50.18 \\
        69A & 72.26 & 72.28 & 72.26 & 72.26 & 72.26 & \\
        4A & 66.43 & 66.44 & 66.52 & 66.43 & 66.40 & -16.94 \\
        26A & 61.30 & 61.49 & 61.35 & 62.21 & 61.60 & \\
        100A & 58.05 & 57.77 & 57.63 & 57.21 & 57.47 & \\
        57A & 55.69 & 55.90 & 55.76 & 55.51 & 55.83 & 40.60 \\
        6A & 52.71 & 52.75 & 52.96 & 52.70 & 52.77 & -68.03 \\
        7A & 50.79 & 51.04 & 50.97 & 51.34 & 51.14 & -42.87 \\
        129A & 50.69 & 50.65 & 50.68 & 50.82 & 50.62 & \\
        120A & 45.45 & 45.55 & 45.79 & 45.46 & 45.80 & \\
        48A & 44.79 & 45.16 & 44.99 & 45.16 & 45.32 & 18.44 \\
        44A & 44.47 & 44.48 & 44.68 & 44.08 & 44.27 & -8.95 \\
        107A & 41.41 & 41.64 & 41.23 & 41.44 & 41.50 & 79.84 \\
        70A & 40.29 & 40.27 & 40.03 & 40.48 & 40.17 & -23.10 \\
        87A & 33.27 & 33.33 & 33.30 & 33.65 & 33.23 & -11.23 \\
        73A & 31.48 & 31.71 & 31.86 & 31.67 & 31.86 & -64.56 \\
        119A & 31.42 & 31.37 & 31.35 & 31.98 & 31.21 & \\
        82A & 25.51 & 25.54 & 25.47 & 25.85 & 25.51 & -69.15 \\
        13A & 17.81 & 18.01 & 18.04 & 17.98 & 18.04 & -35.89 \\
        9A & 17.76 & 17.80 & 17.80 & 18.05 & 18.05 & -20.70 \\
        91A & 17.15 & 17.28 & 17.28 & 17.69 & 17.28 & -23.96 \\
        27A & 16.65 & 16.94 & 16.85 & 16.69 & 16.85 & \\
        83A & 12.20 & 12.27 & 12.15 & 12.07 & 12.20 & -28.65 \\
        89A & 11.62 & 11.59 & 11.65 & 11.81 & 11.54 & -30.62 \\
        86A & 6.82 & 6.99 & 6.87 & 7.07 & 6.99 & -60.98 \\
        53A & 1.23 & 1.19 & 1.10 & 1.29 & 1.13 & -43.21 \\
        \hline
    \end{tabular}
    \caption{Table showing the SUCRA values for each of the 35 rankograms in Fig.~\ref{fig supp: all rankogram} for all theories. Dediu's PC1 values are also shown where possible. The Spearman rank correlation coefficients, $r_{S}$, between each column and Dediu's PC1 is 0.824, except $H_{\rm SPA}(\tau)$ which is 0.832, all with $p$-value$<0.01$.}
    \label{tab supp: SUCRA}
\end{table}

\begin{figure}[htbp]
    \centering
    \includegraphics[scale=0.8]{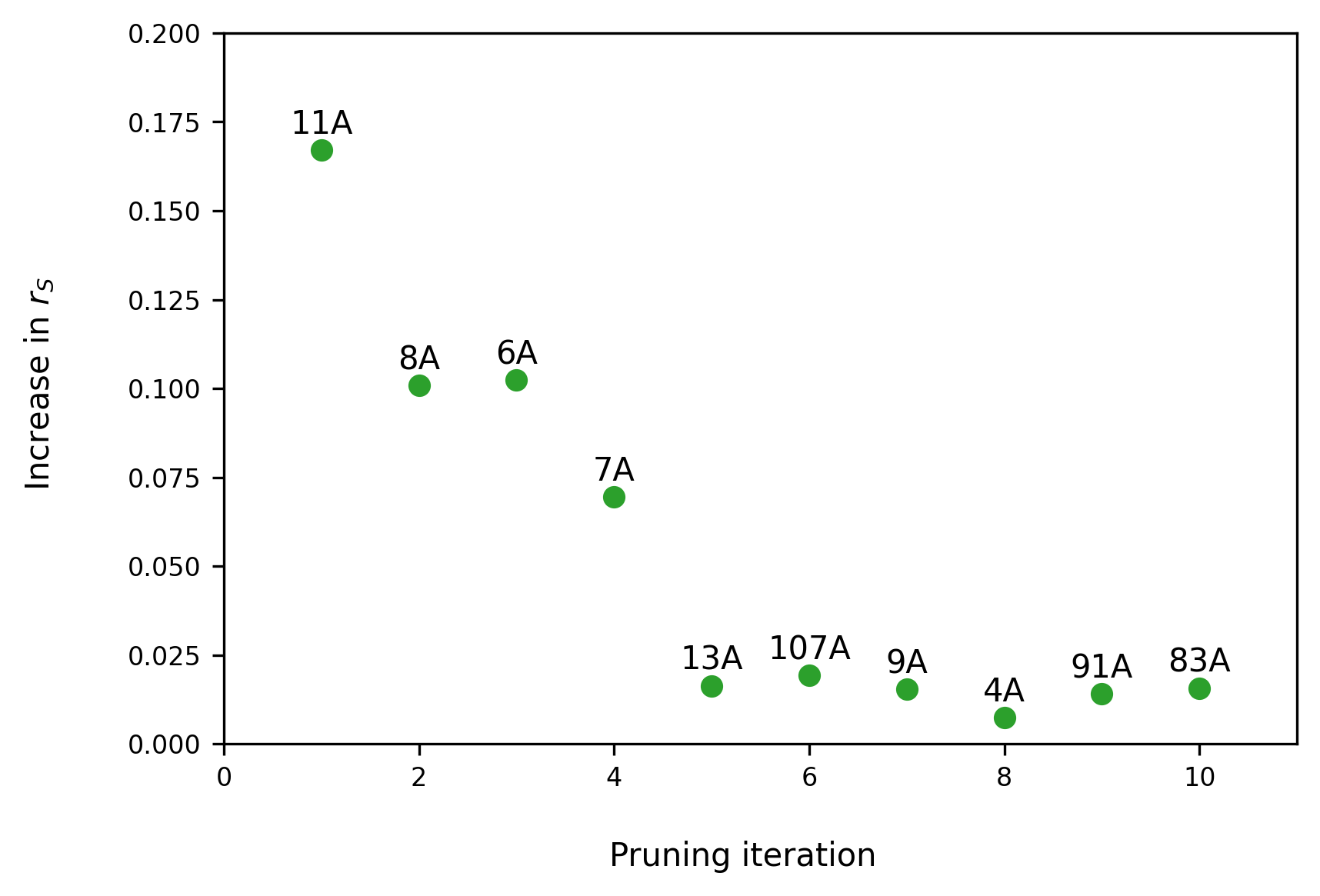}
    \caption{Visual representation of outlier removal when determining the Spearman rank correlation coefficient, $r_{S}$, between the linguistic traits ranked by SUCRA value and Dediu's PC1 [see Tab.~\ref{tab supp: SUCRA}]. We recursively prune the data and remove the feature which would give the largest improvement in $r_{S}$.}
    \label{fig supp: outliers}
    \end{figure}

To determine the outliers we recursively prune the sample. At each step we remove the trait that would give the largest increase in $r_{s}$. We show this in Fig.~\ref{fig supp: outliers}. We find that the four traits 11A, 8A, 6A and 7A give significant improvement when removed. This is a different ordering than in Fig.~\ref{fig supp: dediu all theory}, here it seems that 6A and 7A are outliers whereas 107A and 57A are not. This is likely simply due to using different pruning methods. It is encouraging however that 11A and 8A are still outliers which is the same as before. In \cite{kauhanen2021geospatial} there are strong linguistics arguments as to why Dediu incorrectly classifies 11A. 

\FloatBarrier
\newpage
\section{Further mathematical background} \label{appendix: proofs}

\subsection{Exponential generating function for modified Bessel functions of the first kind} \label{appendix: bessel generating function}

The exponential generating function for the modified Bessel function of the first kind is \cite[49:6:1]{spanier1987atlas}
\begin{equation}
    I_{0}(2x)=\sum_{k=0}^{\infty}\frac{x^{2k}}{(k!)^{2}} = \frac{x^{0}}{(0!)^{2}}+\frac{x^{2}}{(1!)^{2}}+\frac{x^{4}}{(2!)^{2}}+....
\end{equation}
Squaring this equation provides us with the exponential generating function for $I_{0}(2x)^{2}$,
\begin{align}
    I_{0}(2x)^{2}&=\left(\frac{1}{(0!)^{2}}\frac{1}{(0!)^{2}}\right)x^{0}+\left(2\frac{1}{(0!)^{2}}\frac{1}{(1!)^{2}}\right)x^{2} \nonumber \\
    &\qquad +\left(2\frac{1}{(0!)^{2}}\frac{1}{(2!)^{2}}+\frac{1}{(1!)^{2}}\frac{1}{(1!)^{2}}\right)x^{4}+... \nonumber \\
    &= \sum_{k=0}^{\infty}\sum_{a_{1}+a_{2}=k}\frac{x^{2k}}{(a_{1}!)^{2}(a_{2}!)^{2}}.
\end{align}
To generalise, first define the combinatoric sum
\begin{equation}
    \sum_{a_{1}+...+a_{d}=k}\equiv\sum_{A_{d}^{k}}.
\end{equation}
In general then we can form the exponential generating function for $I_{0}(2x)^{d}$,
\begin{align}
    I_{0}(2x)^{d}&=\sum_{k=0}^{\infty}\sum_{A_{d}^{k}}\frac{x^{2k}}{\prod_{i}^{d}(a_{i}!)^{2}} \nonumber \\
    &= \sum_{k=0}^{\infty}\sum_{A_{k}^{d}}\frac{(2k)!}{\prod_{i}^{d}(a_{i}!)^{2}}\frac{x^{2k}}{(2k)!}.
\end{align}
Now recall the expression for $W_{d}^{(2k)}$ from equation (\ref{eq supp: d-Dim lattice multiplicy of origin}),
\begin{equation}
    W_{d}^{(2k)}=\sum_{A_{k}^{d}}\frac{(2k)!}{\prod_{i=1}^{d}(a_{i}!)^{2}}, 
\end{equation}
which gives us the final expression
\begin{equation}
    I_{0}(2x)^{d}=\sum_{k=0}^{\infty}W_{d}^{(2k)}\frac{x^{2k}}{(2k)!}.
\end{equation}

\subsection{Determining standard model parameters for a given \texorpdfstring{$\tau$}{} and \texorpdfstring{$\avg{m}_{\rm st}$}{}} \label{appendix: determining model parameters tau and m}
In this section we describe the algorithm we used to generate model parameters in the simulation. The challenge is to generate sets of non-negative $h_\pm, h'_\pm, v'_\pm$ leading to different stationary trait frequencies $\avg{\rho}_{\rm st}$ but at the same time keeping $\tau = 2(h_{+}'+h_{-}'+v_{+}'+v_{-}')/[(h_{+}+h_{-})-(h_{+}'+h_{-}')]$ fixed. Fixing $\avg{\rho}_{\rm st}$ is equivalent to fixing the stationary magnetisation $\avg{m}_{\rm st}=2\avg{\rho}_{\rm st}-1$.

We apply the restriction in Eq.~(\ref{eq supp: gamma restriction}), which we write in the form
\begin{equation}
    h_{+}  = h_{+}'-h_{-}' + h_{-}.\label{eq supp: algo aux 1}
\end{equation}
Equations (\ref{eq supp: complete network steady-state magnetisation}) and (\ref{eq supp: tau}) can be solved for $v_{+}'$ and $v_{-}'$, assuming $\tau \neq 0$
\begin{subequations}
\begin{gather}
    v_{+}' = \frac{\tau}{2}(1+\avg{m}_{\rm st})(h_{-}-h_{-}')-h_{+}', \label{eq supp: algo aux 2} \\
    v_{-}' = \frac{\tau}{2}(1-\avg{m}_{\rm st})(h_{-}-h_{-}')-h_{-}'. \label{eq supp: algo aux 3}
\end{gather}
\end{subequations}

\subsubsection{The case \texorpdfstring{$\tau>0$}{}}
For $\tau>0$ and $-1<\avg{m}_{\rm st}<1$, we set $h_{+}'$ and $h_{-}'$ to small random values and then choose a value of $h_{-}$ large enough so that $v_{+}'$, $v_{-}'$ and $h_{+}$ are non-negative. This amounts to choosing $h_{-}$ such that 
\be
    h_{-} \geq \text{max}\left(0, h_{-}'-h_{+}', h_{-}'+\frac{2}{\tau(1+\avg{m}_{\rm st})}h_{+}', \left[1+\frac{2}{\tau(1-\avg{m}_{\rm st})}\right]h_{-}'\right).
\ee

For the case $\tau > 0$ and $\avg{m}_{\rm st} = - 1$, Eq.~(\ref{eq supp: algo aux 2}) means that we need to set $v_{+}'=h_{+}'=0$. Eq.~(\ref{eq supp: algo aux 3}) then gives $v_{-}'=\tau h_{-}-(1+\tau)h_{-}'$. So to ensure $v_{-}'\geq 0$ we draw random values of $h_{-}$ and $h_{-}'$ such that $h_{-} \geq \left(1+\frac{1}{\tau}\right)h_{-}'$. This allows us to calculate $v_{-}'$. Finally $h_{+}$ is determined from Eq.~(\ref{eq supp: algo aux 1}).

For the case $\tau > 0$ and $\avg{m}_{\rm st} = + 1$, Eq.~(\ref{eq supp: algo aux 3}) means that we should set $v_{-}'=h_{-}'=0$. Eq.~(\ref{eq supp: algo aux 2}) then gives $v_{+}'=\tau h_{-}-h_{+}'$. So to ensure $v_{-}'\geq 0$ we draw random values of $h_{-}$ and $h_{+}'$ such that $h_{-} \geq \tau h_{+}'$. This allows us to calculate $v_{+}'$. The coefficient $h_{+}$ is again determined from Eq.~(\ref{eq supp: algo aux 1}).

\subsubsection{The case \texorpdfstring{$\tau=0$}{} (standard voter model)}
For $\tau=0$ Eqs.~(\ref{eq supp: algo aux 2}) and (\ref{eq supp: algo aux 3}) dictate that we need to set the rates of all unfaithful events to zero ($h_{+}'= h_{-}'=v_{+}'=v_{-}'=0$). We then have $h_{+}=h_{-}$ from Eq.~(\ref{eq supp: algo aux 1}). The model therefore reduces to the standard voter model. The common value of the rates $h_\pm$ can be drawn at random (the only significance of the value it to set a time scale). We note however that in this case the expression for $\avg{m}_{\rm st}$, Eq.~(\ref{eq supp: complete network steady-state magnetisation}), is not defined. In the voter model average magnetisation is conserved in time, and every initial magnetisation is therefore a steady-state value  ($\avg{m}_{\rm st }=\avg{m}_{0}$).

\subsubsection{Negative values of \texorpdfstring{$\tau$ ($\tau\leq -2$)}{}}
 As discussed in Sec.~\ref{appendix: H and tau} the quantity $\tau$ cannot take values in the interval $-2<\tau<0$, that is, the only negative values of $\tau$ permitted by the model setup are those in the range $\tau\leq -2$. We can confirm this through numerical search for model parameters. More precisely, we conducted searches of parameter space, enforcing the constrain in Eq.~(\ref{eq supp: gamma restriction}) and fixing $\avg{m}_{\rm st}$ and $\tau$, in Eqs.~(\ref{eq supp: complete network steady-state magnetisation}) and (\ref{eq supp: tau}) respectively. For $-2<\tau<0$ the search returns no solutions (see the Mathematica notebook at the following GitHub repository \cite{github} for a demonstration).

\begin{figure}[htbp]
    \centering
    \includegraphics[scale = 0.9]{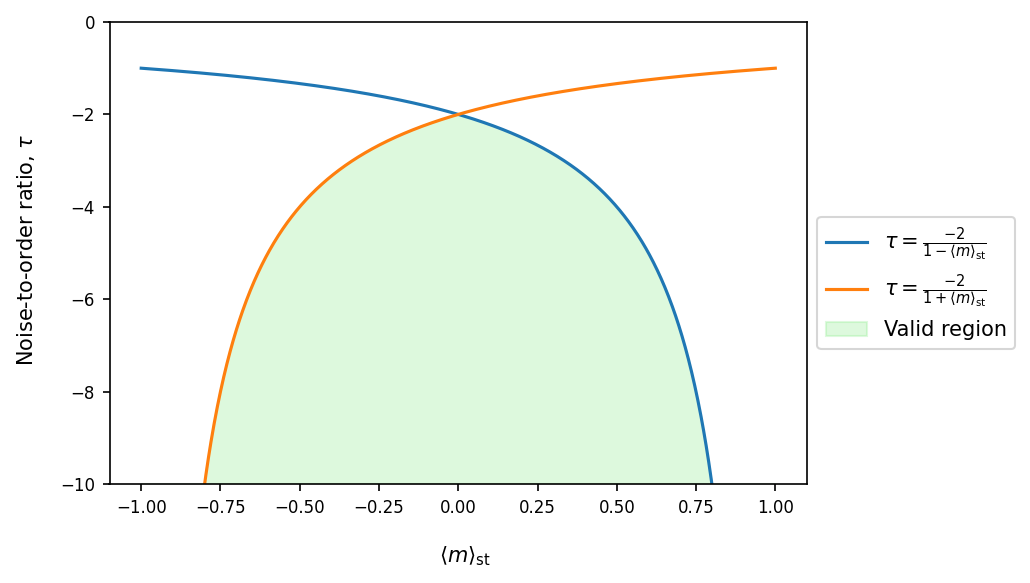}
    \caption{Diagram showing the values of $\tau\leq -2$ and $\avg{m}_{\rm st}$ for which it is possible to generate model parameters. The boundaries are given by Eq.~(\ref{eq supp: algo tau bounds}).}
    \label{fig supp: tau bounds}
    \end{figure}

Focusing on $\tau\leq -2$, a simple method to generate model parameters is to set $h_{+}'=h_{-}'$ and then draw this number at random. Consequently, from Eq.~(\ref{eq supp: algo aux 1}) we have $h_{+}=h_{-}$. Then, to ensure $v_{+}'\geq$ and $v_{-}\geq 0$, Eqs.~(\ref{eq supp: algo aux 2}) and (\ref{eq supp: algo aux 3}) give the following bounds on $h_{-}$:
\be
    0 \leq h_{-} \leq \text{min}\left(1+\frac{2}{\tau(1+\avg{m}_{\rm st})}, 1+\frac{2}{\tau(1-\avg{m}_{\rm st})}\right)h_{-}'.
\ee
To ensure the upper bound here is at least zero we require
\be
    \tau \leq \text{min}\left(\frac{-2}{1+\avg{m}_{\rm st}}, \frac{-2}{1-\avg{m}_{\rm st}}\right). \label{eq supp: algo tau bounds}
\ee
This means that for a given $\tau\leq -2$, $\avg{m}_{\rm st}$ must be within the following bounds.
\be
    -1-\frac{2}{\tau} \leq \avg{m}_{\rm st} \leq 1+\frac{2}{\tau}. \label{eq supp: negative tau bounds}
\ee
A visual representation of these bounds is shown in Fig.~\ref{fig supp: tau bounds}. 

The inequalities in Eq.~(\ref{eq supp: negative tau bounds}) bound the possible values of $\avg{m}_{\rm st}$ that can be reached for a given $\tau$ by model parameters {\em generated using our algorithm}. In particular we used the restriction $h_{+}=h_{-}$ to derive these bounds. We can therefore not exclude that models with  $h_{+}\neq h_{-}$ can cover a wider range of $\avg{m}_{\rm st}$ for a given $\tau$. However, numerical searches (allowing for $h_+\neq h_-$) find that the property from Eq.~(\ref{eq supp: negative tau bounds}) seems to continue to hold \cite{github}.

We here provide further intuitive support for these bounds. From Sec.~\ref{appendix: H and tau} we know that $\tau\to -2^{-}$ corresponds to increasing amounts of anti-correlation between the spin states of nearest-neighbour nodes. This requires an increasingly equal proportion of up and down spin nodes [see Fig.~\ref{fig supp: H grid}(e)]. At $\tau=-2$ this requires $\avg{m}_{\rm st} =0$, and as $\tau$ becomes more negative from there the range of possible $\avg{m}_{\rm st}$ values increases.

\end{document}